%% file: main.tex
\documentclass[aps,prx,twocolumn,notitlepage,showkeys,showpacs,superscriptaddress,preprintnumbers]{revtex4-2}
\usepackage{epsf}
\usepackage{here,times}
\usepackage[colorlinks=true,citecolor=blue,linkcolor=blue]{hyperref}
\usepackage{slashed}
\usepackage{amsthm}
\usepackage{thmtools}
\usepackage{blindtext}
\usepackage{physics, bm}
\usepackage{mathrsfs}
\usepackage{amsfonts, mathtools, amsmath,amssymb,dsfont}
\usepackage{esvect}
\usepackage{enumerate,dcolumn,multirow,bbold, enumitem}
\usepackage{longtable}
\usepackage{array}
\usepackage{tikz-cd}
\usepackage{bbding}
\usepackage[normalem]{ulem}
\usepackage{mathrsfs}
\usepackage{calligra}

\newcommand{\ii}{\mathrm{i}}

\newcommand{\mZ}{\mathbb{Z}}
\newcommand{\bk}{{\bm{k}}}
\newcommand{\bR}{\bm{R}}
\newcommand{\kstar}{\bm{k}_{\star}}

\newcommand{\sgn}{\text{sgn}}
\newcommand{\calT}{\mathcal{T}}
\newcommand{\calM}{\mathcal{M}}
\newcommand{\calA}{\mathcal{A}}

\newcommand{\calC}{\mathcal{C}}

\newcommand{\clb}{\color{blue!75!black!}}

\newcommand{\calD}{\mathcal{D}}

\newcommand{\calGk}{\mathcal{G}_{\bm{k}}}
\newcommand{\calG}{\mathcal{G}}

\newcommand{\scrD}{\mathscr{D}}

\newcommand{\UBdG}{U^{\text{BdG}}}

\begin{document}
\title{Fermi-surface diagnosis for topological superconductivity with  $s$-wave-like pairing symmetries}
\author{Zhongyi Zhang}
\affiliation{Department of Physics, Hong Kong University of Science and Technology, Clear Water Bay, Hong Kong, China}

\author{Ken Shiozaki}
\affiliation{Center for Gravitational Physics and Quantum Information, Yukawa Institute for Theoretical Physics, Kyoto University, Kyoto 606-8502, Japan}

\author{Chen Fang}
\affiliation{Beijing National Laboratory for Condensed Matter Physics, and Institute of Physics, Chinese Academy of Sciences, Beijing 100190, China}
\affiliation{Kavli Institute for Theoretical Sciences, Chinese Academy of Sciences, Beijing 100190, China}

\author{Seishiro Ono}
\thanks{Corresponding authors:~\href{mailto:seishiro.ono@riken.jp}{seishiro.ono@riken.jp}}
\affiliation{RIKEN Center for Interdisciplinary Theoretical and Mathematical Sciences (iTHEMS), RIKEN, Wako 351-0198, Japan}

\preprint{RIKEN-iTHEMS-Report-24, YITP-24-67}

\begin{abstract}
Theoretical prediction of topological superconductivity is key to their discovery.
Recently, it is proved that in 199 out of 230 space groups, topological superconductivity coexists with an $s$-wave-like pairing symmetry, raising the hope of finding more candidates for this exotic phase.
However, a comprehensive and efficient method for diagnosing topological superconductivity in realistic materials remains elusive.
Here, we derive Fermi-surface formulas for gapped and gapless topological phases of time-reversal symmetric superconductors with $s$-wave-like pairing symmetries in all layer and space groups, applicable to thin-film and bulk materials. 
Our diagnosis uses only the sign of the pairing and the Fermi velocity at several Fermi points, and yields complete (partial) diagnosis for gapped topological superconductivity in 159 (40) out of the 199 space groups.
This provides a fundamental basis for the first-principles prediction of new topological superconductors.
\end{abstract}
\maketitle

\section{Introduction}
Over the past decades, the discovery of materials exhibiting bulk topological superconductivity has been one of the central issues in condensed matter physics~\cite{kitaev2001unpaired, RevModPhys.80.1083, RevModPhys.83.1057, RevModPhys.80.1083, alicea2012new,sato2017topological}.
Thanks to intensive studies in recent years, it has been shown that a wide variety of topological superconducting phases exist~\cite{das2012zero,machida2019zero,doi:10.1126/science.1259327,zhang2018observation, Morimoto-Furusaki2013,PhysRevX.10.041014,PhysRevB.93.224505,PhysRevB.90.165114,Shiozaki-Sato-Gomi2016,Cornfeld-Chapman,PhysRevB.97.205135,PhysRevB.93.115129,PhysRevX.9.011012,PhysRevB.99.075105,PhysRevResearch.3.013052,CC-Shiozaki,zhang2022symmetry,PhysRevB.106.L121108,PhysRevB.106.035105,zhang2024topological, Ono-Shiozaki2022,zhang2025double}.
Unconventional superconductors, such as $p$- and $d$-wave superconductors, serve as important platforms for realizing topological superconductivity~\cite{PhysRevB.79.214526,PhysRevB.81.220504,PhysRevLett.105.097001,PhysRevB.89.020509,PhysRevB.89.140504,PhysRevB.90.100509,PhysRevB.94.174502,PhysRevX.8.041026,ishizuka2020periodic,Jiao:2020aa, PhysRevX.8.011029, Ahn2019, PhysRevB.92.104511,doi:10.1126/science.abe7518,PhysRevX.12.031001, PhysRevB.105.104515}.
However, such non-$s$-wave superconductors are rare in nature.
As a consequence, we have not yet verified the existence of intrinsic topological superconductivity in real materials.

Recently, a class of time-reversal symmetric spinful superconductors has been extensively studied~\cite{PhysRevLett.102.187001,PhysRevLett.111.056402,Fe-based_TSC, Ono-Shiozaki-Watanabe2022, Ono-Shiozaki2023,PhysRevB.81.134515,PhysRevB.89.220504,PhysRevLett.108.036803,PhysRevB.93.174509,haim2019time,PhysRevB.90.045130,PhysRevB.90.054503,PhysRevLett.126.137001,PhysRevB.94.174508,PhysRevLett.111.116402}.
In these superconductors, the Cooper pairs are invariant under all spatial symmetry operations for a given space group. We refer to a time-reversal symmetric spinful superconductor with this property as ``superconductor with $s$-wave-like paring symmetry.'' (More precise definition is provided below.)
Notably, most ever-discovered superconductors fall into this class.
It is natural to ask how many topological phases exist in this class.
In fact, this question is answered by classification of gapped topological superconducting phases in all space groups~\cite{Ono-Shiozaki-Watanabe2022}.
It turns out that topological superconductivity is compatible with $s$-wave-like pairing symmetries in 199 out of the 230 space groups.
Importantly, all nontrivial gapped phases in this class exhibit topologically protected gapless states on boundaries.
More recently, Ref.~\cite{Ono-Shiozaki2023} has found topological invariants defined for Bogoliubov-de Gennes (BdG) Hamiltonians, which can characterize gapped and gapless superconductors in this class.
Despite these advances, it remains challenging to compute topological invariants for realistic materials following these definitions, as full BdG Hamiltonians are often unavailable for realistic superconductors.

\begin{figure}[b]
	\begin{center}
		\includegraphics[width=0.99\columnwidth]{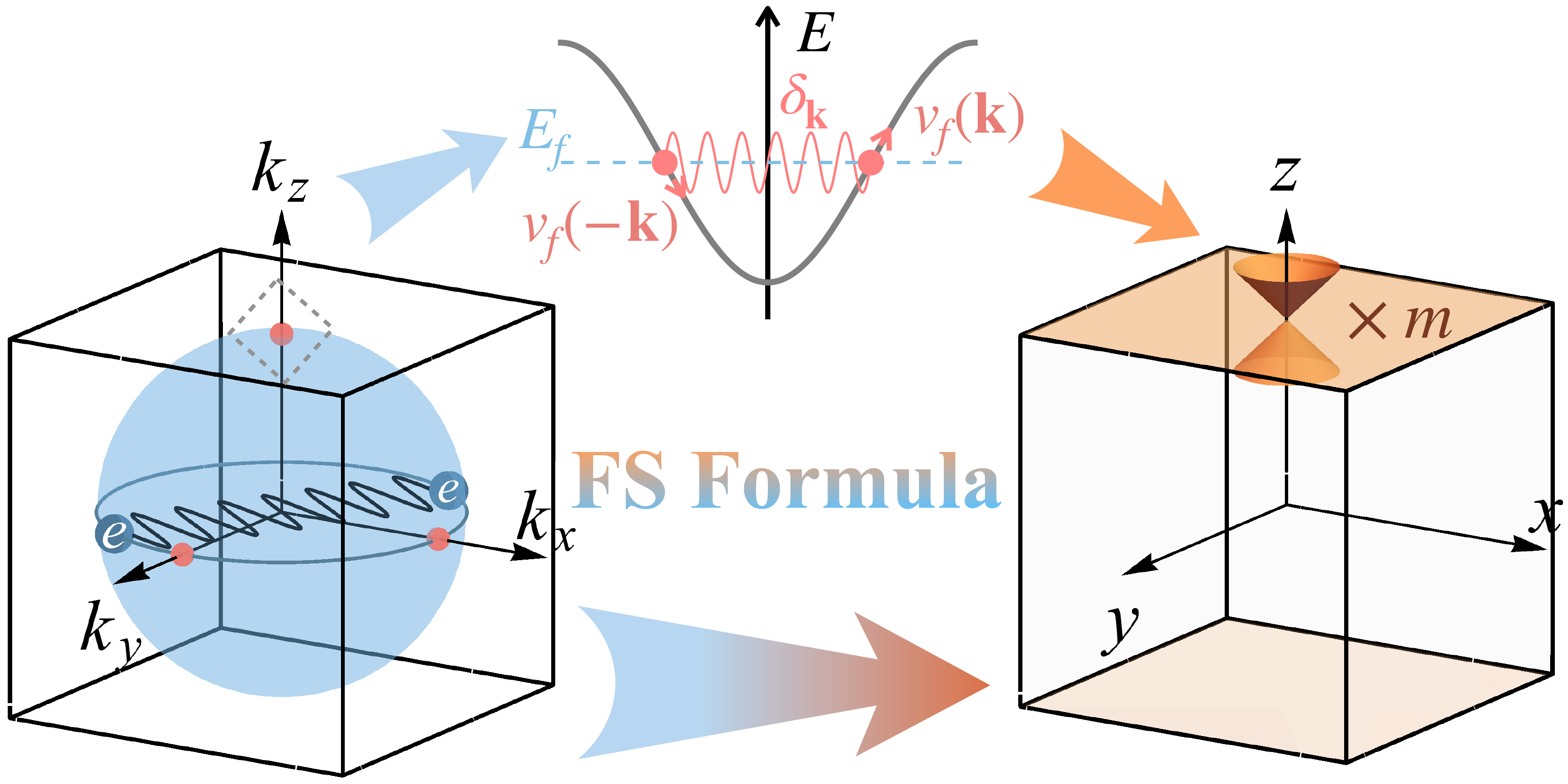}
		\caption{\label{fig:intro}
			Illustration of our Fermi-surface formulas.
			On the left-hand side, the red dots represent intersections of the Fermi surface with certain line segments in the Brillouin zone.
			All we need for the formulas are the signs of Fermi velocities $v_f$ and intraband pairings $\delta_{\bk}$ at the intersections.
			Our formulas enable us to compute topological invariants only from the information at the intersections.
			Nonzero values of the invariants indicate the existence of some gapless states on boundaries, as illustrated by the example on the right-side~\cite{PhysRevB.103.184502, PhysRevB.106.L121108}.
		}
	\end{center}
\end{figure}

The theories of symmetry indicators and topological quantum chemistry have been introduced to efficiently detect nontrivial topology in insulators and semimetals~\cite{SI_NC_Po, TQC, PhysRevX.7.041069, zhang2019catalogue, catalogue2,vergniory2019complete}.
Moreover, symmetry‐indicator theory has been generalized to superconductors~\cite{SI_TSC_Skurativska, Ono-Po-Watanabe2020, Ono-Po-Shiozaki2021, SI_TSC_Luka}, enabling the diagnosis of topology in realistic superconductors using only irreducible representations of normal-state wave functions and pairing symmetries under certain assumptions~\cite{catalogue_sc}.
However, this approach fails for $s$-wave–like pairings:~all topologically nontrivial phases with $s$-wave-like pairing symmetries cannot be detected by existing symmetry indicators.
Thus, an efficient diagnostic framework for the topology with $s$-wave-like pairing symmetries remains elusive.

In this work, we present Fermi-surface formulas of topological invariants in all layer and space groups, applicable to both thin-film and bulk materials.
Importantly, all our formulas require only signs of the Fermi velocities and intraband pairings on line segments connecting two high-symmetry points---no information about the entire Fermi surfaces is needed (See Fig.~\ref{fig:intro} for an illustration).
In addition to the efficiency, our formulas can detect a wide range of nontrivial topological phases---including both gapped and gapless---with $s$-wave-like pairing symmetries.
For gapped phases, our formulas provide a complete diagnosis in 159 space groups and partial coverage in the remaining 40.
Moreover, for nodal superconductors, Fermi-surface formulas exist in all space groups except for $P1$ and $P\bar{1}$.

\section{Result}
\subsection{Fermi surface formulas}
In this work, we focus on $s$-wave-like pairing symmetries, formally defined by the symmetry properties of the order parameter $\Delta_{\bk}$.
Let $U_{\bk}(g)$ be a representation of a space group symmetry $g$.
We say that the pairing symmetry is $s$-wave-like if $U_{\bk}(g)\Delta_{\bk}U_{-\bk}^{\top}(g) = \Delta_{g\bk}$ for all symmetries in a given space group.
It should be noted that the $s$-wave-like pairing symmetry does not necessarily mean that the system is a conventional $s$-wave superconductor.
For example, consider a single orbital superconducting order $\Delta_{\bk} = \ii \sin k_x \sigma_y$ under the mirror symmetry along $z$-axis, where the Pauli matrices $\sigma_{i=x,y,z}$ represent the spin degree of freedom.
While this is $p$-wave like, the pairing symmetry is $s$-wave like, as one can see $\ii\sigma_z\Delta_{\bk} (\ii\sigma_z)^\top=\Delta_{\bk}$.
Another representative example is $s_{\pm}$-wave pairing.

Topological invariants are often defined on subregions in momentum space.
For convenience, we decompose the Brillouin zone into points, line segments, and polygons in a symmetric manner~\cite{G-CWcomplex,Shiozaki-Ono2023}.
See Fig.~\ref{fig:example1}(b) for an illustration of the decomposition.
We assign an orientation to each component.
For example, line segment $a$ in Fig.~\ref{fig:example1}(b) is oriented from $\Gamma$ to $\text{X}$.
In this work, we consider topological invariants defined on the line segments.

Before presenting our main results on Fermi surfaces of the topological invariants, we introduce our basic assumptions on target systems.
We always assume that superconductors are in the weak coupling regime.
More precisely, we assume that the target superconducting system can be continuously deformed into a superconductor that possesses the following properties:
\begin{enumerate}
	\setlength{\itemsep}{-2pt}
	\item[(i)] intraband pairings dominate, i.e. all interband pairings are negligible;
	\item[(ii)] the superconducting gap is negligible except on Fermi surfaces.
\end{enumerate}
In addition, we further assume that the following conditions are satisfied:
\begin{enumerate}
	\setlength{\itemsep}{-2pt}
	\item[(iii)] all Fermi surfaces are minimally degenerate as allowed by symmetries;
	\item[(iv)] normal state energies at high-symmetry points are not at the Fermi level;
	\item[(v)] there is no superconducting node on all line segments including their endpoints.
\end{enumerate}
The assumptions (iii) and (iv) usually hold for realistic systems.
The assumption (v) must be satisfied to define topological invariants on line segments.

Under these five assumptions, intraband pairing potentials and Fermi velocities can determine the values of topological invariants defined on the line segments.
To define intraband pairing potentials and Fermi velocities, let $\Phi_{\bk}^{\alpha}$ be a matrix whose columns are eigenvectors of the normal conducting Hamiltonian $h_{\bk}$, where $\alpha$ is a label of an irreducible representation (irrep) $u^{\alpha}_{\bk}(g)$.
The matrix $\Phi_{\bk}^{\alpha}$ satisfies the relations $h_{\bk}\Phi_{\bk}^{\alpha} = \varepsilon_{\bk, \alpha}\Phi_{\bk}^{\alpha}\ (\varepsilon_{\bk, \alpha} \in \mathbb{R})$ and $U_{\bk}(g)\Phi_{\bk}^{\alpha} = \Phi_{\bk}^{\alpha}u^{\alpha}_{\bk}(g)$.
Then, the intraband pairing potential is defined by a projected superconducting order
\begin{align}
	\tilde{\Delta}_{\bk}^{\alpha} \coloneqq [\Phi_{\bk}^{\alpha}]^{\dagger}\Delta_{\bk}[U(\calT)]^*\Phi_{\bk}^{\alpha}.
\end{align}
Here, $U(\calT)$ is a unitary representation of time-reversal symmetry (TRS) satisfying $U(\calT)[U(\calT)]^*=-\mathds{1}$, where $\mathds{1}$ is the identity matrix.
The above intraband pairing satisfies the relations
\begin{align}
	&[\tilde{\Delta}^{\alpha}_{\bk}]^{\dagger} = \tilde{\Delta}^{\alpha}_{\bk};\ \ u_{\bk}^{\alpha}(g)\tilde{\Delta}^{\alpha}_{\bk}[u_{\bk}^{\alpha}(g)]^{\dagger} = \tilde{\Delta}^{\alpha}_{\bk}.
\end{align}
Since $u_{\bk}^{\alpha}(g)$ is irreducible, the intraband pairing is proportional to the identity matrix, namely, 
\begin{align}
	\label{eq:effective_potential}
	\tilde{\Delta}_{\bk}^{\alpha} &= \delta_{\bk}^{\alpha} \mathds{1}\ \ \ (\delta_{\bk}^{\alpha}\in \mathbb{R}).
\end{align}

The Fermi velocity is defined by the directional derivative of the energy. 
Suppose that we consider a line segment $l$ connecting from $\bk_0$ to $\bk_1$ and that there is a Fermi point $\bk_\star$ on the line. 
The Fermi velocity of the energy $\varepsilon_{\bk, \alpha}$ is given by
\begin{align}
	\label{eq:Fermi_velocity}
	v^{(l, \alpha)} \coloneqq \bm{\nabla}_{\bk}\varepsilon_{\bk, \alpha}\vert_{\bk = \kstar} \cdot \hat{\bm{e}}_l,
\end{align}
where $\hat{\bm{e}}_l = (\bk_1 - \bk_0)/\|\bk_1 - \bk_0\|$.

For each of topological invariants and space groups, the line segments and irreps considered in expressions of topological invariants are different. 
For a given momentum-space decomposition in a space group, Ref.~\cite{Ono-Shiozaki2023} proposes a method to identify a set of line segments and irreps involved in topological invariants. 
This information is encoded in an integer-valued tuple denoted by
\begin{align}
	\label{eq:bm_x}
	\bm{x}_{i} = \left(x_{i,(l_1, \alpha_1)}, x_{i,(l_1, \alpha_2)}, \cdots, x_{i,(l_2, \beta_1)}, \cdots \right),
\end{align}
where $x_{i, (l, \alpha)}$ is a contribution weight of irrep $\alpha$ on the line segment $l$ for the $i$-th topological invariant.

Now, we are ready to provide general forms of Fermi-surface formulas of topological invariants. If the $i$-th topological invariant $\mathcal{W}_i$ is $\mZ$-valued, the Fermi-surface formula is given by
\begin{align}
	\label{eq:FS_Z}
	\mathcal{W}{_i} &= \sum_{l, \alpha}x_{i,(l, \alpha)}\sum_{m=1}^{n_{(l, \alpha)}}\text{sgn}(v^{(l, \alpha)}_{m})\frac{1-\text{sgn}(\delta^{(l, \alpha)}_{m})}{2},
\end{align}
where $n_{(l, \alpha)}$ is the number of minimally degenerate Fermi points with irrep $\alpha$ on the line segment $l$, $v^{(l, \alpha)}_{m}$ denotes the Fermi velocity \eqref{eq:Fermi_velocity} of the $m$-th Fermi point with irrep $\alpha$, and $\delta^{(l, \alpha)}_{m}$ is the coefficient in Eq.~\eqref{eq:effective_potential} defined for the eigenvectors at the $m$-th Fermi point on the line segment $l$.

For a given $\bm{x}_i$ corresponding to a $\mZ_\lambda$-valued invariant $\mathcal{X}_i$, our Fermi-surface formula is
\begin{align}
    \label{eq:FS_Zn}
    &\mathcal{X}{_i} = \sum_{l, \alpha}x_{i,(l, \alpha)}\sum_{m=1}^{n_{(l, \alpha)}}\text{sgn}(v^{(l, \alpha)}_{m})\frac{1-\text{sgn}(\delta^{(l, \alpha)}_{m})}{2}\  \text{ mod } \lambda.
\end{align}
In particular, when $\lambda=2$, we can drop $\text{sgn}(v_{m}^{(l, \alpha)})$ since $n = -n$ mod $2$ for an integer $n$.
As a result, the formulas can be further simplified to
\begin{align}
    \label{eq:FS_Z2}
    &\mathcal{X}{_i} = \sum_{l, \alpha}x_{i,(l, \alpha)}\sum_{m=1}^{n_{(l, \alpha)}}\frac{1-\text{sgn}(\delta^{(l, \alpha)}_{m})}{2} \  \text{ mod } 2.
\end{align}

These formulas immediately give us an insight.
The quantities \eqref{eq:FS_Z}-\eqref{eq:FS_Z2} are always trivial for superconductors whose intraband pairings have the same sign everywhere.
This means that superconductors in the BCS limits, which can be connected to vacuum without closing gap, cannot have nontrivial values of Eqs.~\eqref{eq:FS_Z}-\eqref{eq:FS_Z2}.

We make three remarks on our formulas.
First, several topological invariants can be defined for a given space group, as indicated by the subscripts in Eqs.~(6)-(8).
For example, in space group $P4_1$, there exist a $\mathbb{Z}_2$-valued, a $\mathbb{Z}_4$-valued, and a $\mathbb{Z}_8$-valued invariants, which are denoted by $\mathcal{X}_1$, $\mathcal{X}_2$, and $\mathcal{X}_3$, respectively.
Second, our Fermi-surface formulas are well-defined for a given decomposition, including the orientations of line segments, of the Brillouin zone in a space group.
In Supplemental Materials, we provide our formulas together with all line segments we used for all layer groups and space groups.
If we adopt a different decomposition, we may have physically equivalent but different forms of the formulas.
This is because changing the decomposition can change the elements of $\bm{x}$~\eqref{eq:bm_x} in general.
Last, among $\mZ$-valued invariants, there exist invariants that detect gapless points on two-dimensional subregions but not on the line segments.
Topological invariants for gapped phases are meaningful only when these invariants for gapless points are all trivial.
We distinguish gapless topological invariants from gapped ones by using $\mathcal{W}^{\text{gapless}}_i$ and $\mathcal{W}^{\text{gapped}}_i$ (See Supplemental~Materials~II for more details).

\subsection{Examples}
\label{sec:example}
Here, we demonstrate how it works through three theoretical models constructed by real-space construction method and one realistic model from real material. 
\begin{figure}[t]
	\begin{center}
		\includegraphics[width=0.99\columnwidth]{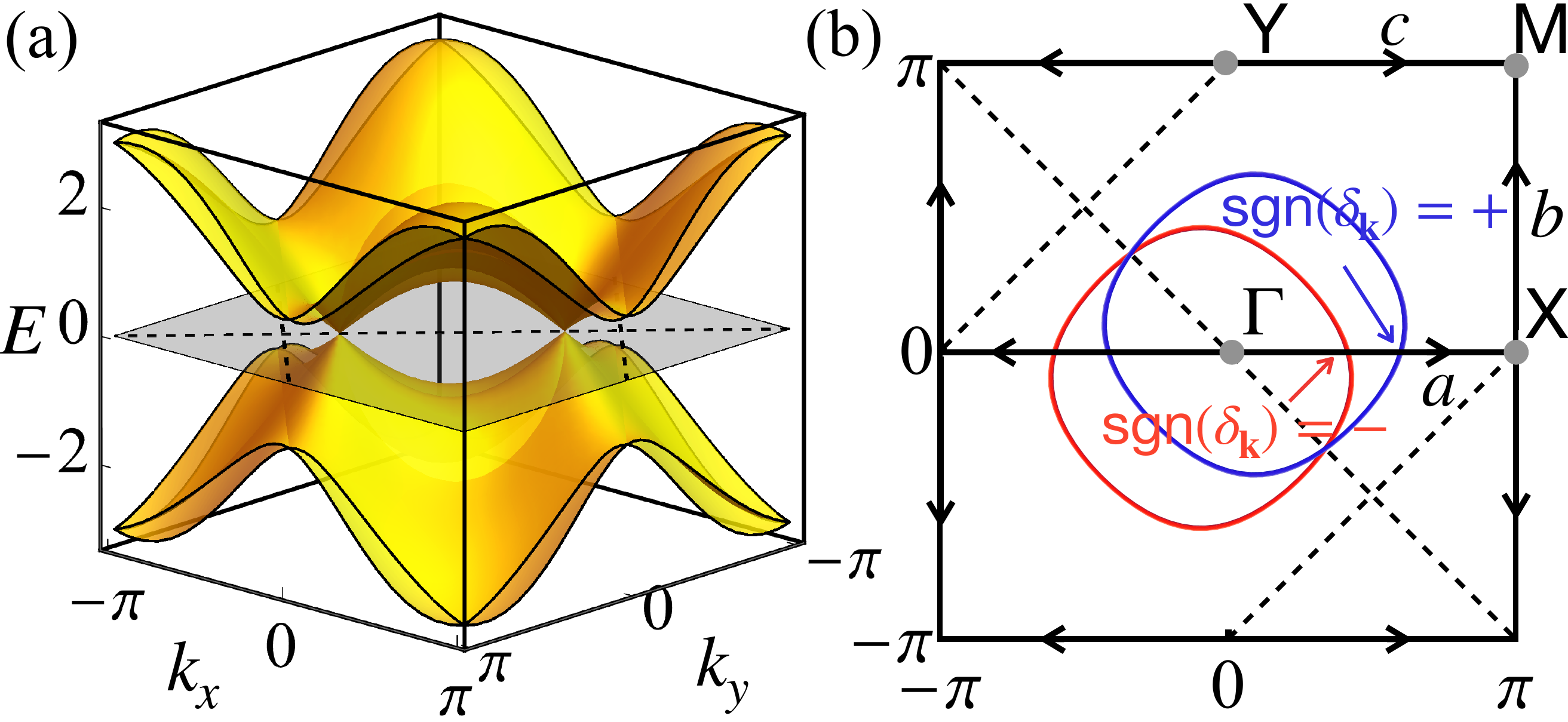}
		\caption{\label{fig:example1}
			Illustration of the model~\eqref{eq:example1normal} and \eqref{eq:example1bdg}.
			(a) shows the energy dispersion of BdG Hamiltonian~\eqref{eq:example1bdg}.
			(b) shows decomposition of the Brillouin zone, Fermi surfaces (colored solid lines), and the pairing nodes (dashed lines).
			The solid black lines and arrows denote the line segments specified by the gray endpoints and their orientations.
			The red and blue colors in the figure indicate Fermi surfaces with mirror eigenvalues $+\ii$ and $-\ii$, respectively.
			The intersections of the Fermi surface and the pairing nodes correspond to the gapless points in (a).
			The parameters are set to be  $\{t,\lambda,\mu,\Delta\}=\{-1,0.3,-1,0.3\} $.
		}
	\end{center}
\end{figure}

First, we consider a quasi-two-dimensional gapless superconductor in layer group $p11m$.
This layer group is generated by the mirror symmetry $\mathcal{M}_z=\{M_z|(000)\}$ and translation symmetries along $x$- and $y$-direction.
The decomposition of the Brillouin zone is shown in Fig.~\ref{fig:example1}(b).
The normal state Hamiltonian for the superconductor is
\begin{equation}\label{eq:example1normal}
	h_{\bk}=t(\cos k_x+\cos k_y) s_0+\lambda(\sin k_x+\sin k_y) s_3,
\end{equation}
and the corresponding BdG Hamiltonian takes the form
\begin{equation}\label{eq:example1bdg}
	H^{\text{BdG}}_{\bk}=[h_{\bk}-\mu]\kappa_3+\Delta_{\text{sc}}(\sin k_x+\sin k_y) s_1\kappa_1,
\end{equation}
where the Pauli matrices $s_j$ and $\kappa_j$ $(j=0,1,2,3)$ stand for spins and the Nambu spinor.
In addition to TRS and mirror symmetry $\calM_z$, the BdG Hamiltonian inherently possesses particle-hole symmetry $\calC$.
For the above BdG Hamiltonian, their representations are given by
\begin{equation}
	U^{\text{BdG}}(\mathcal{T})=\ii s_2,\ U^{\text{BdG}}(\mathcal{C})=\kappa_1,\ U^{\text{BdG}}(\mathcal{M}_z)=\ii \kappa_3 s_3.
\end{equation}
The pairing term $\Delta_{\bk}=\Delta_{\text{sc}}(\sin k_x+\sin k_y) s_1$ remains invariant under mirror symmetry $\mathcal{M}_z$: $U_{\bk}(\mathcal{M}_z) \Delta_{\bk}U^{\top}_{\bk}(\mathcal{M}_z)= \Delta_{\mathcal{M}_z\bk}$, where $U_{\bk}(\mathcal{M}_z)$ is the representation of the mirror symmetry in the normal phase. Therefore, the pairing exhibits the $s$-wave-like pairing symmetry.
As shown in Fig.~\ref{fig:example1}(a), the energy spectrum of $H^{\text{BdG}}_{\bk}$~\eqref{eq:example1bdg} has two superconducting gapless points.

Our formula for $\mZ$-valued invariants can detect these gapless points.
Using Eq.~\eqref{eq:FS_Z}, we find the formula
\begin{align}
	\mathcal{W}^{\text{gapless}}_1 &= \sum_{\alpha=\pm} \alpha\ \left[\sum_{m=1}^{n_{(c, \alpha)}}\text{sgn}(v_{m}^{(c, \alpha)})\frac{1-\text{sgn}(\delta_{m}^{(c, \alpha)})}{2}\right.\nonumber\\
	&\quad\quad\quad \left. -\sum_{m=1}^{n_{(a, \alpha)}}\text{sgn}(v_{m}^{(a, \alpha)}) \frac{1-\text{sgn}(\delta_{m}^{(a, \alpha)})}{2}\right],
\end{align}
where $\alpha=\pm$ means the mirror eigenvalue $\pm \ii$ for eigenstates of the normal state Hamiltonian~\eqref{eq:example1normal}.
The signs of Fermi velocities $v^{(a,\pm)}$ on the line segment $a$ are positive.
According to the sign of $\delta_{\bk}$ shown in Fig.~\ref{fig:example1}(b), we have $\mathcal{W}^{\text{gapless}}=-1$.
In fact, the invariant is equivalent to the mirror winding number defined on a loop $(-\pi, 0)\rightarrow(\pi, 0)\rightarrow(\pi, \pi)\rightarrow(-\pi, \pi)\rightarrow(-\pi, 0)$.
The nontrivial mirror winding number indicates the existence of gapless points inside the loop~\cite{PhysRevLett.111.056403}.

\begin{figure}[t]
	\begin{center}
		\includegraphics[width=0.99\columnwidth]{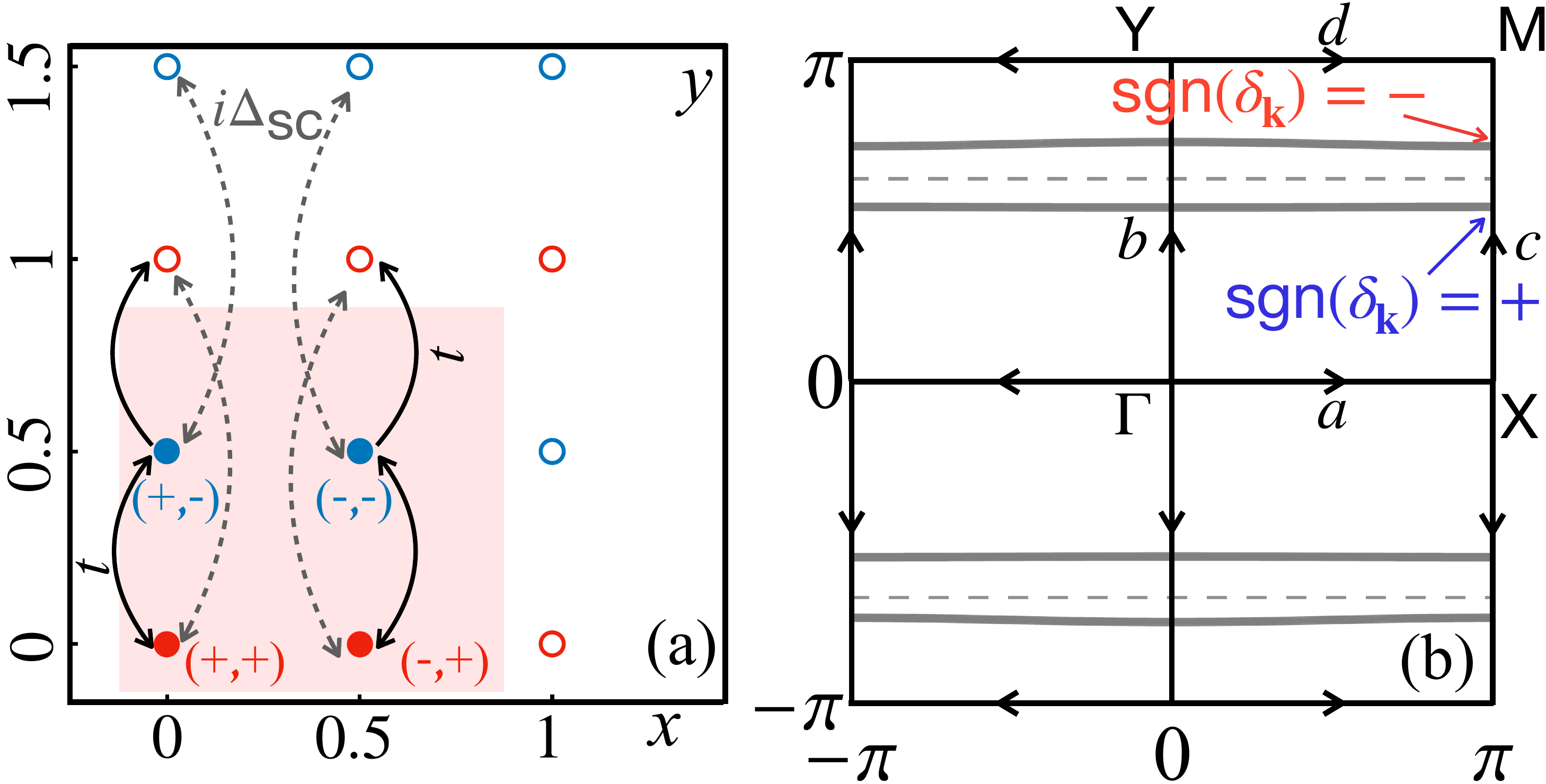}
		\caption{\label{fig:example2}
		Illustration of the model in layer group $p2_1/b11$.
		(a) Real-space description of the model. 
        The red shaded region indicates unit cell.
		The red and blue solid circles denote the four sublattices labeled by $(\tau_z,\sigma_z)=(\pm1,\pm1)$ in the unit cell.
		The solid (dashed) arrows represent the hopping between the sublattice degrees in the normal (pairing) part of BdG Hamiltonian~\eqref{eq:HBdGexample2}.
		(b) Fermi surfaces for the Hamiltonian~\eqref{eq:Hnormalexample2}.
		The gray solid (dashed) lines show the normal-state Fermi surfaces with (without) symmetry-allowed perturbations. 
		The red and blue colors mean states with glide eigenvalue $+\ii e^{-\ii k_y/2}$ and $-\ii e^{-\ii k_y/2}$, respectively.
		The parameters are set to be $\{t,\mu\}=\{-1,1\}$.
		}
	\end{center}
\end{figure}

Next, we consider a gapped topological superconductor (TSC) in layer group $p2_1/b11$ to show our formulas can be used even when the formula in Ref.~\cite{FS_TSC_Qi} cannot.
Generators of this layer group are glide symmetry $\mathcal{G}_x =\{M_x\vert (\frac{1}{2} \frac{1}{2} 0)\}$, inversion symmetry $\mathcal{I} =\{I\vert (00 0)\}$, and translation symmetry along $x$-direction.
In the presence of inversion symmetry and TRS, Fermi surfaces in spinful electronic systems must be at least twofold degenerate.
Furthermore, topological invariants for class DIII in the tenfold classification~\cite{PhysRevB.78.195125, Kitaev_bott, Ryu_2010} are also trivial for even-parity pairings. 
Therefore, this symmetry setting is out of the scope of Ref.~\cite{FS_TSC_Qi}.

Here, we describe our model in which there are four sublattice degrees of freedom labeled by $(\tau_z,\sigma_z)=(\pm1,\pm1)$.
Their coordinates in a unit cell are specified by $(x, y) = \frac{1}{4}[(2-\tau_z)(1, 0)+(1-\sigma_z)(0, 1)]$ [See Fig.~\ref{fig:example2}(a)].
According to Ref.~\cite{Ono-Shiozaki-Watanabe2022}, the classification of gapped topological phases is $\mZ_2$, whose generator is constructed by placing one-dimensional TSCs along $x=1/4$ and $x=3/4$ in each unit cell.
To realize this, we consider a model whose normal state Hamiltonian is given by
\begin{equation}\label{eq:Hnormalexample2}
	h_{\bk}=t(\Gamma_{010}+\cos k_y\Gamma_{010}+\sin k_y \Gamma_{020}),
\end{equation}
and the BdG Hamiltonian is
\begin{equation}\label{eq:HBdGexample2}
	H^{\text{BdG}}_{\bk}=[h_{\bk}-\mu]\kappa_3+2\Delta_{\text{sc}}\sin k_y\Gamma_{301}\kappa_1,
\end{equation}
where $\Gamma_{ijk} \coloneqq \tau_i\otimes \sigma_j\otimes s_k\ (i,j,k = 0,1,2,3)$.
As mentioned above, the minimal degeneracy of states is twofold at each momentum.
However, the Fermi surfaces of $h_{\bk}$ in Eq.~\eqref{eq:Hnormalexample2} are fourfold degenerate, which violates the assumption (iii), as indicated by the gray dashed lines in Fig.~\ref{fig:example2}(b).
This violation can be easily resolved by introducing symmetry-allowed perturbations that are small enough not to change any topology, as shown by the solid gray lines in Fig.~\ref{fig:example2}(b).

The $\mZ_2$ topology can be diagnosed by our formula for $\mZ_2$-valued invariant in Eq.~\eqref{eq:FS_Z2}.
The decomposition of the Brillouin zone is shown in Fig.~\ref{fig:example2}(b).
Then, the formula is given by
\begin{align}
    \mathcal{X}_1 &= \sum_{\alpha=\pm}\sum_{m=1}^{n_{(c,\alpha)}}\frac{1-\text{sgn}(\delta^{(c, \alpha)}_{n})}{2} +\sum_{\beta=\pm}\sum_{m=1}^{n_{(d,\beta)}}\frac{1-\text{sgn}(\delta^{(d, \beta)}_{n})}{2},
\end{align}
where $\alpha=\pm$ represents the glide eigenvalue $\pm \ii e^{-\ii k_y/2}$ on the line segment $c$, and $\beta=\pm$ denotes the screw symmetry $\mathcal{S}_x=\mathcal{G}_x\mathcal{I}$ eigenvalue $\pm \ii e^{-\ii k_x/2}$ on the line segment $d$.
According to the sign of $\delta_{\bk}$ shown in Fig.~\ref{fig:example2}(b), we have $\mathcal{X}=1$ mod $2$.

Last, we discuss a strong TSC in space group $P4_1$ to show that our formula can inform the three-dimensional winding number $w_{\text{3D}}$ modulo four.
This space group is generated by fourfold screw symmetry $\mathcal{S}_{4z}=\{C_{4z}|(00\frac{1}{4})\}$ and translation symmetries along $x$- and $y$-directions.

Our model is described as follows. 
There are four sublattice degrees of freedom in a unit cell, whose coordinates in a unit cell are $(x,y,z) = (0,0,0), (0,0,1/4), (0,0,1/2),$ and $(0,0,3/4)$, labeled as 1, 2, 3, and 4, respectively.
The normal state Hamiltonian is constructed as
\begin{align}\label{eq:examplenor3}
	\hat{\mathcal{H}}_{0}=&\sum\limits_{s ,\bR }\sum_{i=1}^4 -\frac{t}{2}(\hat{c}^\dagger_{si\bR+{\bm{a}}}\hat{c}_{si\bR}+\hat{c}^\dagger_{si\bR +{\bm{b}}}\hat{c}_{si\bR})+\mathrm{h.c.}\nonumber\\
	&+\sum\limits_{s ,\bR }-\frac{t}{2}(\hat{c}^\dagger_{s1\bR}\hat{c}_{s2\bR}+\hat{c}^\dagger_{s2\bR}\hat{c}_{s3\bR}+\hat{c}^\dagger_{s4\bR}\hat{c}_{s3\bR})+\mathrm{h.c.}\nonumber\\
	&+\sum\limits_{s ,\bR }-\frac{t}{2}\hat{c}^\dagger_{s4\bR+{\bm{c}}}\hat{c}_{s1\bR}+\mathrm{h.c.}, \sum_{s, \bR, i}\hat{c}^\dagger_{si\bR+{\bm{c}}}\hat{c}_{si\bR},
\end{align}
where the $\hat{c}_{si\bR}\ (\hat{c}^\dagger_{si\bR})$ is the annihilation (creation) operator of an electron at the $i$-th sublattice with spin $s$ in the unit cell at $\bR$.
Also, ${\bm{a}}, {\bm{b}}$, and ${\bm{c}}$ are the primitive lattice vectors along $x$-, $y$-, and $z$-direction, respectively.
The pairing term is given by
\begin{align}
\hat{\Delta}=&\sum\limits_{s ,\bR }-\frac{\ii}{2}\Delta_{\text{sc}}(\hat{c}^\dagger_{s1\bR}\hat{c}^\dagger_{\bar{s}2\bR}+\hat{c}^\dagger_{s4\bR}\hat{c}^\dagger_{\bar{s}3\bR}+\hat{c}^\dagger_{s2\bR}\hat{c}^\dagger_{\bar{s}3\bR})+\mathrm{h.c.}\nonumber\\
&+\sum\limits_{s ,\bR }\frac{\ii}{2}\Delta_{\text{sc}}\left(\hat{c}^\dagger_{s1\bR}\hat{c}^\dagger_{\bar{s}4\bR+{\bm{c}}}-\sum_{i=1}^4s\hat{c}^\dagger_{si\bR}\hat{c}^\dagger_{si\bR+{\bm{a}}}\right)+\mathrm{h.c.}\nonumber\\
&+\sum\limits_{s ,\bR }\sum_{i=1}^4\frac{1}{2}\Delta_{\text{sc}}\hat{c}^\dagger_{si\bR}\hat{c}^\dagger_{si\bR+{\bm{b}}}-\mathrm{h.c.},
\end{align}
with $s$ and $\bar{s}$ being the opposite spin.
It is easy to check the system described by $\hat{\mathcal{H}}_{\text{BdG}}=\hat{\mathcal{H}}_{0}+\hat{\Delta}$ has the nonzero three-dimensional winding number $w_{\text{3D}}=1$.

\begin{figure}[t]
	\begin{center}
		\includegraphics[width=0.95\columnwidth]{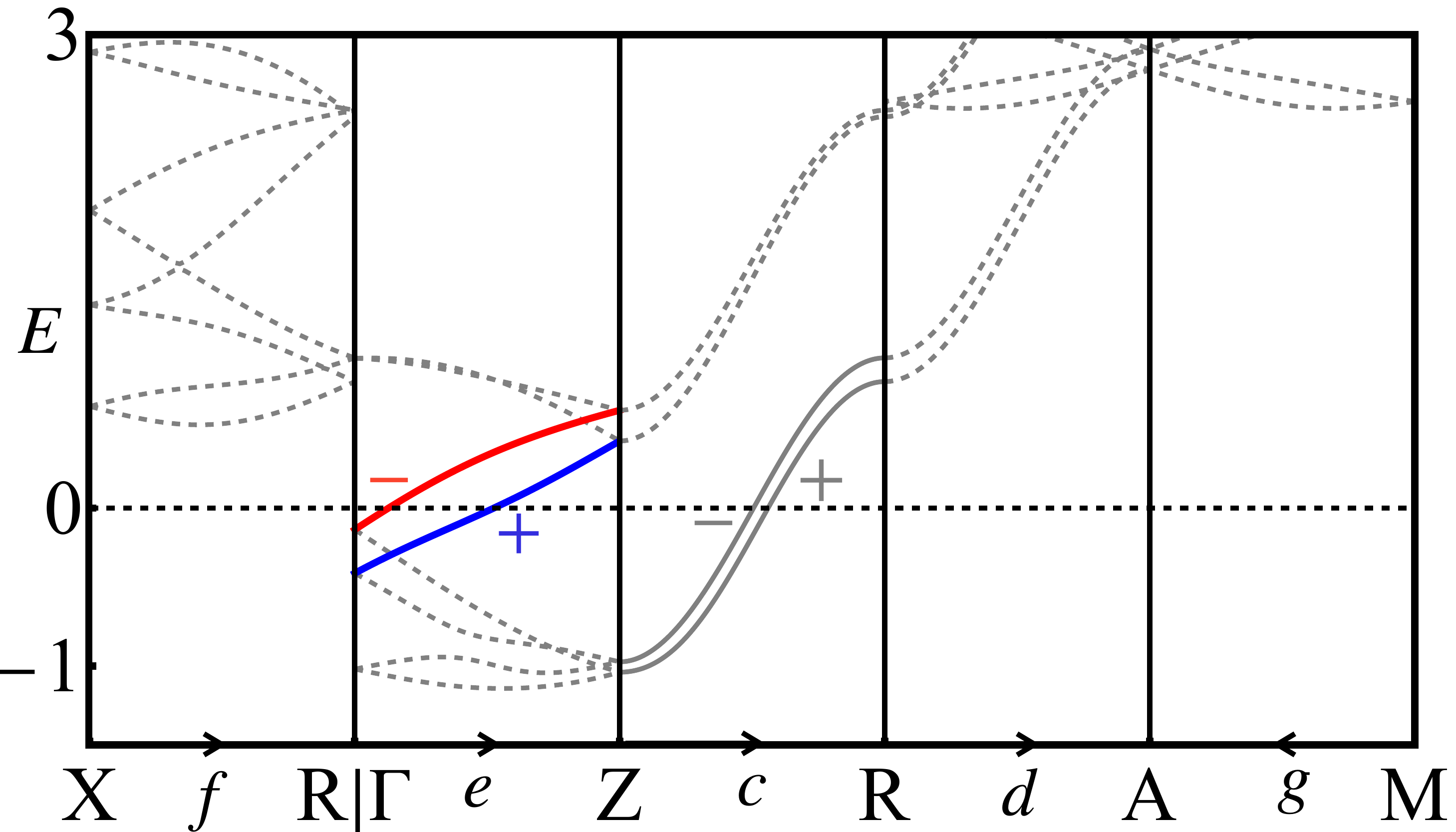}
		\caption{\label{fig:example3}
		The band structure of the model~\eqref{eq:examplenor3} with symmetry-allowed perturbations in space group $P4_1$.
		Solid lines (dashed lines) represent bands crossing (not crossing) the Fermi level.
		The arrows represent the positive direction $\hat{\bm{e}}_l$ of the line segments $c,d,e,f$, and $g$.
		The red and blue colors mean states with screw eigenvalue $e^{-i(3\pi+k_z)/4}$ and $e^{-i(\pi+k_z)/4}$, respectively.
		The $+$ and $-$ represent the sign of $\delta_{\bk}$ at the corresponding Fermi point.
		The parameters are set to be $\{t,\mu \}=\{1,1.8\}$.}
	\end{center}
\end{figure}

Here, we compute the $\mZ_8$-valued invariant by our formula~\eqref{eq:FS_Zn}:
\begin{align}
	\mathcal{X}_3&=\sum_{l=e,g}\sum_{\alpha=1,3,5,7}(4-\alpha)\sum_{m=1}^{n_{(l,\alpha)}}\sgn(v_m^{(l,\alpha)})\frac{1-\sgn(\delta_m^{(l,\alpha)})}{2}\nonumber\\
	&\quad +\sum_{l=c,d}\sum_{m=1}^{n_{l}}\sgn(v_{m}^{l})\left(1-\sgn(\delta_{m}^{l})\right) \nonumber \\
    &\quad + \sum_{\beta = \pm} \beta \sum_{m=1}^{n_{(f, \beta)}}\sgn(v_{m}^{(f, \beta)})\left(1-\sgn(\delta_{m}^{(f, \beta)})\right), 
\end{align}
where for line segment $l=e$ and $g$, $v_{m}^{(l, \alpha)}$ and $\delta_{m}^{(l, \alpha)}\ (\alpha=1,3,5,7)$ are defined for screw eigenvalue $e^{-\ii (\alpha \pi+k_z)/4}$;~for line segment $f$, $v_{m}^{(f, \beta)}$ and $\delta_{m}^{(l, \beta)}\ (\beta=\pm)$ are defined for twofold rotation eigenvalues $ \beta \ii$.
After adding symmetry allowed perturbations to lift the accidental degeneracy in $\hat{\mathcal{H}}_0$, the energy spectrum is as shown in Fig.~\ref{fig:example3}.
According to the sign of $\delta_{\bk}$ shown in Fig.~\ref{fig:example3}, we find $\mathcal{X}_3=3$.
In Sec.~\ref{sec:method_1}, we show that $\mathcal{X}_3$ mod $4$ serves as a $\mZ_4$-valued indicator of $w_{\text{3D}}$.

\subsection{Application to realistic materials}
\label{sec:material}
We discuss CaFeAs$_2$, whose space group is $P2_1$, to demonstrate how to apply our formula to the realistic materials based on first-principles calculations.
\begin{figure}[b]
	\begin{center}
		\includegraphics[width=0.9\columnwidth]{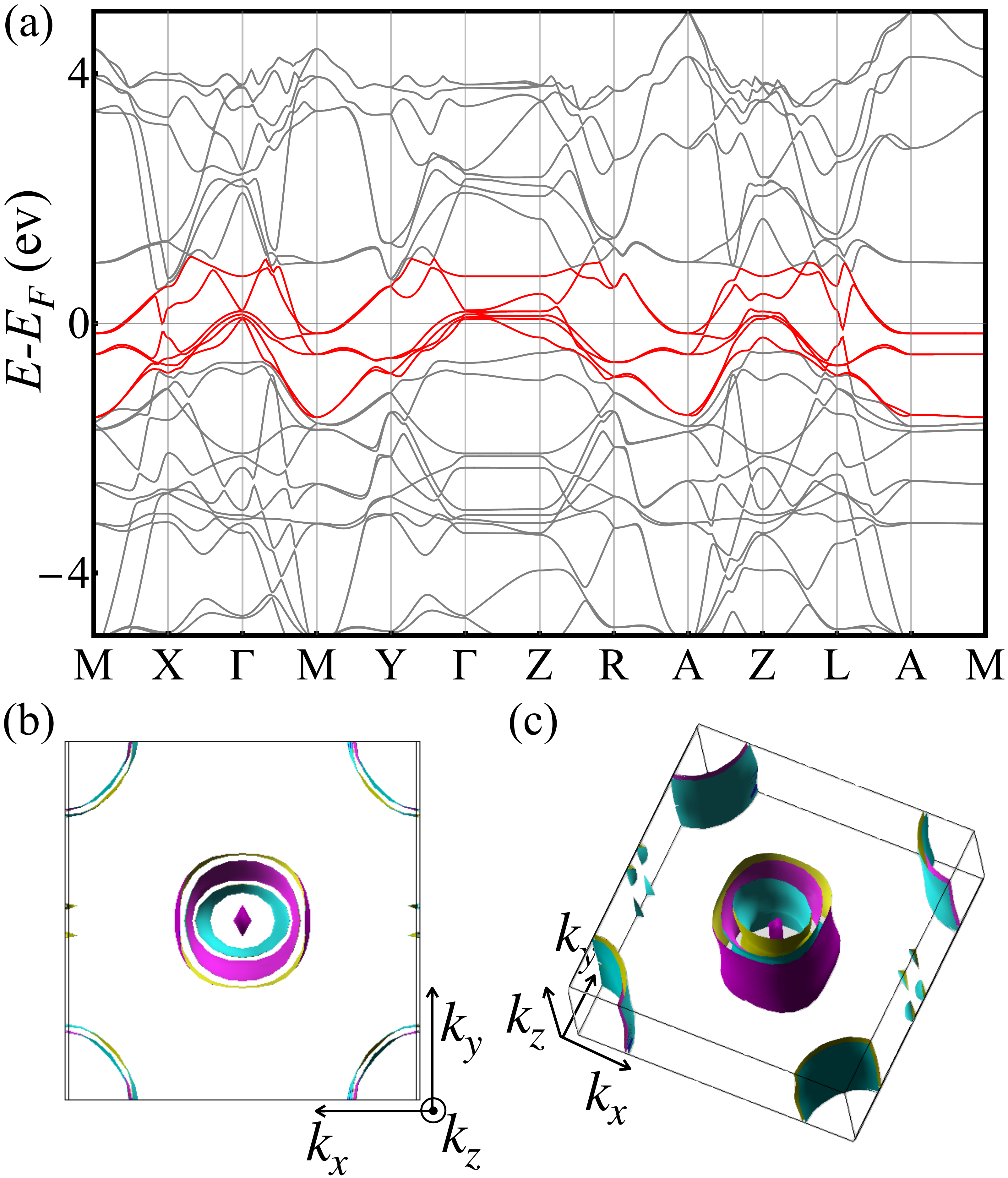}
		\caption{\label{fig_Reply_CaFeAs2_DFT}
		(a) The band structure of CaFeAs$_2$.
		The bands near the Fermi energy $E_F$ are marked in red.
		M, X, $\Gamma$, Y, Z, R, A, L represent $(\pi/2,\pi/2,0)$, $(\pi/2,0,0)$, $(0,0,0)$, $(0,\pi/2,0)$, $(0,0,\pi/2)$, $(0,\pi/2,\pi/2)$, $(\pi/2,\pi/2,\pi/2)$, $(\pi/2,0,\pi/2)$ in the Brillouin zone, respectively.
		(b) and (c) The Fermi surface of CaFeAs$_2$ with different viewpoints.
		The Fermi surfaces are visualized by XCrySDen~\cite{xcrysden}.
		}
	\end{center}
\end{figure}

\begin{figure*}[]
	\begin{center}
		\includegraphics[width=2\columnwidth]{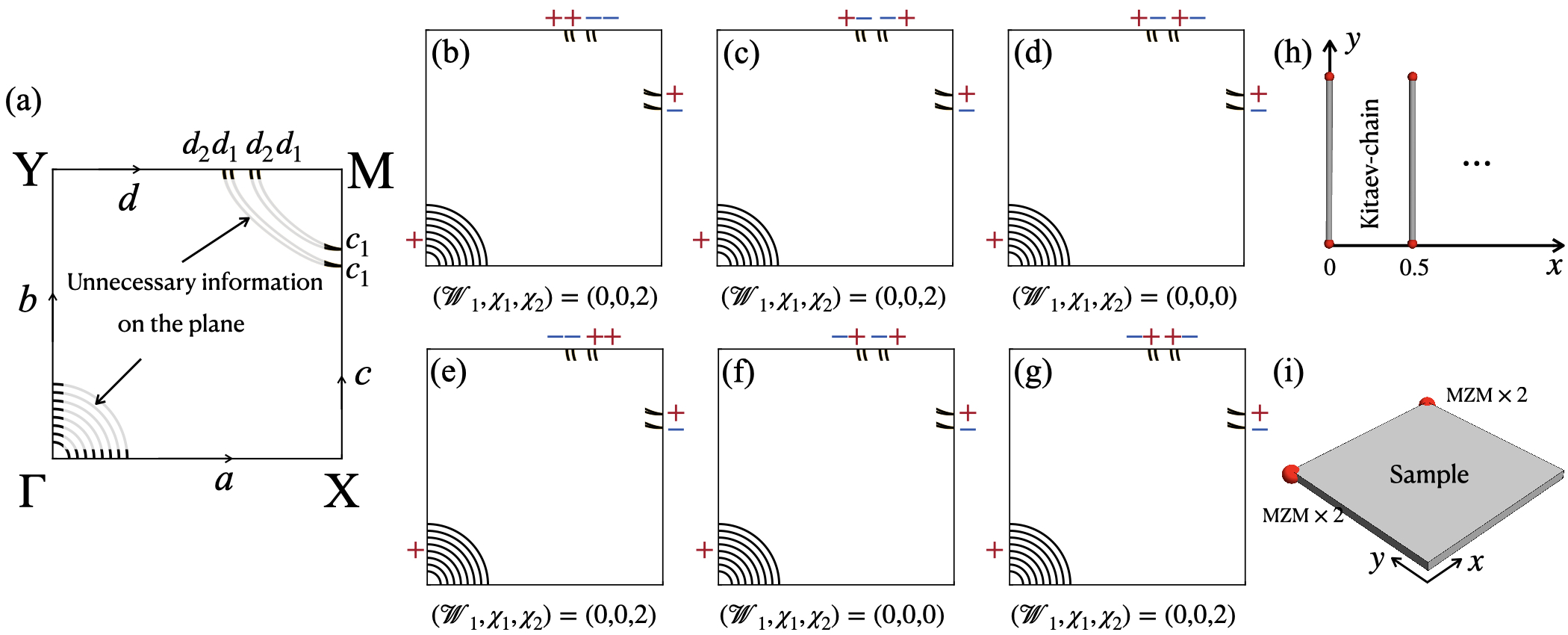}
		\caption{\label{fig_Reply_CaFeAs2_FSformula}
			(a) The illustration of Fermi surfaces of quasi-2D CaFeAs$_2$ based on the DFT-calculated data.
			$a,b,c,d$ are the one-cell, whose directions are label by arrows.
			$c_i$ and $d_i$ represents the irreducible representations on the one-cell $c$ and $d$.
			$\Gamma$, X, Y, M represents $(0,0)$, $(\pi,0)$, $(0,\pi)$ and $(\pi,\pi)$.
			(b)-(g) The six inequivalent pairing sign configurations satisfying $\mathcal{W}^{\text{gapless}}_1 = 0$.
			$+$ and $-$ represent the pairing sign $\delta^{l}_i$ of $i$-th Fermi points on the line segment $l$.
			$\mathcal{W}$, $\mathcal{X}_1$ and $\mathcal{X}_2$ represent the value of the topological invariants defined in the Eq.~\eqref{eq_1}, \eqref{eq_2} and \eqref{eq_3}.
			(h) The real-space construction of the topological nontrivial state with $(\mathcal{W},\mathcal{X}_1,\mathcal{X}_2)=(0,0,2)$.
			(i) The corner modes of the topological nontrivial state with $(\mathcal{W},\mathcal{X}_1,\mathcal{X}_2)=(0,0,2)$.
		}
	\end{center}
\end{figure*}

In  Fig.~\ref{fig_Reply_CaFeAs2_DFT}, we show the band structure and Fermi surfaces obtained by DFT calculations.
From the raw data of Fig.~\ref{fig_Reply_CaFeAs2_DFT}(a), we notice that there are eight Fermi surfaces around the $\Gamma$ point and four Fermi surfaces around the M point.
This implies that although the Fermi surfaces in Fig.~\ref{fig_Reply_CaFeAs2_DFT}(b) and (c) seem to be degenerate, there are no Fermi surface degeneracy.
This is consistent with the symmetry analysis.

One can see that the $k_z$-dependence of Fermi surfaces is sufficiently weak, and topological properties inherit the quasi-two-dimensional nature of its electronic structure.
Therefore, we focus on $k_z=0$ plane with the exchange of $k_x$- and $k_y$-axes to illustrate the workflow of our diagnostic scheme, which results in Layer group $p2_111$ (No.~9).

According to our Fermi surface formula, there exist a $\mathbb{Z}$-valued invariant for gapless points, a $\mathbb{Z}_2$-valued invariant, and a $\mathbb{Z}_4$-valued invariant, whose expressions are given by

\begin{widetext}
\begin{align}
	\label{eq_1}
	&\mathcal{W}_1^{\text{gapless}} =\sum_{m=1}^{n_{a_1}}\sgn(v_m^{a_1})\frac{1-\sgn(\delta_m^{a_1})}{2}
	+\sum_{m=1}^{n_{a_2}}\sgn(v_m^{a_2})\frac{1-\sgn(\delta_m^{a_2})}{2}
	-\sum_{m=1}^{n_{b_1}}\sgn(v_m^{b_1})\frac{1-\sgn(\delta_m^{b_1})}{2}\\
	&\quad\quad\quad\quad +2\sum_{m=1}^{n_{c_1}}\sgn(v_m^{c_1})\frac{1-\sgn(\delta_m^{c_1})}{2}
	-\sum_{m=1}^{n_{d_1}}\sgn(v_m^{d_1})\frac{1-\sgn(\delta_m^{d_1})}{2}
	-\sum_{m=1}^{n_{d_2}}\sgn(v_m^{d_2})\frac{1-\sgn(\delta_m^{d_2})}{2};\nonumber\\
	\label{eq_2}
	&\mathcal{X}_1 = \sum_{m=1}^{n_{b_1}}\frac{1-\sgn(\delta_m^{b_1})}{2} +\sum_{m=1}^{n_{d_1}}\frac{1-\sgn(\delta_m^{d_1})}{2} +\sum_{m=1}^{n_{d_2}}\frac{1-\sgn(\delta_m^{d_2})}{2} \ \bmod 2;\\
	\label{eq_3}
	&\mathcal{X}_2 =\sum_{m=1}^{n_{a_2}}\sgn(v_m^{a_2})(1-\sgn(\delta_m^{a_2}))
	-\sum_{m=1}^{n_{b_1}}\sgn(v_m^{b_1})\frac{1-\sgn(\delta_m^{b_1})}{2}
	-\sum_{m=1}^{n_{d_1}}\sgn(v_m^{d_1})(1-\sgn(\delta_m^{d_1})) \ \bmod 4.
\end{align}
\end{widetext}

Here, $a_1$ and $d_1$ ($a_2$ and $d_2$) are symbols of the irreducible representations whose the character of screw symmetry along $x$-axis is $-\mathrm{i} e^{-\mathrm{i}k_x/2} \ (+\mathrm{i} e^{-\mathrm{i}k_x/2})$ (see Section~{\clb L9} in Supplementary Material II for more detailed information).

As a demonstration, here we discuss some scenarios out of all possibilities, in which sign changes happen between the Fermi surfaces near the M point while fixing the pairing signs of the Fermi surfaces near the $\Gamma$ point to be positive.
Although a global gauge transformation can give rise to the global minus sign of the superconducing order parameter, this does not affect the topology.
As a result, there are six inequivalent pairing sign choices for possible fully gapped superconducting states (whose $\mathcal{W}_1^{\text{gapless}}$ must be trivial), as shown in Fig.~\ref{fig_Reply_CaFeAs2_FSformula}(b)-(g).
According to the formulas of the invariants, we find that four of these configurations correspond to topologically nontrivial states, $(\mathcal{W}^{\text{gapless}}_1, \mathcal{X}_1, \mathcal{X}_2) = (0,0,2)$.
This phase is equivalent to $x$-direction stacking copies of one-dimensional class DIII topological superconductors along the $y$-direction, as shown in Fig.~\ref{fig_Reply_CaFeAs2_FSformula}(h) (More details in Sec.~\textcolor{blue}{C1} of Supplementary Material I.).
The model in Fig.~\ref{fig_Reply_CaFeAs2_FSformula}(h) with a small perturbation is a second-order topological superconductor with Majorana corner modes, as shown in Fig.~\ref{fig_Reply_CaFeAs2_FSformula}(i).

Finally, let us return to the discussion of the three-dimensional nature.
Due to its quasi-two-dimensional electronic nature, three-dimensional topological properties can be understood by stacking the aforementioned two-dimensional topological states along the $z$-direction.
As a result, we conclude that this three-dimensional topological phase exhibits hinge modes along the $z$-direction.

\section{Discussion}
We introduce Fermi-surface formulas of topological invariants for time-reversal symmetric superconductors with $s$-wave-like pairing symmetries in all layer and space groups.
Most of the superconductors ever discovered belong to this class, and within it, many potential candidates for topological superconductors remain to be verified, such as iron-based superconductors~\cite{katayama2013superconductivity}, nickelate-based superconductors~\cite{sun2023signatures, Ko:2025aa}, highly overdoped CuO$_2$ monolayer \cite{PhysRevLett.121.227002} and heavy fermion superconductors~\cite{RevModPhys.56.755}.
This new method requires minimal information on the material, namely the sign of pairing and Fermi velocity at several Fermi points, and yet provides complete topological properties for 159 space groups.
While our diagnosis for gapped phases remains partial for 40 space groups, a large part of nontrivial phases can still be detected.

We emphasize the distinction between our formulas and the existing formulas introduced in Refs.~\cite{FS_TSC_Qi, FS_TSC_Sato}. 
They introduce formulas for class DIII without any spatial symmetries and for the one-dimensional winding number, respectively.
In their derivation, they assume that there is no Fermi surface degeneracy.
However, this assumption is often invalid in the presence of space group symmetries, where such Fermi surface degeneracy occurs.
In addition, space group symmetries often prevent certain topological invariants from being nontrivial. For example, topological invariants in the scope of Ref.~\cite{FS_TSC_Qi} are always trivial for even-parity pairings with inversion symmetry.
On the other hand, our formulas can detect nontrivial topology even in such cases, as shown in Sec.~\ref{sec:example}.
Thus, our formulas have more broad applicability than Refs.~\cite{FS_TSC_Qi, FS_TSC_Sato}.

Furthermore, the diagnostic scheme proposed in this work identifies a broader range of topological phases based on much less information on the superconductor than all previous works.
In Sec.~\ref{sec:material}, we actually demonstrate the effectiveness of our formulas by applying them to the density functional~theory~(DFT) calculation for a realistic material.
Given the success of large-scale investigations of topological insulators using symmetry indicators and topological quantum chemistry~\cite{zhang2019catalogue, catalogue2,vergniory2019complete}, we anticipate that our formulas will play a crucial role in the discovery of realistic topological superconductors.

\begin{acknowledgments}
Z.Z.~and S.O.~thank Hoi Chun Po for the support and encouragement for the project.
Z.Z.~thanks Shengshan Qin and Xianxin Wu for helpful discussions.
K.S.~and S.O.~thank Masatoshi Sato for helpful discussions.
K.S., C.F., and S.O.~thank the Yukawa Institute for Theoretical Physics at Kyoto University. Discussions during the YITP workshop YITP-T-24-03 on ``Recent Developments and Challenges in Topological Phases'' were useful to complete this work.
Z.Z.~was supported by National Key R$\&$D Program of China (Grant No.~2021YFA1401500, Hoi Chun Po). K.S.~was supported by JST CREST (Grant No.~JPMJCR19T2) and JSPS KAKENHI (Grant No.~22H05118 and 23H01097).
C.F.~was supported by Chinese Academy of Sciences (Grant No.~XDB33020000), National Natural Science Foundation of China (Grant No.~12325404) and National Key R$\&$D Program of China (Grant No.~2022YFA1403800 and 2023YFA1406700).
S.O.~was supported by RIKEN Special Postdoctoral Researchers Program and KAKENHI Grant No.~23K19043 from the Japan Society for the Promotion of Science (JSPS).

\textbf{Data availability:}~All data to derive the conclusions are present in the manuscript and/or Supplementary Materials.

\end{acknowledgments}

{
\appendix
\section{Relation between $\mathcal{X}_3$ and $w_{\text{3D}}$}
\label{sec:method_1}
Here, we argue that an odd value of $\mathcal{X}_3$ indicates oddness of $w_{\text{3D}}$.
To show this, we compute $\mathcal{X}_3$ for all representatives of topological phases. 
In addition to the strong TSC, we have $(\mZ_2)^2$ phases constructed from one-dimensional TSCs and $(\mZ_2)^2$ phases constructed from two-dimensional TSCs~\cite{Ono-Shiozaki-Watanabe2022}, whose illustrations are shown in Fig.~\ref{fig_P41}.
In the Supplemental Materials, we calculate all the gapless and gapped topological invariants for all representative models for these phases, which are summarized in Table.~\ref{tab:1}.
We find that all models with $w_{\text{3D}}=0$ have $\mathcal{X}_3 = 0 \bmod 4$.
Also, we find $\mathcal{X}_3 = -w_{\text{3D}} \bmod 4$ for the generator of strong TSCs. 
Therefore, we conclude that $\mathcal{X}_3 \bmod 4$ faithfully reflects $w_{\text{3D}} \bmod 4$.
This argument can be generalized to $Pn_1\ (n = 2, 3, 4,6)$. 
In Supplemental Materials, we show that our invariants provide a $\mZ_2$-valued indicator of $w_{\text{3D}}$ in $P2_1$ and a $\mZ_6$-valued indicator for $P3_1$ and $P6_1$.

\begin{figure*}[t]
	\begin{center}
		\includegraphics[width=2\columnwidth]{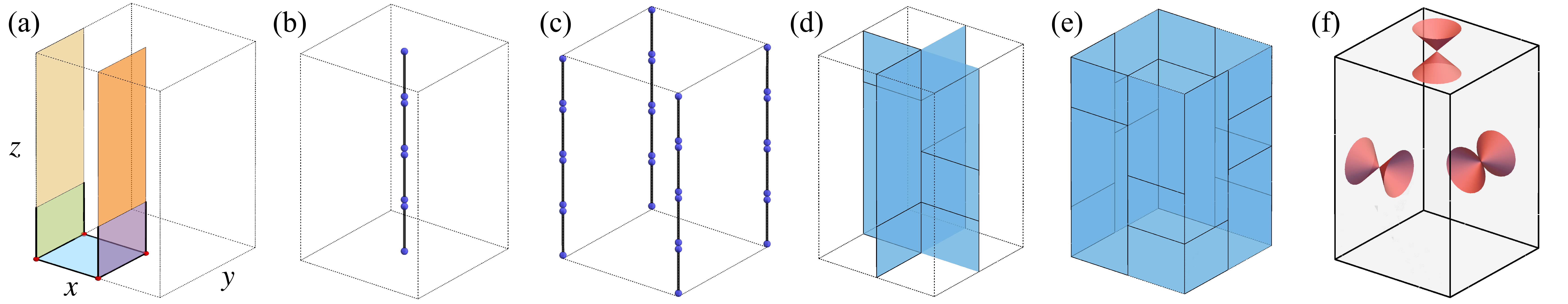}
		\caption{\label{fig_P41}
			(a) Unit cell and its cell complex for space group $P4_1$ (No.~76).
			The dashed lines denote the unit cell.
			The symmetry-inequivalent 2-cells, 1-cells and 0-cells are represented as colored faces, bold line segments, and red dots, respectively.
			The other cells can be obtained from them by acting with symmetry operations in SG $P4_1$.
			(b)-(c) Two generators of $\mZ^{2}_{2}$-topological phases constructed from 1D TSCs. The black bold lines represent the nontrivial 1D TSCs in DIII class, and the blue dots represent the corresponding Majorana zero modes at the endpoints of 1D TSCs.
			(d)-(e) Two generators of $\mZ^{2}_{2}$ constructed from 2D TSCs. The blue faces represent the nontrivial 2D TSCs in DIII class. The regions enclosed by solid lines represent the decorations on the symmetry-inequivalent 2-cells.
			(f) The illustration for the strong TSC in DIII class.
		}
	\end{center}
\end{figure*}

\begin{table}[b]
	\caption{\label{tab:1}
		The quantitative mappings between all representative models and the topological invariants $(\mathcal{W}_1^{\text{gapless}},\mathcal{W}_2^{\text{gapless}}, \mathcal{X}_1,\mathcal{X}_2,\mathcal{X}_3)$ computed by our Fermi surface formula.
		Here, $\mathcal{W}_1^{\text{gapless}}$ and $\mathcal{W}_2^{\text{gapless}}$ are $\mZ$-valued invariants;~$\mathcal{X}_1,\mathcal{X}_2$, and $\mathcal{X}_3$ are $\mZ_2$-, $\mZ_4$- and $\mZ_8$-valued ones.
		The formulas are listed in Supplemental Materials.
	}
	\centering
	\renewcommand\arraystretch{1.3}
	\begin{tabular}{l|c|c|c}
		\hline\hline
		Classification of Ref.~\cite{Ono-Shiozaki-Watanabe2022} & $w_{\text{3D}}$ & $(\mathcal{W}^{\text{gapless}}_1,\mathcal{W}^{\text{gapless}}_2)$& $(\mathcal{X}_1,\mathcal{X}_2,\mathcal{X}_3)$\\ \hline
		1D TSCs: $(1, 0) \in \mZ_2 \times \mZ_2$ & $0$ & $(0,0)$ &$( 1,2,4)$\\ \hline
		1D TSCs: $(0, 1) \in \mZ_2 \times \mZ_2$ & $0$ & $(0,0)$ &$( 1,0,4)$\\ \hline
		2D TSCs: $(1, 0) \in \mZ_2 \times \mZ_2$ & $0$ & $(0,0)$ &$( 0,3,4)$\\ \hline
		2D TSCs: $(0, 1) \in \mZ_2 \times \mZ_2$ & $0$ & $(0,0)$ &$( 0,1,0)$\\ \hline
		Strong TSC: $1 \in \mZ$ & $1$ & $(0,0)$ &$( 1,1,3)$\\ \hline\hline
	\end{tabular}
\end{table}

\section{Detailed information on DFT calculations}
In our DFT calculations, we employ \textsc{Quantum ESPRESSO}~\cite{QE-2009,QE-2017} and \texttt{qeirreps}~\cite{qeirreps} to obtain irreducible representations defined on line segments.
The $k$-mesh is chosen as $24 \times 24 \times 24$ in our self-consistent field calculations.
The lattice structure is obtained from Cambridge Crystallographic Data Centre (CCDC 961741)~\cite{CCDC} for La$_{x}$Ca$_{1-x}$FeAs$_2$~\cite{katayama2013superconductivity} by taking $x=0$.
}

\bibliography{ref}
\clearpage
\onecolumngrid
\begin{center}
	\large
	\textbf{Supplementary Materials for ``Fermi-surface diagnosis for topological superconductivity with  $s$-wave-like pairing symmetries''}
\end{center}
\setcounter{section}{0}
\setcounter{equation}{0}
\setcounter{figure}{0}
\setcounter{table}{0}
\renewcommand{\thesection}{S\arabic{section}}
\renewcommand{\theequation}{S\arabic{equation}}
\renewcommand{\thefigure}{S\arabic{figure}}
\renewcommand{\thetable}{S\arabic{table}}
\addtocontents{toc}{\protect\setcounter{tocdepth}{0}}
\input{SM}

\end{document}

%% file: SM.tex
\section{Detailed derivation of Fermi surface formulas}
\label{sec:main_result}
In the main text, we present the Fermi surface formulas for topological invariants in time-reversal symmetric superconductors with trivial pairing symmetries. 
The derivation of the formulas is based on the full expressions of topological invariants proposed in Ref.~\cite{Ono-Shiozaki2023}.
Here, we derive our Fermi surface formulas in detail.
In \ref{sec:setup}, we introduce systems we focus on in this work. 
In \ref{sec:review}, we discuss topological invariants written in terms of the $q$-matrix.
In \ref{sec:derivation}, we derive Fermi surface formulas from the expressions discussed in \ref{sec:review}.

\subsection{Set up}
\label{sec:setup}
Let $\hat{c}^{\dagger}_{\bm{k}, \sigma}$ and $\hat{c}_{\bm{k}, \sigma}\ (\sigma = 1, 2, \cdots, N_{\text{orb}})$ be fermionic creation and annihilation operators with degree of freedom $\sigma$ at momentum $\bm{k}$. 
Time-reversal symmetry (TRS) $\calT$ transforms the creation operator $\hat{c}^{\dagger}_{\bm{k}, \sigma}$ into
\begin{align}
	\hat{\calT}\hat{c}^{\dagger}_{\bm{k}, \sigma}\hat{\calT}^{-1} = \sum_{\sigma'}\hat{c}^{\dagger}_{-\bm{k}, \sigma'}[U(\calT)]_{\sigma'\sigma};\quad \hat{\calT}\ii \hat{\calT}^{-1} = -\ii;\quad U(\calT)[U(\calT)]^*=-\mathds{1},
\end{align}
where $U(\calT)$ is a unitary representation of TRS and $\mathds{1}$ is the identity matrix.
We assume that superconducting systems can be described by the following mean-field Hamiltonian:
\begin{align}
	\label{eq:BdG ham}
	\hat{H} &= \sum_{\bm{k}, \sigma\sigma'}\left[\hat{c}^{\dagger}_{\bm{k}, \sigma}[h_{\bk}]_{\sigma\sigma'}\hat{c}_{\bm{k},\sigma'} + \frac{1}{2}\hat{c}^{\dagger}_{\bm{k}, \sigma}[\Delta_{\bm{k}}]_{\sigma\sigma'}\hat{c}^{\dagger}_{-\bm{k},\sigma'} + \text{h.c.}\right]\nonumber \\
	&=\frac{1}{2}\Upsilon_{\bk}^{\dagger}\begin{pmatrix}
		h_{\bk} & \Delta_{\bk} \\
		\Delta_{\bk}^{\dagger} & -h_{-\bk}^{\top}
	\end{pmatrix}\Upsilon_{\bk} + \text{const}.,
\end{align}
where $h_{\bm{k}}$ and $\Delta_{\bm{k}} = -\Delta^{\top}_{-\bk}$ are a normal conducting Hamiltonian and a superconducting order parameter, respectively. 
Also, we introduce the Nambu spinor $\Upsilon_{\bk}^{\dagger} = (\hat{c}_{\bk, 1}^{\dagger}\ \hat{c}_{\bk, 2}^{\dagger}\ \cdots \hat{c}_{\bk, N_{\text{orb}}}^{\dagger}\ \hat{c}_{-\bk, 1} \cdots \hat{c}_{-\bk, N_{\text{orb}}})$ and the Bogoliubov-de Gennes (BdG) Hamiltonian
\begin{align}
	H_{\bk}^{\text{BdG}} = \begin{pmatrix}
		h_{\bk} & \Delta_{\bk} \\
		\Delta_{\bk}^{\dagger} & -h_{-\bk}^{\top}
	\end{pmatrix}.
\end{align}
The BdG Hamiltonian inherently has the following symmetry:
\begin{align}
	\label{eq:PHS}
	&\UBdG(\calC)(H_{\bk}^{\text{BdG}})^{\top}[\UBdG(\calC)]^{\dagger} = -H_{-\bk}^{\text{BdG}};\\
	&\UBdG(\calC) = \begin{pmatrix}
		O & \mathds{1}_{N_{\text{orb}}}\\
		\mathds{1}_{N_{\text{orb}}} & O
	\end{pmatrix}.
\end{align}
We refer to $\calC$ as particle-hole symmetry (PHS).
The BdG Hamiltonian is also invariant under time-reversal symmetry:
\begin{align}
	\label{eq:TRS}
	&\UBdG(\calT)(H_{\bk}^{\text{BdG}})^{*}[\UBdG(\calT)]^{\dagger} = H_{-\bk}^{\text{BdG}};\\
	&\UBdG(\calT) = \begin{pmatrix}
		U(\calT) & O\\
		O & U^*(\calT)
	\end{pmatrix}.
\end{align}
From Eqs.~\eqref{eq:PHS} and \eqref{eq:TRS}, the combination of $\calC$ and $\calT$ is also a symmetry of the BdG Hamiltonian:
\begin{align}
	\label{eq:chiral}
	&\UBdG(\gamma)H_{\bk}^{\text{BdG}}[\UBdG(\gamma)]^{\dagger} = -H_{\bk}^{\text{BdG}};\\
	&\UBdG(\gamma) = \ii \UBdG(\calT)[\UBdG(\calC)]^*.
\end{align}

Let $\calG$ be a space group. 
Suppose that the normal conducting phase possesses a magnetic symmetry group $\calM = \calG + \calG\calT$, whose elements transform the creation operators into
\begin{align}
	\hat{g}\hat{c}^{\dagger}_{\bm{k}, \sigma}\hat{g}^{-1} = \sum_{\beta}\hat{c}^{\dagger}_{g\bm{k}, \sigma'}[U_{\bk}(g)]_{\sigma'\sigma};\ \ \hat{g}\ii\hat{g}^{-1} = \phi_g \ii,
\end{align}
where $U_{\bk}(g)$ is a unitary matrix, and $\phi: \calM \rightarrow \{\pm 1\}$ tell us which symmetry $g \in \calM$ is unitary $(\phi_g = +1)$ or antiunitary $(\phi_g = -1)$.
To satisfy $\hat{g}\hat{H}\hat{g}^{-1} = \hat{H}$, matrices $h_{\bm{k}}$ and $\Delta_{\bm{k}}$
\begin{align}
	U_{\bm{k}}(g)h^{\phi_g}_{\bk}U^{\dagger}_{\bm{k}}(g) &= h_{g\bm{k}}; \\
	U_{\bm{k}}(g)\Delta^{\phi_g}_{\bk}U^\top_{-\bm{k}}(g) &= \chi_{g}\Delta_{g\bm{k}}\ \ (\chi_g \in \text{U}(1)),
\end{align}
where we introduce a notation for matrix $M$ such that $M^{\phi_g} = M$ for $\phi_g=+1$ and $M^{\phi_g} = M^*$ for $\phi_g=-1$.
Correspondingly, the BdG Hamiltonian satisfies
\begin{align}
	&\UBdG_{\bk}(g)[H_{\bk}^{\text{BdG}}]^{\phi_g} = H_{g\bk}^{\text{BdG}}\UBdG_{\bk}(g); \\
	&\UBdG_{\bk}(g) = \begin{pmatrix}
		U_{\bk}(g) & \\
		& \chi_g U_{-\bk}^{*}(g)
	\end{pmatrix}.
\end{align}
We say the pairing symmetry is trivial if $\chi_g = +1$ for all elements in $\calM$.

It is often convenient to consider a symmetry group that keeps $\bk$ unchanged, which is defined by
\begin{align}
	\calM_{\bk} \coloneq \{g \in \calM \ \vert\ g\bk = \bk + \bold{G}\ (\bold{G}:\text{reciprocal lattice vector})\}.
\end{align}
This symmetry group is called \textit{little group}.
We can decompose the little group into two parts:
\begin{align}
	\calM_{\bk} = \calGk + \calA_{\bk},
\end{align}
where 
\begin{align}
	\calG_{\bk} &= \{g \in \calM_{\bk} \ \vert\ \phi_g = +1\};\\
	\calA_{\bk} &= \{g \in \calM_{\bk} \ \vert\ \phi_g = -1\}.
\end{align}

\subsection{A brief review of topological invariants}
\label{sec:review}
\subsubsection{$q$-matrix}
Before moving to the expressions of topological invariants, we introduce the \textit{$q$-matrix} to formulate topological invariants.
In the following, let $\alpha$ and $\chi_{\bk}^{\alpha}(g)$ denote a symbol of an irreducible representation (irrep) of $\calG_{\bk}$ and the character of irrep $\alpha$ of $g \in \calGk$. 
\begin{align}
	\label{eq:projection}
	P_{\alpha} &= \frac{d_{\alpha}}{\vert \calGk/\Pi \vert} \sum_{g \in  \calGk/\Pi} \left[\chi_{\bk}^{\alpha}(g)\right]^{*}\UBdG_{\bk}(g),
\end{align}
where $d_{\alpha}$ is the dimension of irrep $\alpha$, and $\Pi$ is a translation subgroup of $\calG$.
The eigenvalues of $P_{\alpha}$ are either $1$ or $0$, and $V^{\alpha}$ denotes a matrix whose columns are eigenvectors of $P_{\alpha}$ with eigenvalue $1$. 
By using $V^{\alpha}$, we project the Hamiltonian to each irrep sector:
\begin{align}
	H^{\alpha}_{\bk} \coloneqq [V^{\alpha}]^{\dagger}H_{\bk}^{\text{BdG}}V^{\alpha}.
\end{align}
In our symmetry settings, every momentum is invariant under chiral symmetry $\gamma = \calT \calC$. 
Furthermore, the chiral symmetry $\gamma$ does not transform each irrep of $\calGk$ into another irrep. 
Therefore, we can define the chiral matrix for each irrep sector by
\begin{align}
	\Gamma^{\alpha} \coloneqq [V^{\alpha}]^{\dagger}\UBdG_{\bk}(\gamma)V^{\alpha}.
\end{align}
This chiral matrix can be diagonalized as
\begin{align}
	\Gamma^{\alpha}V_{\pm} = \pm V_{\pm}.
\end{align}
Then, we define  \textit{q-matrix} by 
\begin{align}
	\label{eq:qmatrix}
	q^{\alpha}_{\bk} &= [V_{-}]^{\dagger} H_{\bk}^{\alpha} V_{+}.
\end{align}

\subsubsection{Topological invariants in Ref.~\cite{Ono-Shiozaki2023}}

As we introduce in the main text, for each of topological invariants and space groups, the line segments and irreps considered in expressions of topological invariants are different.
For a given momentum-space decomposition in a space group, the necessary information about irreps on line segments (without endpoints) is encoded in an integer-valued tuple denoted by
\begin{align}
	\label{eq:bm_x}
	\bm{x} = \left(x_{(l_1, \alpha_1)}, x_{(l_1, \alpha_2)}, \cdots, x_{(l_2, \beta_1)}, \cdots \right),
\end{align}
where $x_{(l, \alpha)}$ is a contribution weight of irrep $\alpha$ on the line segment $l$ for a topological invariant. 
In addition, information about irreps at endpoints is also used for the $\mZ_\lambda$-valued invariants.
This is also encoded in another integer-valued tuple denoted by
\begin{align}
	\label{eq:bm_v}
	\bm{v} = \left(v_{(p_1, \beta_1)}, v_{(p_1, \beta_2)}, \cdots, v_{(p_2, \gamma_1)}, \cdots \right),
\end{align}
where $v_{(p, \beta)}$ is a contribution weight of irrep $\beta$ at point $p$ for a topological invariant.

There exist important relations among $\bm{v}$, $\bm{x}$, and compatibility relations. 
Compatibility relations are the relationships among irreps at different points. 
The compatibility relations among all line segments and their endpoints can also be encoded in an integer-valued matrix $M$. 
For $\bm{x}$ corresponding to a $\mZ$-valued invariant, 
\begin{align}
	\bm{x}^\top M = \bm{0}.
\end{align}
For $\bm{x}$ and $\bm{v}$ corresponding to a $\mZ_\lambda$-valued invariant, 
\begin{align}
	\bm{x}^\top M = \lambda\bm{v}.
\end{align}
These relations are used in the derivation of Fermi surface formulas. 
Later, we elaborate on compatibility relations.

We explain how to use $\bm{x}$ and $\bm{v}$ to construct topological invariants defined on line segments.
For later convenience, we use the following abbreviation:
\begin{align}
	\mathscr{D}_{\alpha} \coloneqq \begin{cases}
		d_{\alpha} \\
		2d_{\alpha} 
	\end{cases},
\end{align}
where we use the second line if irrep $\alpha$ of $\calGk$ forms a Kramers pair by itself; otherwise, we use the first line.
We define the following key quantities:
\begin{align}
	\label{eq:w_alpha}
	w_{(l, \alpha)} &\coloneq  \frac{1}{2\pi\ii \scrD_\alpha}\int_l d(\log \det q_{\bk}^{\alpha}-\log \det (q^{\alpha}_{\bk})^{\text{vac}}).
\end{align}
where $q_{\bk}^{\alpha}$ is a $q$-matrix defined for irrep $\alpha$ on line segment $l$. 
It should be emphasized that $w_{(l, \alpha)}$ is invariant under gauge transformations.
This is because $w_{(l, \alpha)}$ is defined by a difference between the winding of $q$-matrices for the target and a reference system. 
Then, the effect of gauge transformations is canceled out.
We can always consider the vacuum (i.e., the infinite chemical potential limit of the system in which we are interested) as the reference, and $(q^{\alpha}_{\bk})^{\text{vac}}$ denotes the $q$-matrix of the vacuum.

Actually, $\bm{x}$ and $\bm{v}$ tell us how to take the combinations to construct topological invariants.
\begin{itemize}[leftmargin=0.3 cm]
	\item $\mZ$-valued invariant for gapless points on polygons
	\begin{align}
		\label{eq:gapless_Z}
		&\mathcal{W}^{\text{gapless}} = \sum_{l, \alpha}x_{(l, \alpha)}w_{(l, \alpha)} \in \mZ.
	\end{align}
	\item $\mZ$-valued invariant for gapped phases
	\begin{align}
		\label{eq:gapped_Z}
		&\mathcal{W}^{\text{gapped}} =  \sum_{l, \alpha}x_{(l, \alpha)}w_{(l, \alpha)} \in \mZ.
	\end{align}
	\item $\mZ_{\lambda}$-valued invariant for gapped phases
	\begin{align}
		\label{eq:Zn}
		&\exp\hspace{-0.5mm}\left[\frac{2\pi \mathrm{i}}{\lambda}\mathcal{X}\right] = \frac{\exp[-2\pi \ii \sum_{l, \alpha}x_{(l, \alpha)}w_{(l, \alpha)}/\lambda]}{\prod_{p, \beta}\left(\mathcal{Z}[q_{\bk_p}^{\beta}]/\mathcal{Z}[(q^{\beta}_{\bk_p})^{\text{vac}}]\right)^{v_{(p, \beta)}}},
	\end{align}
\end{itemize}
where $\bk_p$ is a point in momentum space. Also, $\mathcal{Z}[q_{\bk}^{\beta}]$ is a ``generalized'' Pfaffian  defined by a product of duplicated eigenvalues of $q_{\bk_p}^{\beta}$ for irrep $\beta$ at $\bk_p$, i.e.,
\begin{align}
	\mathcal{Z}[q_{\bk_p}^{\beta}]= \prod_{m} \pi_m,
\end{align}
where $\pi_m$ is the $m$-th duplicated eigenvalues of $q_{\bk_p}^{\beta}$.
One can find
\begin{align}
	\label{eq:Z_beta}
	\det q^{\beta}_{\bk_p} = \left(\mathcal{Z}[q^{\beta}_{\bk_p}]\right)^{\scrD_\beta}.
\end{align}

Here, we argue that the above quantities in Eqs.~\eqref{eq:gapless_Z}-\eqref{eq:Zn} are topological invariants.
They have the following properties:
\begin{enumerate}
	\item[(i)] They are quantized;
	\item[(ii)] They are invariant under gauge transformations;
	\item[(iii)] They are zeros for the vacuum. 
\end{enumerate}
Here, we do not prove the first property.
A proof of this is presented in Ref.~\cite{Ono-Shiozaki2023}.
As mentioned above, the second property is ensured by the difference structure of $w_{(l, \alpha)}$. 
The third property is also confirmed by the definition of $w_{(l, \alpha)}$.
When we substitute $\det q_{\bk}^{\alpha}$ of the first term in $w_{(l, \alpha)}$ by $\det (q_{\bk}^{\alpha})^{\text{vac}}$, it is obvious zero.
Then, if the target is in the trivial phase, the above three properties ensure that these quantities must be zero.
Namely, when either~\eqref{eq:gapless_Z}, \eqref{eq:gapped_Z}, or \eqref{eq:Zn} is nonzero, the system must be nontrivial. 
It should be noted that the converse is not true in general. 
That is, all of them are zeros for some nontrivial phases sometimes when our invariants are not complete.

\subsection{Derivation of Fermi surface formulas}
\label{sec:derivation}

\subsubsection{Effective Hamiltonian in general symmetry settings}
\label{sec:derivation_Heff}
Let $\kstar$ and $\bm{p}$ be a Fermi point and a displacement from $\kstar$, respectively. 
Here, we focus on the following term in the mean-field Hamiltonian:
\begin{align}
	\frac{1}{2}\left[
	\hat{\bm{c}}_{\kstar + \bm{p}}^{\dagger}h_{\kstar+\bm{p}}\hat{\bm{c}}_{\kstar + \bm{p}}
	+\hat{\bm{c}}_{-(\kstar + \bm{p})}^{\top}[-h^{\top}_{-(\kstar+\bm{p})}](\hat{\bm{c}}_{-(\kstar + \bm{p})}^{\dagger})^\top
	+\hat{\bm{c}}_{\kstar + \bm{p}}^{\dagger}[\Delta_{\kstar+\bm{p}}](\hat{\bm{c}}_{-(\kstar + \bm{p})}^{\dagger})^\top
	+\hat{\bm{c}}_{-(\kstar + \bm{p})}^{\top}[\Delta^{\dagger}_{\kstar+\bm{p}}]\hat{\bm{c}}_{\kstar + \bm{p}}
	\right].
\end{align}
We consider only energy bands closest to the Fermi energy around $\kstar$ and assume that there is no accidental degeneracy.
Let $\Phi_{\kstar+\bm{p}}^{\alpha}$ and $\Phi_{-(\kstar+\bm{p})}^{\alpha'}$ be matrices composed of eigenvectors of $h_{\kstar+\bm{p}}$ and $h_{-(\kstar+\bm{p})}$, respectively. 
Then, we have
\begin{align}
	\label{eq:eigenvector}
	&h_{\kstar+\bm{p}} \Phi^{\alpha}_{\kstar+\bm{p}} = \varepsilon_{\kstar+\bm{p}, \alpha}\Phi^{\alpha}_{\kstar+\bm{p}};\quad
	U_{\kstar+\bm{p}}(g) (\Phi^{\alpha}_{\kstar+\bm{p}})^{\phi_g} = \Phi^{\alpha}_{\kstar+\bm{p}} u_{\kstar+\bm{p}}^{\alpha}(g),\\
	&h_{-\kstar-\bm{p}} \Phi^{\alpha'}_{-\kstar-\bm{p}} = \varepsilon_{-(\kstar+\bm{p}), \alpha'}\Phi^{\alpha'}_{-(\kstar+\bm{p})};\quad
	U_{-(\kstar+\bm{p})}(g) (\Phi^{\alpha'}_{-(\kstar+\bm{p})})^{\phi_g} = \Phi^{\alpha'}_{-(\kstar+\bm{p})} u_{-(\kstar+\bm{p})}^{\alpha'}(g),
\end{align}
where $\varepsilon_{\bk, \alpha}$ is an eigenenergy and $u_{\bk}^{\alpha}(g)$ is an irrep of $\calM_{\bk}$. The dimension of irrep $\alpha$ of $\calM_{\bk}$ is denoted by $\calD_{\alpha}$. 
%

%
Let us move to the band basis $\hat{\bm{f}}_{\cdot, \alpha}^{\dagger} = (\hat{f}_{\cdot, 1}^{\dagger}\ \hat{f}_{\cdot, 2}^{\dagger}\ \cdots\ \hat{f}_{\cdot, \calD_{\alpha}}^{\dagger}) = \hat{\bm{c}}_{\cdot}^{\dagger}\Phi^{\alpha}_{\cdot}$. 
In fact, $\Phi^{\alpha}_{\kstar+\bm{p}}$ and $\Phi^{\alpha'}_{-(\kstar+\bm{p})}$ are isometries. 
Namely, $(\Phi_{\cdot}^{\alpha})^{\dagger}\Phi^{\alpha}_{\cdot} = \mathds{1}_{\calD_{\alpha}}$ and $P^{\alpha}_{\cdot} = \Phi^{\alpha}_{\cdot}(\Phi^{\alpha}_{\cdot})^{\dagger}$ is a projection. 
As a result, we have
\begin{align}
	&\hat{\bm{c}}_{\kstar + \bm{p}}^{\dagger}h_{\kstar+\bm{p}}\hat{\bm{c}}_{\kstar + \bm{p}} \rightarrow \hat{\bm{c}}_{\kstar + \bm{p}}^{\dagger}(P^{\alpha}_{\kstar+\bm{p}}h_{\kstar+\bm{p}}P^{\alpha}_{\kstar+\bm{p}})\hat{\bm{c}}_{\kstar + \bm{p}}  =  \varepsilon_{\kstar+\bm{p}, \alpha}\hat{\bm{f}}_{\kstar+\bm{p}, \alpha}^{\dagger}\hat{\bm{f}}_{\kstar+\bm{p}, \alpha}\\
	&\hat{\bm{c}}_{\kstar + \bm{p}}^{\dagger}[\Delta_{\kstar+\bm{p}}](\hat{\bm{c}}_{-(\kstar + \bm{p})}^{\dagger})^\top\rightarrow \hat{\bm{c}}_{\kstar + \bm{p}}^{\dagger}[P^{\alpha}_{\kstar+\bm{p}}\Delta_{\kstar+\bm{p}}(P^{\alpha'}_{-\kstar-\bm{p}})^{*}](\hat{\bm{c}}_{-(\kstar + \bm{p})}^{\dagger})^\top =
	 \hat{\bm{f}}_{\kstar+\bm{p}, \alpha}^{\dagger}[(\Phi^{\alpha}_{\kstar+\bm{p}})^{\dagger}\Delta_{\kstar+\bm{p}}(\Phi^{\alpha'}_{-\kstar-\bm{p}})^{*}](\hat{\bm{f}}_{-\kstar-\bm{p}, \alpha'}^{\dagger})^\top\\                  
	 &\hat{\bm{c}}_{-(\kstar + \bm{p})}^{\top}[\Delta_{\kstar+\bm{p}}^{\dagger}]\hat{\bm{c}}_{\kstar + \bm{p}} \rightarrow \hat{\bm{c}}_{-(\kstar + \bm{p})}^{\top}[(P^{\alpha'}_{-\kstar-\bm{p}})^{*}\Delta^{\dagger}_{\kstar+\bm{p}}P^{\alpha}_{\kstar+\bm{p}}]\hat{\bm{c}}_{\kstar + \bm{p}} =
	 \hat{\bm{f}}_{-\kstar-\bm{p}, \alpha'}^{\top}[(\Phi^{\alpha'}_{-\kstar-\bm{p}})^{\top}\Delta^{\dagger}_{\kstar+\bm{p}}\Phi^{\alpha}_{\kstar+\bm{p}}]\hat{\bm{f}}_{\kstar+\bm{p}, \alpha}\\
	 &\hat{\bm{c}}_{-(\kstar + \bm{p})}^{\top}h^{\top}_{-\kstar-\bm{p}}(\hat{\bm{c}}_{-(\kstar + \bm{p})}^{\dagger})^\top \rightarrow
	  \hat{\bm{c}}_{-(\kstar + \bm{p})}^{\top}[(P^{\alpha'}_{-\kstar-\bm{p}})^{*}h^{\top}_{-\kstar-\bm{p}}(P^{\alpha'}_{-\kstar-\bm{p}})^{*}](\hat{\bm{c}}_{-(\kstar + \bm{p})}^{\dagger})^\top
	 = \varepsilon_{-(\kstar+\bm{p}), \alpha'}\hat{\bm{f}}_{-\kstar-\bm{p}, \alpha'}\hat{\bm{f}}^{\dagger}_{-\kstar-\bm{p}, \alpha'}.
\end{align}
Then, the projected BdG Hamiltonian is 
\begin{align}
	&\frac{1}{2}\begin{pmatrix}
		\hat{\bm{f}}_{\kstar+\bm{p}, \alpha}^\dagger & 	\hat{\bm{f}}_{-\kstar-\bm{p}, \alpha'}^{\top}
	\end{pmatrix}
	\begin{pmatrix}
		\varepsilon_{\kstar+\bm{p}, \alpha}\mathds{1}_{\calD_{\alpha}} & \tilde{\Delta}_{\kstar+\bm{p}} \\ 
		\tilde{\Delta}_{\kstar+\bm{p}}^{\dagger} & -\varepsilon_{-(\kstar+\bm{p}), \alpha'}\mathds{1}_{\calD_{\alpha}}
	\end{pmatrix}
	\begin{pmatrix}
			\hat{\bm{f}}_{\kstar+\bm{p}, \alpha} \\	\left(\hat{\bm{f}}^{\dagger}_{-\kstar-\bm{p}, \alpha'}\right)^{\top}
	\end{pmatrix},\\
	&H^{\text{eff}}_{\bm{p}} \coloneq \begin{pmatrix}
		\varepsilon_{\kstar+\bm{p}, \alpha}\mathds{1}_{\calD_{\alpha}} & \tilde{\Delta}_{\kstar+\bm{p}} \\ 
		\tilde{\Delta}_{\kstar+\bm{p}}^{\dagger} & -\varepsilon_{-(\kstar+\bm{p}), \alpha'}\mathds{1}_{\calD_{\alpha}}
	\end{pmatrix},
\end{align}
where $\tilde{\Delta}_{\kstar+\bm{p}} = (\Phi^{\alpha}_{\kstar+\bm{p}})^{\dagger}\Delta_{\kstar+\bm{p}}(\Phi_{-\kstar-\bm{p}}^{\alpha'})^{*}$.

We discuss symmetry constraints on the above effective Hamiltonian. 
The symmetry group is $\calM_{\bk}$.
First, we see the symmetry properties $\hat{\bm{f}}^{\dagger}_{\bk}$:
\begin{align}
	\hat{g}\hat{\bm{f}}^{\dagger}_{\bk, \alpha}\hat{g}^{-1} &= \hat{g}\hat{\bm{c}}^{\dagger}_{\bk}\hat{g}^{-1}\hat{g}\Phi^{\alpha}_{\bk}\hat{g}^{-1}
	= \hat{\bm{c}}^{\dagger}_{\bk}U_{\bk}(g)(\Phi^{\alpha}_{\bk})^{\phi_g} = \hat{\bm{c}}^{\dagger}_{\bk}\Phi^{\alpha}_{\bk}u_{\bk}^{\alpha}(g) = \hat{\bm{f}}^{\dagger}_{\bk, \alpha}u_{\bk}^{\alpha}(g);\\
	\hat{g}\hat{\bm{f}}^{\dagger}_{-\bk, \alpha'}\hat{g}^{-1} &= \hat{g}\hat{\bm{c}}^{\dagger}_{-\bk}\hat{g}^{-1}\hat{g}\Phi^{\alpha'}_{-\bk}\hat{g}^{-1}
	= \hat{\bm{c}}^{\dagger}_{-\bk}U_{-\bk}(g)(\Phi^{\alpha'}_{-\bk})^{\phi_g} = \hat{\bm{c}}^{\dagger}_{-\bk}\Phi^{\alpha'}_{-\bk}u_{-\bk}^{\alpha'}(g) = \hat{\bm{f}}^{\dagger}_{-\bk, \alpha'}u_{-\bk}^{\alpha'}(g).
\end{align}
It is easy to confirm that the diagonal part of $H_{\bm{p}}^{\text{eff}}$ is symmetric under the above transformations. 
Next, we consider the off-diagonal part. 
Recall that $U_{\bk}(g)\Phi_{\bk}^{\phi_g} = \Phi_{\bk}u_{\bk}^{\alpha}(g)$.
The relations $\Phi_{\bk}^{\phi_g}[u_{\bk}^{\alpha}(g)]^{\dagger} = U_{\bk}^{\dagger}(g)\Phi_{\bk}$ and  $u_{\bk}^{\alpha}(g)[\Phi_{\bk}^{\dagger}]^{\phi_g} = \Phi^{\dagger}_{\bk}U_{\bk}$ immediately follow from the transformation of eigenvectors. 
Then, we can derive symmetry transformations of $\tilde{\Delta}_{\bk}$ as 
\begin{align}
	\hat{g}(\hat{\bm{f}}^{\dagger}_{\bk, \alpha}\tilde{\Delta}_{\bk}(\hat{\bm{f}}^{\dagger}_{-\bk, \alpha'})^{\top})\hat{g}^{-1}
	&= \hat{\bm{f}}^{\dagger}_{\bk, \alpha} u_{\bk}^{\alpha}(g)\tilde{\Delta}^{\phi_g}_{\bk}[u_{-\bk}^{\alpha'}(g)]^\top(\hat{\bm{f}}^{\dagger}_{-\bk, \alpha'})^{\top}  \nonumber \\
	&=  \hat{\bm{f}}^{\dagger}_{\bk, \alpha}u_{\bk}^{\alpha}(g)[(\Phi^{\alpha}_{\bk})^{\dagger}]^{\phi_g}\Delta^{\phi_g}_{\bk}[(\Phi^{\alpha'}_{-\bk})^{*}]^{\phi_g}[u_{-\bk}^{\alpha'}(g)]^\top(\hat{\bm{f}}^{\dagger}_{-\bk, \alpha'})^{\top} \nonumber \\
	&= \hat{\bm{f}}^{\dagger}_{\bk, \alpha}(\Phi^{\alpha}_{\bk})^{\dagger}U_{\bk}(g)\Delta_{\bk}^{\phi_g}U_{-\bk}^{\top}(g)(\Phi^{\alpha'}_{-\bk})^{*}(\hat{\bm{f}}^{\dagger}_{-\bk, \alpha'})^{\top} \nonumber \\
	&= \hat{\bm{f}}^{\dagger}_{\bk}\tilde{\Delta}_{\bk}(\hat{\bm{f}}^{\dagger}_{-\bk})^{\top},
\end{align}
where we used $\chi_g = +1$ for $^\forall g \in \calM_{\bk}$.
The above transformation implies the relation
\begin{align}
	\label{eq:tilde_delta}
	u_{\bk}^{\alpha}(g)\tilde{\Delta}^{\phi_g}_{\bk}[u_{-\bk}^{\alpha'}(g)]^\top = \tilde{\Delta}_{\bk}
\end{align}
must hold.
%

Here we prove that $\tilde{\Delta}_{\kstar} = \delta_{\kstar} \mathds{1}_{\calD_{\alpha}}\ (\delta_{\kstar} \in \mathbb{R})$.
Since time-reversal symmetry exists, we choose the gauge at $-\bk$ such that $\Phi^{\alpha'}_{-\bk} = U(\calT)(\Phi^{\alpha}_{\bk})^{*}$. 
As a result, we have
\begin{align}
	\tilde{\Delta}_{\kstar+\bm{p}} = (\Phi^{\alpha}_{\kstar+\bm{p}})^{\dagger}\left[\Delta_{\kstar+\bm{p}}U^*(\calT)\right]\Phi^{\alpha}_{\kstar+\bm{p}}.
\end{align}
Correspondingly, we also obtain effective chiral symmetry. 
\begin{align}
	u(\gamma) &\coloneqq \begin{pmatrix}
		\Phi^{\alpha}_{\bk} & O \\
		O & U^{*}(\calT)\Phi^{\alpha}_{\bk}
	\end{pmatrix}^{\dagger}U(\gamma)\begin{pmatrix}
	\Phi^{\alpha}_{\bk} & O \\
	O & U^{*}(\calT)\Phi^{\alpha}_{\bk}
	\end{pmatrix}\nonumber \\
	&= \begin{pmatrix}
		\Phi^{\alpha}_{\bk} & O \\
		O & U^{*}(\calT)\Phi^{\alpha}_{\bk}
	\end{pmatrix}^{\dagger}
	\begin{pmatrix}
		O & \ii U(\calT) \\
		\ii U^*(\calT) & O
	\end{pmatrix}
	\begin{pmatrix}
		\Phi^{\alpha}_{\bk} & O \\
		O & U^{*}(\calT)\Phi^{\alpha}_{\bk}
	\end{pmatrix}\nonumber \\
	&=\begin{pmatrix}
		&  -\ii \mathds{1}_{\calD_{\alpha}} \\
		\ii \mathds{1}_{\calD_{\alpha}} & 
	\end{pmatrix}.
\end{align}
Suppose that $\kstar$ is a Fermi point on a line where the $\Delta_{\kstar}$ is nonzero. 
In addition to Eq.~\eqref{eq:tilde_delta}, the relation $u(\gamma)H^{\text{eff}}_{\bm{p}} = - H^{\text{eff}}_{\bm{p}}u(\gamma)$ holds.
These symmetry conditions give rise to the following constraints on $\tilde{\Delta}_{\kstar}$
\begin{align}
	\label{eq:delta_constraint1}
	&u_{\kstar}^{\alpha}(g)\tilde{\Delta}_{\kstar}^{\phi_g}[u_{\kstar}^{\alpha}(g)]^{\dagger} = \tilde{\Delta}_{\kstar};\\
	\label{eq:delta_constraint2}
	&\tilde{\Delta}_{\kstar} = \tilde{\Delta}_{\kstar}^{\dagger},
\end{align}
where we used the fact that $u_{-\bk}^{\alpha'}(g) = [u_{\bk}^{\alpha}(g)]^*$ hold.
Notice that constraints~\eqref{eq:delta_constraint1} and \eqref{eq:delta_constraint2} are the same as those for Hamiltonians. 
As a Hamiltonian is proportional to the identity matrix with an eigenenergy, 
\begin{align}
	\label{eq:intra_pairing}
	\tilde{\Delta}_{\kstar} = \delta_{\kstar}\mathds{1}_{\calD_{\alpha}}\ (\delta_{\kstar} \in \mathbb{R})
\end{align}
is a unique solution~%
\footnote{
The proof is almost the same as the Schur’s lemma.
Since $\Delta_{\kstar}$ is Hermitian, there exists a real eigenvalue $\delta_{\kstar} \in \mathbb{R}$ and its eigenvector $\bm{u}$ such that $\Delta_{\kstar}\bm{u} = \delta_{\kstar}\bm{u}$.
In other words,
\begin{align}
	\det \left(\Delta_{\kstar} - \delta_{\kstar}\mathds{1}\right) = 0.
\end{align}
From Eqs.~\eqref{eq:delta_constraint1} and \eqref{eq:delta_constraint2}, one can see that
\begin{align}
	\left(\Delta_{\kstar} - \delta_{\kstar}\mathds{1}\right)^{\phi_g}[u_{\bk}^{\alpha}(g)]^{\dagger}\bm{u} 
	= [u_{\bk}^{\alpha}(g)]^{\dagger} \left(\Delta_{\kstar} - \delta_{\kstar}\mathds{1}\right) \bm{u} =\bm{0} \quad \text{for } ^\forall g \in \calM_{\kstar}.
\end{align}
Since $u_{\bk}^{\alpha}(g)$ is an irreducible representation, $\{[u_{\bk}^{\alpha}(g)]^\dagger\bm{u}\}_{g\in\calM_{\bk}}$ can span the invariant subspace of the vector space.
Therefore, $\left(\Delta_{\kstar} - \delta_{\kstar}\mathds{1}\right)^{\phi_g} = O$, which immediately implies $\Delta_{\kstar} = \delta_{\kstar}\mathds{1}$.
}%
.

Finally, we obtain the $q$-matrix of the effective Hamiltonian. 
When we expand the projected Hamiltonian near $\kstar$, we have
\begin{align}
	\begin{pmatrix}
		\hat{\bm{f}}_{\kstar+\bm{p}, \alpha}^\dagger & 	\hat{\bm{f}}_{-\kstar-\bm{p}, \alpha'}^{\top}
	\end{pmatrix}
	\begin{pmatrix}
		(\bm{v}_f \cdot \bm{p})\mathds{1}_{\calD_{\alpha}} & \delta_{\kstar} \mathds{1}_{\calD_{\alpha}} \\ 
		\delta_{\kstar} \mathds{1}_{\calD_{\alpha}} & -(\bm{v}_f\cdot \bm{p})\mathds{1}_{\calD_{\alpha}}
	\end{pmatrix}
	\begin{pmatrix}
		\hat{\bm{f}}_{\kstar+\bm{p}, \alpha} \\	\left(\hat{\bm{f}}^{\dagger}_{-\kstar-\bm{p}, \alpha'}\right)^{\top}
	\end{pmatrix},
\end{align}
where $\bm{v}_f \coloneq \nabla_{\bk}\varepsilon_{\bk, \alpha}\vert_{\bk=\kstar}$ and we used the relations $\varepsilon_{\bk, \alpha} = \varepsilon_{-\bk, \alpha'}$.
To define the $q$-matrix, we diagonalize $u(\gamma)$ by $\mathcal{U}_{\pm} = \frac{1}{\sqrt{2}}\begin{pmatrix}
	\mathds{1}_{\calD_{\alpha}} \\
	\pm \ii \mathds{1}_{\calD_\alpha}
\end{pmatrix}$. That is,
\begin{align}
	(\mathcal{U}_{+}\  \mathcal{U}_{-})^{\dagger}u(\gamma)(\mathcal{U}_{+}\  \mathcal{U}_{-}) = \text{diag}(\mathds{1}_{\calD_{\alpha}}\ -\mathds{1}_{\calD_{\alpha}}).
\end{align}
Correspondingly, the Hamiltonian is also transformed as
\begin{align}
	&(\mathcal{U}_{+}\  \mathcal{U}_{-})^{\dagger}H^{\text{eff}}_{p}(\mathcal{U}_{+}\  \mathcal{U}_{-}) = \begin{pmatrix}
		& (\tilde{q}^\alpha)^{\dagger}\\
		\tilde{q}^{\alpha} & 
	\end{pmatrix};\\
	\label{eq:qeff}
	&\tilde{q}^\alpha = \mathcal{U}_{-}^{\dagger}H^{\text{eff}}_{p}\mathcal{U}_{+} = (\bm{v}_f \cdot \bm{p} + \ii \delta) \mathds{1}_{\calD_\alpha}.
\end{align}

We remark that $\tilde{q}^{\alpha}$ is defined for an irrep of $\calM_{\bk}$, while the $q$-matrices are defined for irreps of $\calGk$ in \ref{sec:review}.
When antiunitary symmetries in $\calA_{\bk}$ do not change an irrep or $\calA_{\bk}$ is the empty set, we can use $\tilde{q}^{\alpha}$ as the $q$-matrix of the effective Hamiltonian. 
On the other hand, if they form a pair of an irrep of $\calGk$ and another irrep of $\calGk$, we obtain the effective $q$-matrix for an irrep of $\calGk$ from $\tilde{q}^{\alpha}$.
The unitary representation of $g \in \calM_{\bk}$ for $H^{\text{eff}}_{p}$ is given by
\begin{align}
	U^{\text{eff}}(g) = \begin{pmatrix}
		u_{\bk}^{\alpha}(g) & O \\
		O & u_{\bk}^{\alpha}(g)
	\end{pmatrix}
\end{align}
in our gauge choice.
One can see that the representation is the same even in the basis such that the effective chiral symmetry is diagonalized. 
Then, from symmetry relations, we find
\begin{align}
	\begin{pmatrix}
		u_{\bk}^{\alpha}(g) & O \\
		O & u_{\bk}^{\alpha}(g)
	\end{pmatrix}
	\begin{pmatrix}
		& (\tilde{q}^\alpha)^{\dagger}\\
		\tilde{q}^{\alpha} & 
	\end{pmatrix}
	\begin{pmatrix}
		u_{\bk}^{\alpha}(g) & O \\
		O & u_{\bk}^{\alpha}(g)
	\end{pmatrix}^{\dagger}\quad (g \in \calGk),
\end{align}
which leads to the symmetry constraints on $\tilde{q}^{\alpha}$:
\begin{align}
	u_{\bk}^{\alpha}(g)\tilde{q}^{\alpha}[u_{\bk}^{\alpha}(g)]^{\dagger} = \tilde{q}^{\alpha} \ \ (g\in \calGk).
\end{align}
Suppose irrep $\alpha$ of $\calM_{\bk}$ is composed of two irreps of $\calGk$ related to each other by antiunitary symmetries in $\calA_{\bk}$. (Here, $\alpha'$ and $\calT\alpha'$ denote these two irreps). 
In this case, the above equation is further written as
\begin{align}
	\begin{pmatrix}
		u_{\bk}^{\alpha'}(g) & O \\
		O & u_{\bk}^{\calT\alpha'}(g)
	\end{pmatrix}
	\begin{pmatrix}
		(\bm{v}_f \cdot \bm{p}+ \ii \delta)\mathds{1}_{d_{\alpha'}}  & O\\
		O & (\bm{v}_f \cdot \bm{p}+ \ii \delta)\mathds{1}_{d_{\alpha'}} 
	\end{pmatrix}
	\begin{pmatrix}
		u_{\bk}^{\alpha'}(g) & O \\
		O & u_{\bk}^{\calT\alpha'}(g)
	\end{pmatrix}^\dagger = 	\begin{pmatrix}
	(\bm{v}_f \cdot \bm{p}+ \ii \delta)\mathds{1}_{d_{\alpha'}}  & O\\
	O & (\bm{v}_f \cdot \bm{p}+ \ii \delta)\mathds{1}_{d_{\alpha'}} 
	\end{pmatrix},
\end{align}
Therefore, we can consider $q^{\alpha'} \coloneqq (\bm{v}_f \cdot \bm{p})\mathds{1}_{d_{\alpha'}}$ as the effective $q$-matrix for irrep $\alpha'$ of $\calGk$.
In summary, we define the effective $q$-matrices for irreps of $\calGk$, which correspond to the $q$-matrices in \ref{sec:review}, by
\begin{align}
	\label{eq:eff_qmatrix}
	q^{\alpha} &= \begin{cases}
		(\bm{v}_f \cdot \bm{p}+ \ii \delta)\mathds{1}_{d_{\alpha}} \\
		(\bm{v}_f \cdot \bm{p}+ \ii \delta)\mathds{1}_{2d_{\alpha}}
	\end{cases}\\
	&=(\bm{v}_f \cdot \bm{p}+ \ii \delta)\mathds{1}_{\scrD_{\alpha}}
\end{align}

\subsubsection{Detailed derivation of Fermi surface formulas}
\label{app:derivation}
As the first step, using the assumptions we introduce in the main text, we compute the $w_{(l, \alpha)}$. 
We consider a line segment $l$ connecting to two momenta $\bk_0$ and $\bk_1$.
The positive direction of the line segment $l$ is given by $\hat{\bm{e}}_l = (\bk_1 - \bk_0)/\|\bk_1 - \bk_0\|$.
The displacement from the Fermi point $\kstar$ on the line segment is represented by $\bm{p} = p \hat{\bm{e}}_l$.
As assumption (ii), we assume that the superconducting gap is negligible except at Fermi points.
Therefore, $\det \tilde{q}^\alpha$ is real away from Fermi points, and we focus on near a Fermi point. 
The determinant of the effective $q$-matrix $q^\alpha$ is expressed by
\begin{align}
	\det q^\alpha = (v^{(l, \alpha)} p + \ii \delta^{(l, \alpha)})^{\scrD_\alpha} = (r e^{\ii \theta})^{\mathscr{D}_{\alpha}},
\end{align}
where $v^{(l, \alpha)} \coloneqq \bm{v}_f \cdot \hat{\bm{e}}_l = \nabla_{\bk}\varepsilon_{\bk, \alpha}\vert_{\bk=\kstar} \cdot \hat{\bm{e}}_l$ and $\delta^{(l, \alpha)}$ is the coefficient of Eq.~\eqref{eq:intra_pairing} for the Fermi point on the line segment. 
Also, $r = \sqrt{(v^{(l, \alpha)})^2p^2 + (\delta^{(l, \alpha)})^2}$, $\cos \theta = \frac{v^{(l, \alpha)} p}{r}$, and $\sin \theta = \frac{\delta^{(l, \alpha)}}{r}$.
Then, one can find that $\Im \log \det  q^\alpha$ acquires $-\scrD_{\alpha} \pi \text{sgn}(v^{(l, \alpha)})\text{sgn}(\delta^{(l, \alpha)})$ when a single Fermi point exists on the line segment.
When there are $n_{(l, \alpha)}$ minimally degenerate Fermi points on the line segment $l$, the winding of $q$-matrix for the target Hamiltonian is computed by
\begin{align}
	\label{eq:winding_target}
	\int_{0}^{1} ds \partial_s \log \det q_{\bk_s}^{\alpha} = \log \frac{\det q_{\bk_1}^{\alpha}}{\det q_{\bk_0}^{\alpha}} - \ii \pi \mathscr{D}_{\alpha} \sum_{m=1}^{n_{(l, \alpha)}}\text{sgn}(v_{m}^{(l, \alpha)})\text{sgn}(\delta_{m}^{(l, \alpha)})
\end{align}
where each momentum in the line segment is given by $\bk_s = \bk_{0} + s (\bk_1 - \bk_0)\ (0 \leq s \leq 1)$.  
On the other hand, the vacuum Hamiltonian does not have any Fermi point. Thus, 
\begin{align}
	\label{eq:winding_vac}
	\int_{0}^{1} ds \partial_s \log \det (q_{s}^{\alpha})^{\text{vac}} =  \log \frac{\det (q_{\bk_1}^{\alpha})^{\text{vac}}}{\det (q_{\bk_0}^{\alpha})^{\text{vac}}}.
\end{align}
Combining Eqs.~\eqref{eq:winding_target} and \eqref{eq:winding_vac}, we find
\begin{align}
	w_{(l, \alpha)} &= \frac{1}{2\pi \ii \mathscr{D}_{\alpha}}	\int_{0}^{1} ds (\partial_s \log \det q_{\bk_s}^{\alpha} - \partial_s \log \det (q_{\bk_s}^{\alpha})^{\text{vac}})  \nonumber \\
	\label{eq:wa_weak}
	&=\frac{1}{2 \pi \ii \mathscr{D}_{\alpha}}\log \frac{\det q_{\bk_1}^{\alpha}/\det (q_{\bk_1}^{\alpha})^{\text{vac}}}{\det q_{\bk_0}^{\alpha}/\det (q_{\bk_1}^{\alpha})^{\text{vac}}} - \frac{1}{2} \sum_{m=1}^{n_{(l, \alpha)}}\text{sgn}(v_{m}^{(l, \alpha)})\text{sgn}(\delta_{m}^{(l, \alpha)}).
\end{align}
It should be noted that the $q$-matrices in the first term are defined for irreps on the line segment at this moment.

Due to compatibility relations, the first term of Eq.~\eqref{eq:wa_weak} can be rewritten in terms of $q$-matrices for irreps at the endpoints.
Recall $M$ is a matrix that contains compatibility relations between all line segments and their endpoints. 
As a result, we have 
\begin{align}
	\label{eq:q_decomp}
	\left(\frac{\det q_{\bk_1}^{\alpha}/\det (q_{\bk_1}^{\alpha})^{\text{vac}}}{\det q_{\bk_0}^{\alpha}/\det (q_{\bk_0}^{\alpha})^{\text{vac}}}\right)^{1/\mathscr{D}_{\alpha}} = \prod_{\beta}\left({\det q_{\bk_\beta}^{\beta}/\det (q_{\bk_\beta}^{\beta})^{\text{vac}}}\right)^{-M_{\alpha\beta}/\mathscr{D}_\beta},
\end{align}
where $\bk_\beta = \bk_0 \text{ or }\bk_1$ depending on momentum on which $\beta$ is defined, and $q_{\bk_{\beta}}^{\beta}$ is a $q$-matrix for an irreducible representation $\beta$ on $\bk_0$ or $\bk_1$. 
The proof can be found in Ref.~\cite{Ono-Shiozaki2023}. 
Alternatively, here, we verify the above expression based on our assumptions in the main text.
Since we assume that the superconducting gap is negligible except at Fermi points, $q_{\bk_s}^{\alpha} \sim h_{\bk_s}^{\alpha}$ at momentum $\bk_s$ away from Fermi points, where $h_{\bk_s}^{\alpha}$ is an irrep sector of the normal conducting Hamiltonian. 
Then, the determinant of $q_{\bk_s}^{\alpha}$ is just product of eigenvalues of $h_{\bk_s}^{\alpha}$, i.e.,
\begin{align}
	(\det q_{\bk_s}^{\alpha})^{1/\mathscr{D}_{\alpha}}={{\prod_{z_\alpha}}} \vert \varepsilon_{\bk_s, z_\alpha} \vert,
\end{align}
where $\varepsilon_{\bk_s, z_\alpha}$ is the $z_\alpha$-th energy level of $h_{\bk_s}^{\alpha}$.
We can consider the limits $s\rightarrow 0$ and $s\rightarrow 1$. 
In this limit, there are additional degeneracy in the sense of irreps on the line segment $(s\neq 0, 1)$.
This additional degeneracy is nothing but the absolute value of the elements of the matrix $M$ of compatibility relations.
Then, the following relation holds
\begin{align}
	(\det q_{\bk_{i=0,1}}^{\alpha})^{1/\mathscr{D}_{\alpha}} = \prod_{\beta}^{'}\prod_{z_\beta} \vert \varepsilon_{\bk_i, z_\beta} \vert^{\vert M_{\alpha\beta} \vert}, 
\end{align}
where $\overset{'}{\prod}$ indicates that indices run only over irreducible representations at $\bk_i$.
Taking the orientation of the line segment into account, we find
\begin{align}
	(\det q_{\bk_1}^{\alpha}/\det q_{\bk_0}^{\alpha})^{1/\mathscr{D}_{\alpha}} &= \prod_{\beta}\prod_{z_\beta} \vert \varepsilon_{\bk_\beta, z_\beta} \vert^{-M_{\alpha\beta}} \nonumber \\
	&= \prod_\beta (\det q_{\bk_\beta})^{-M_{\alpha\beta}/\scrD_\beta}.
\end{align}

Recall that the information about topological invariants is encoded in two tuples $\bm{x}$ and $\bm{v}$.
They must hold the relations
\begin{align}
	\label{eq:CR_Z}
	\bm{x}^\top M = \bm{0}
\end{align}
for $\mZ$-valued invariants and 
\begin{align}
	\label{eq:CR_Zn}
	\bm{x}^\top M = \lambda\bm{v}
\end{align}
for $\mZ_\lambda$-valued invariants. 
For later discussions, we discuss more details of the compatibility relations.
Let $N_{\textsf{b}=(\bk_p, \beta)}$ be the number of occupied states with irrep $\beta$ of $\calG_{\bk_p}$.
Again, we consider a line segment described by $\bk_s = \bk_{0} + s (\bk_1 - \bk_0)\ (0 < s < 1)$.
In general, $\{N_{(\bk_0, \beta)}\}_\beta$ and $\{N_{(\bk_1, \beta)}\}_\beta$ are not independent due to the band connectivity.
Actually, the relations among these irreps on all line segments and those of their endpoints are exactly what we call compatibility relations. 
When the system is gapped, the following conditions must be satisfied:
\begin{align}
	\sum_{\textsf{b}}M_{{\textsf{ab}}} N_{\textsf{b}} = 0\quad (\text{for } ^\forall {\textsf{a}}=(l, \alpha)).
\end{align}
Conversely, if there exists ${\textsf{a}}=(l, \alpha)$ such that
\begin{align}
	\sum_{\textsf{b}}M_{{\textsf{ab}}} N_{\textsf{b}} \neq 0,
\end{align}
there must exist Fermi points with irrep $\alpha$ on the line segment $l$.
In particular, we can rewrite the sum of signs of Fermi velocities by
 \begin{align}
 	\label{eq:CR_velocity}
 	\sum_{m=1}^{n_{(l, \alpha)}}\text{sgn}(v_{m}^{(l, \alpha)}) = \sum_{\textsf{b}}M_{{\textsf{ab}}} N_{\textsf{b}}\quad (\text{for } {\textsf{a}}=(l, \alpha)).
 \end{align}

Now, we are ready to derive Fermi surface formulas under the weak-pairing assumption. 
Let us start with the $\mZ$-valued invariants in Eqs.~\eqref{eq:gapless_Z} and \eqref{eq:gapped_Z}. 
\begin{align}
	\mathcal{W}^{\text{gapped}/\text{gapless}} &= \sum_{l, \alpha} x_{(l, \alpha)}w_{(l, \alpha)} \nonumber \\
	&\overset{\text{Eq.~\eqref{eq:wa_weak}}}{=} \sum_{l, \alpha} x_{(l, \alpha)}\left(\frac{1}{2 \pi \ii \scrD_{\alpha}}\log \frac{\det q_{\bk_1}^{\alpha}/\det (q_{\bk_1}^{\alpha})^{\text{vac}}}{\det q_{\bk_0}^{\alpha}/\det (q_{\bk_1}^{\alpha})^{\text{vac}}} - \frac{1}{2} \sum_{m=1}^{n_{(l, \alpha)}}\text{sgn}(v_{m}^{(l, \alpha)})\text{sgn}(\delta_{m}^{(l, \alpha)})\right)\nonumber \\
	&\overset{\text{Eq.~\eqref{eq:q_decomp}}}{=}  \frac{1}{2 \pi \ii}\log \prod_{\beta}\left(\det q_{\bk_\beta}^{\beta}/\det (q_{\bk_\beta}^{\beta})^{\text{vac}}\right)^{-\sum_{{\textsf{a}}}[\bm{x}]_{{\textsf{a}}}M_{{\textsf{ab}}}/\scrD_\beta} - \frac{1}{2}\sum_{l, \alpha}\sum_{m=1}^{n_{(l, \alpha)}}x_{(l, \alpha)}\text{sgn}(v_{m}^{(l, \alpha)})\text{sgn}(\delta_{m}^{(l, \alpha)})\nonumber \\
	&\overset{\text{Eq.~\eqref{eq:CR_Z}}}{=}  - \frac{1}{2}\sum_{l, \alpha}\sum_{m=1}^{n_{(l, \alpha)}}x_{(l, \alpha)}\text{sgn}(v_{m}^{(l, \alpha)})\text{sgn}(\delta_{m}^{(l, \alpha)}) \nonumber \\
	&\overset{\text{Eqs.~\eqref{eq:CR_Z} and \eqref{eq:CR_velocity}}}{=}\sum_{l, \alpha}\sum_{m=1}^{n_{(l, \alpha)}}x_{(l, \alpha)}\text{sgn}(v_{m}^{(l, \alpha)})\frac{1-\text{sgn}(\delta_{m}^{(l, \alpha)})}{2},
\end{align}
where we use sequential labels $\textsf{a} = (l, \alpha)$ and $\textsf{b} = (p, \beta)$.
In the last line, we used
\begin{align}
	\label{eq:velocity_equality}
	\sum_{l, \alpha}x_{(l, \alpha)}\sum_{m=1}^{n_{(l, \alpha)}}\text{sgn}(v_{m}^{(l, \alpha)}) &= \sum_{{\textsf{a}}}[\bm{x}]_{{\textsf{a}}}\sum_{\textsf{b}}M_{{\textsf{ab}}} N_{\textsf{b}} \nonumber \\
	&\overset{\text{Eq.~\eqref{eq:CR_Z}}}{=}0.
\end{align}

For the $\mZ_{\lambda}$-valued invariants in Eq.~\eqref{eq:Zn}, 
\begin{align}
	\exp\hspace{-0.5mm}\left[\frac{2\pi \mathrm{i}}{\lambda}\mathcal{X}\right] &= \frac{\exp[-2\pi \ii \sum_{l, \alpha}x_{(l,\alpha)}w_{(l,\alpha)}/\lambda]}{\prod_{p, \beta}\left(\mathcal{Z}[q_{\bk}^{\beta}]/\mathcal{Z}[(q^{\beta}_{\bk})^{\text{vac}}]\right)^{v_{(p, \beta)}}} \nonumber \\
	&=\frac{\exp[-\frac{1}{\lambda}\log \prod_{p, \beta} \left(\det q_{\bk_p}^{\beta}/\det (q_{\bk_p}^{ \beta})^{\text{vac}}\right)^{-\sum_{{\textsf{a}}}[\bm{x}]_{{\textsf{a}}}M_{{\textsf{ab}}}/\scrD_\beta} -\frac{\ii \pi}{\lambda}\sum_{l, \alpha}\sum_{m=1}^{n_{(l, \alpha)}}x_{(l, \alpha)}\text{sgn}(v_{m}^{(l, \alpha)})\text{sgn}(\delta_{m}^{(l, \alpha)}) ]}{\prod_{\beta}\left(\mathcal{Z}[q_{\bk}^{\beta}]/\mathcal{Z}[(q^{\beta}_{\bk})^{\text{vac}}]\right)^{[\bm{v}_{j}]_{\beta}}} \nonumber\\
	&\overset{\text{Eq.~\eqref{eq:CR_Zn}}}{=}\frac{\exp[-\frac{1}{\lambda}\log \prod_{p, \beta} \left(\det q_{\bk_p}^{\beta}/\det (q_{\bk_p}^{\beta})^{\text{vac}}\right)^{-\lambda v_{(p, \beta)}/\scrD_\beta} -\frac{\ii \pi}{\lambda}\sum_{l, \alpha}\sum_{m=1}^{n_{(l, \alpha)}}x_{(l, \alpha)}\text{sgn}(v_{m}^{(l, \alpha)})\text{sgn}(\delta_{m}^{(l, \alpha)}) ]}{\prod_{p, \beta}\left(\mathcal{Z}[q_{\bk_p}^{\beta}]/\mathcal{Z}[(q^{\beta}_{\bk_p})^{\text{vac}}]\right)^{v_{(p, \beta)}}} \nonumber \\
	&=\frac{\prod_{p, \beta} \left(\det q_{\bk_p}^{\beta}/\det (q_{\bk_p}^{\beta})^{\text{vac}}\right)^{v_{(p, \beta)}/\scrD_\beta}}{\prod_{p, \beta}\left(\mathcal{Z}[q_{\bk_p}^{\beta}]/\mathcal{Z}[(q^{\beta}_{\bk_p})^{\text{vac}}]\right)^{v_{(p, \beta)}}}\exp\left[-\frac{\ii \pi}{\lambda}\sum_{l, \alpha}\sum_{m=1}^{n_{(l, \alpha)}}x_{(l, \alpha)}\text{sgn}(v_{m}^{(l, \alpha)})\text{sgn}(\delta_{m}^{(l, \alpha)})\right].
\end{align}
It should be noted that, in the weak-pairing limit,
\begin{align}
	\frac{(\det q_{\bk_p}^{\beta})^{1/\scrD_\beta}}{\mathcal{Z}[q_{\bk_p}^{\beta}]} = \frac{\prod_{z_\beta} \vert \varepsilon_{\bk_p, z_\beta} \vert}{\prod_{z_\beta} \varepsilon_{\bk_p, z_\beta}} = \prod_{z_\beta} \text{sgn}(\varepsilon_{\bk_{\beta}, z_\beta}) = (-1)^{N_{\bk_p}^{\beta}},
\end{align}
where $N_{\bk_p}^{\beta}$ is the number of occupied states with irrep $\beta$ at $\bk_p$.
As a result, we find the following Fermi surface formula:
\begin{align}
	\exp\hspace{-0.5mm}\left[\frac{2\pi \mathrm{i}}{\lambda}\mathcal{X}\right] &= (-1)^{\sum_{p, \beta}N_{\bk_p}^{\beta} v_{(p, \beta)} }\exp\left[-\frac{\ii \pi}{\lambda}\sum_{l, \alpha}\sum_{m=1}^{n_{(l, \alpha)}}x_{(l, \alpha)}\text{sgn}(v_{m}^{(l, \alpha)})\text{sgn}(\delta_{m}^{(l, \alpha)})\right].
\end{align}
We can rewrite the first part $ (-1)^{\sum_{p, \beta}N_{\bk_p}^{\beta} v_{(p, \beta)} }$ using the relation \eqref{eq:CR_Zn}. 
\begin{align}
	(-1)^{\sum_{p, \beta}N_{\bk_p}^{\beta} v_{(p, \beta)} }&= \exp\left[\pm \ii \pi \sum_{p, \beta}N_{\bk_p}^{\beta} v_{(p, \beta)}\right] \nonumber \\
	&= \exp\left[\pm \ii \pi \left(\sum_{\textsf{a}=(l, \alpha)}\sum_{\textsf{b}=(p, \beta)}\frac{1}{\lambda}x_{(l, \alpha)} M_{{\textsf{ab}}}N_{\bk_p}^{\beta} \right)\right] \nonumber \\
	&\overset{\text{Eq.~\eqref{eq:CR_velocity}}}{=}\exp\left[\pm \ii \pi \left(\sum_{\textsf{a}=(l, \alpha)}\frac{1}{\lambda}x_{(l, \alpha)}\sum_{m=1}^{n_{(l, \alpha)}}\text{sgn}(v_{m}^{(l, \alpha)})  \right)\right].
\end{align}
It should be noted that $\left(\cdot\right)$ is ensured to be an integer.
Finally, we arrive at 
\begin{align}\label{SM_eq_formula1}
	\exp\hspace{-0.5mm}\left[\frac{2\pi \mathrm{i}}{\lambda}\mathcal{X}\right] &= \exp\left[\frac{2 \pi\ii}{\lambda}\sum_{l, \alpha}\sum_{m=1}^{n_{(l, \alpha)}}x_{(l, \alpha)}\text{sgn}(v_{m}^{(l, \alpha)})\frac{1-\text{sgn}(\delta_{m}^{(l, \alpha)})}{2}\right].
\end{align}
Therefore, we have
\begin{align}\label{SM_eq_formula2}
	\mathcal{X} = \sum_{l, \alpha}\sum_{m=1}^{n_{(l, \alpha)}}x_{(l, \alpha)}\text{sgn}(v_{m}^{(l, \alpha)})\frac{1-\text{sgn}(\delta_{m}^{(l, \alpha)})}{2} \mod \lambda. 
\end{align}

\subsection{Limitation for the strength of the interband pairing}

\begin{figure}[b]
	\begin{center}
		\includegraphics[width=0.8\columnwidth]{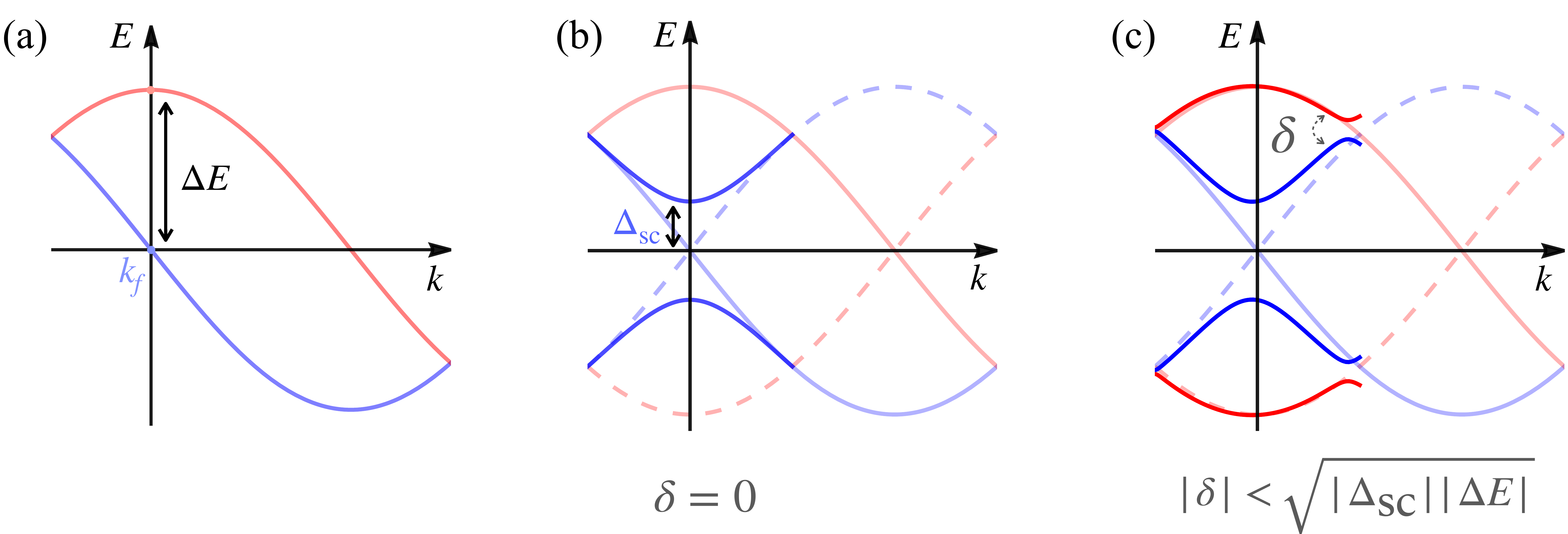}
		\caption{\label{fig_Reply}
		The illustration of the three energy scales.
		(a) The band structure of the normal state Hamiltonian, where $\Delta E$ is the normal state energy closest to the Fermi energ at $k_f$.
		Different bands are marked in different colors.
		(b) The solid blue lines represent the effective BdG Hamiltonian with a gap opened by intraband pairing $\Delta_{\mathrm{sc}}$.
		(c) The BdG band structure with finite interband pairing $\delta$.
		In (b) and (c), $\Delta_{\text{sc}}$ is the energy gap only caused by intraband pairing at the Fermi point $k_f$, and $\delta$ is the interband pairing between two bands. Dashed lines denote hole bands.
		The semi-transparent lines represent the effective BdG Hamiltonian with $\Delta_{\mathrm{sc}} =\delta = 0$.
		}
	\end{center}
\end{figure}

In the main text, we assume that intraband pairings dominate, i.e., all interband pairings are negligible. 
This assumption is explicitly used in the derivation presented above. 
As such, it is natural to ask whether or not our formulas are applicable systems with interband pairings. 
Furthermore, if it is possible, how large can the interband pairings be? 
As shown below, the answer to the first question is affirmative---our formulas can indeed be applied to systems with a certain extent of interband pairings. 
In the following, we provide an argument, based on an effective model, to establish the limitations on the strength of interband pairings.

In our formulas Eqs.~\eqref{SM_eq_formula1} and \eqref{SM_eq_formula2}, each Fermi point contributes to the topological invariants independently.
Let us start with the limit where each Fermi point is completely decoupled, i.e, the assumption is strictly satisfied.
In this limit, our BdG Hamiltonian includes only intraband pairing components, which are nonzero only near the Fermi points.
It is clear that even with significant intraband pairing components, the system does not undergo a topological phase transition in this limit.
Thus, we build an effective Hamiltonian $H^{\mathrm{eff}}(k_f)$ at a Fermi point $\bk_f$.
There are three energy scales: the normal state energy $\Delta E$ closest to the Fermi energy (Fig.~\ref{fig_Reply}(a)), the intraband pairing $\Delta_{\text{sc}}$ of the band crossing the Fermi energy (colored with blue in Fig.~\ref{fig_Reply}(b)), and the interband pairing $\delta$ between them (Fig.~\ref{fig_Reply}(c)).
See Fig.~\ref{fig_Reply} for an illustration.
Following the discussion in Section.~\ref{sec:derivation}, the effective Hamiltonian can be written as
\begin{equation}
	H^{\mathrm{eff}}(k_f)=\begin{pmatrix}
		0 & 0 & \Delta_{\text{sc}} & \delta\\
		0 & \Delta E & \delta^* & 0\\
		\Delta_{\text{sc}} & \delta & 0 & 0\\
		\delta^* & 0 & 0 & -\Delta E\\
	\end{pmatrix}.
\end{equation}
A necessary condition for topological phase transitions is gap closing, i.e., $\det H^{\text{eff}}(k_f) = \Delta_{\text{sc}}^2 (\Delta E)^2 + \vert\delta\vert^4 = 0$. 
From this expression, we find that the gap closes at $\bk_f$ when $\delta = \pm e^{\pm i\pi/4} \sqrt{\vert\Delta_{\text{sc}}\vert \vert\Delta E\vert}$~\footnote{For simplicity, we ignore the energy of the band crossing the Fermi energy. When the energy is taken into account, the gap closing point is shifted from the Fermi point $\bk_f$.}.
Thus, we conclude that our formula remains valid as long as $\vert \delta \vert < \sqrt{\vert\Delta_{\text{sc}}\vert \vert\Delta E\vert}$.

\clearpage
\section{Gapless and gapped topological invariants for all representative models in SG $P4_1$}
In this section, we calculate all the gapless and gapped topological invariant for all representative models of the real-space construction in SG $P4_1$.

\begin{figure}[h]
	\begin{center}
		\includegraphics[width=0.99\columnwidth]{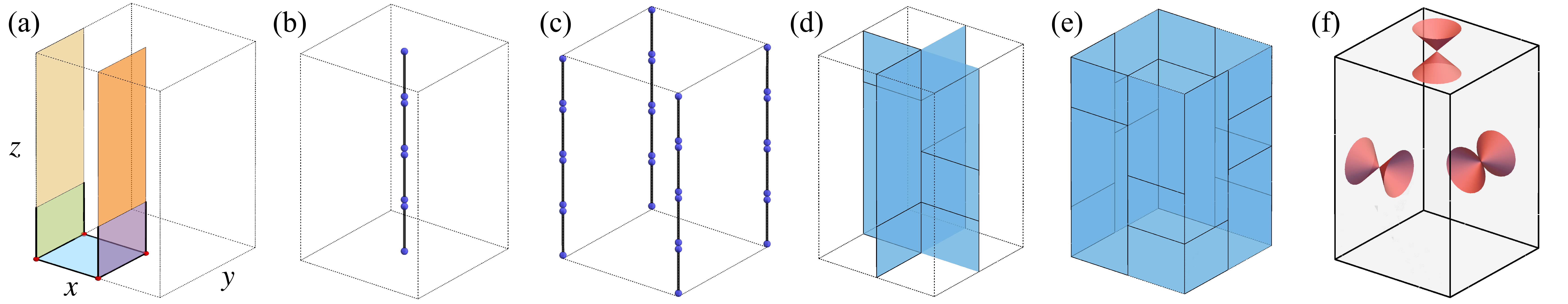}
		\caption{\label{fig:SM_AU}
		(a) Cell complex structure for the SG $P4_1\ (\#.76)$.
		The dashed lines denote the unit cell.
		The symmetry-inequivalent 2-cells, 1-cells and 0-cells are represented as colored faces, bold lines and red dots, respectively.
		The other cells can be obtained from these by acting with symmetry operations in SG $P4_1$.
		(b)-(c) The patchwork for the two $\mZ_2$ phases constructed from 1D TSCs. The black bold lines represent the nontrivial 1D TSCs in DIII class, and the blue dots represent the corresponding Majorana zero modes at the endpoints of 1D TSCs.
		(d)-(e) The patchwork for the two $\mZ_2$ phases constructed from 2D TSCs. The blue faces represent the nontrivial 2D TSCs in DIII class. The regions enclosed by solid lines represent the decorations on the symmetry-inequivalent 2-cells.
		(f) The illustration for the strong TSC in DIII class.
		}
	\end{center}
\end{figure}

Based on the real-space construction, there are two $\mZ_2$ phases constructed from 1D TSCs, two $\mZ_2$ phases constructed from 2D TSCs and a strong TSC characterized by the 3D winding number $w_{\text{3D}}$.
The corresponding patchwork or surface states are shown in Fig.~\ref{fig:SM_AU}(b)-(f).
Due to on the each 1-cell and 2-cell, the on-site symmetries are TRS, PHS and the corresponding chiral symmetry.
The nontrivial decorations on them are $\mZ_2$-type.
It means all gapless edge state can be gapped out with each other, and the non-open-edge condition are satisfied.

Our Fermi surface formulas in the space group $P4_1$ have two $\mZ$-valued gapless topological invariants, one $\mZ_2$-valued gapped topological invariant, one $\mZ_4$-valued gapped topological invariant, and one $\mZ_8$-valued gapped topological invariant.
The expressions of two $\mZ$-valued topological invariants for gapless phases are:
\begin{align}\label{eq:SMgapless}
\mathcal{W}^{\text{gapless}}_{1} &= \sum_{m=1}^{n_{a_1}}\text{sgn}(v_{m}^{a_1})\frac{1-\text{sgn}(\delta^{a_1}_{m})}{2}-2\sum_{m=1}^{n_{c_1}}\text{sgn}(v_{m}^{c_1})\frac{1-\text{sgn}(\delta^{c_1}_{m})}{2}-\sum_{m=1}^{n_{e_1}}\text{sgn}(v_{m}^{e_1})\frac{1-\text{sgn}(\delta^{e_1}_{m})}{2}-\sum_{m=1}^{n_{e_2}}\text{sgn}(v_{m}^{e_2})\frac{1-\text{sgn}(\delta^{e_2}_{m})}{2}\nonumber\\
&-\sum_{m=1}^{n_{e_3}}\text{sgn}(v_{m}^{e_3})\frac{1-\text{sgn}(\delta^{e_3}_{m})}{2}-\sum_{m=1}^{n_{e_4}}\text{sgn}(v_{m}^{e_4})\frac{1-\text{sgn}(\delta^{e_4}_{m})}{2}+\sum_{m=1}^{n_{f_1}}\text{sgn}(v_{m}^{f_1})\frac{1-\text{sgn}(\delta^{f_1}_{m})}{2}+\sum_{m=1}^{n_{f_2}}\text{sgn}(v_{m}^{f_2})\frac{1-\text{sgn}(\delta^{f_2}_{m})}{2}.\\
\mathcal{W}^{\text{gapless}}_{2} &= \sum_{m=1}^{n_{b_1}}\text{sgn}(v_{m}^{b_1})\frac{1-\text{sgn}(\delta^{b_1}_{m})}{2}-2\sum_{m=1}^{n_{d_1}}\text{sgn}(v_{m}^{d_1})\frac{1-\text{sgn}(\delta^{d_1}_{m})}{2}-\sum_{m=1}^{n_{f_1}}\text{sgn}(v_{m}^{f_1})\frac{1-\text{sgn}(\delta^{f_1}_{m})}{2}-\sum_{m=1}^{n_{f_2}}\text{sgn}(v_{m}^{f_2})\frac{1-\text{sgn}(\delta^{f_2}_{m})}{2}\nonumber\\
&+\sum_{m=1}^{n_{g_1}}\text{sgn}(v_{m}^{g_1})\frac{1-\text{sgn}(\delta^{g_1}_{m})}{2}+\sum_{m=1}^{n_{g_2}}\text{sgn}(v_{m}^{g_2})\frac{1-\text{sgn}(\delta^{g_2}_{m})}{2}+\sum_{m=1}^{n_{g_3}}\text{sgn}(v_{m}^{g_3})\frac{1-\text{sgn}(\delta^{g_3}_{m})}{2}+\sum_{m=1}^{n_{g_4}}\text{sgn}(v_{m}^{g_4})\frac{1-\text{sgn}(\delta^{g_4}_{m})}{2}
\end{align}
The expression of the $\mZ_2$-valued topological invariant  for gapped phase is:
\begin{align}
\mathcal{X}_{1} = \sum_{m=1}^{n_{e_1}}\frac{1-\text{sgn}(\delta^{e_1}_{m})}{2} +\sum_{m=1}^{n_{e_2}}\frac{1-\text{sgn}(\delta^{e_2}_{m})}{2} +\sum_{m=1}^{n_{e_3}}\frac{1-\text{sgn}(\delta^{e_3}_{m})}{2} +\sum_{m=1}^{n_{e_4}}\frac{1-\text{sgn}(\delta^{e_4}_{m})}{2}\quad \text{ mod }\ 2.
\end{align}
The expression of the $\mZ_4$-valued gapped topological invariant is:
\begin{align}
\mathcal{X}_{2}=& -2\sum_{m=1}^{n_{c_1}}\text{sgn}(v_{m}^{c_1})\frac{1-\text{sgn}(\delta^{c_1}_{m})}{2}-\sum_{m=1}^{n_{e_1}}\text{sgn}(v_{m}^{e_1})\frac{1-\text{sgn}(\delta^{e_1}_{m})}{2}+\sum_{m=1}^{n_{e_2}}\text{sgn}(v_{m}^{e_2})\frac{1-\text{sgn}(\delta^{e_2}_{m})}{2}-3\sum_{m=1}^{n_{e_3}}\text{sgn}(v_{m}^{e_3})\frac{1-\text{sgn}(\delta^{e_3}_{m})}{2}\nonumber\\
&+3\sum_{m=1}^{n_{e_4}}\text{sgn}(v_{m}^{e_4})\frac{1-\text{sgn}(\delta^{e_4}_{m})}{2}-\sum_{m=1}^{n_{f_1}}\text{sgn}(v_{m}^{f_1})\frac{1-\text{sgn}(\delta^{f_1}_{m})}{2}+\sum_{m=1}^{n_{f_2}}\text{sgn}(v_{m}^{f_2})\frac{1-\text{sgn}(\delta^{f_2}_{m})}{2}\quad \text{ mod }\ 4.
\end{align}
The expression of the $\mZ_8$-valued topological invariant for gapped phase is:
\begin{align}\label{eq:SMZ8}
\mathcal{X}_{3} &= 2\sum_{m=1}^{n_{c_1}}\text{sgn}(v_{m}^{c_1})\frac{1-\text{sgn}(\delta^{c_1}_{m})}{2}-2\sum_{m=1}^{n_{d_1}}\text{sgn}(v_{m}^{d_1})\frac{1-\text{sgn}(\delta^{d_1}_{m})}{2}+\sum_{m=1}^{n_{e_1}}\text{sgn}(v_{m}^{e_1})\frac{1-\text{sgn}(\delta^{e_1}_{m})}{2}-\sum_{m=1}^{n_{e_2}}\text{sgn}(v_{m}^{e_2})\frac{1-\text{sgn}(\delta^{e_2}_{m})}{2}\nonumber\\
&+3\sum_{m=1}^{n_{e_3}}\text{sgn}(v_{m}^{e_3})\frac{1-\text{sgn}(\delta^{e_3}_{m})}{2}-3\sum_{m=1}^{n_{e_4}}\text{sgn}(v_{m}^{e_4})\frac{1-\text{sgn}(\delta^{e_4}_{m})}{2}-2\sum_{m=1}^{n_{f_1}}\text{sgn}(v_{m}^{f_1})\frac{1-\text{sgn}(\delta^{f_1}_{m})}{2}+2\sum_{m=1}^{n_{f_2}}\text{sgn}(v_{m}^{f_2})\frac{1-\text{sgn}(\delta^{f_2}_{m})}{2}\nonumber\\
&+\sum_{m=1}^{n_{g_1}}\text{sgn}(v_{m}^{g_1})\frac{1-\text{sgn}(\delta^{g_1}_{m})}{2}-\sum_{m=1}^{n_{g_2}}\text{sgn}(v_{m}^{g_2})\frac{1-\text{sgn}(\delta^{g_2}_{m})}{2}+3\sum_{m=1}^{n_{g_3}}\text{sgn}(v_{m}^{g_3})\frac{1-\text{sgn}(\delta^{g_3}_{m})}{2}-3\sum_{m=1}^{n_{g_4}}\text{sgn}(v_{m}^{g_4})\frac{1-\text{sgn}(\delta^{g_4}_{m})}{2} \text{ mod }\ 8.
\end{align}
The line segments and representations in the above expressions are summarized in Table.~\ref{tab:76celldec}.
\begin{table}[htbp]
    \caption{\label{tab:76celldec} The decomposition of the Brillouin zone for SG $P4_1$.
    }
    \centering
    \renewcommand\arraystretch{1.8}
\begin{tabular}{p{3.5cm}<{\centering}|p{4.5cm}<{\centering}||p{3.5cm}<{\centering}|p{4.5cm}<{\centering}}
\hline\hline
 \multicolumn{1}{c|}{line segment} & \multicolumn{1}{c||}{character of irrep (generators only)}&\multicolumn{1}{c|}{line segment} & \multicolumn{1}{c}{character of irrep (generators only)} \\
\hline
$a\ \ \left(\pi t,0,0\right)$ & $
\begin{array}{cc}
 \text{} & \left\{E\vert 000
\right\} \\
 a_1 & 1 \\
\end{array}
$ &
$b\ \ \left(\pi ,\pi t,0\right)$ & $
\begin{array}{cc}
 \text{} & \left\{E\vert
000
\right\} \\
 b_1 & 1 \\
\end{array}
$ \\
\hline
$c\ \ \left(\pi t,0,\pi \right)$ & $
\begin{array}{cc}
 \text{} & \left\{E\vert
000
\right\} \\
 c_1 & 1 \\
\end{array}
$ &
$d\ \ \left(\pi ,\pi t,\pi \right)$ & $
\begin{array}{cc}
 \text{} & \left\{E\vert
000
\right\} \\
 d_1 & 1 \\
\end{array}
$ \\
\hline
$e\ \ \left(0,0,\pi t\right)$ & $
\begin{array}{cc}
 \text{} & \left\{C_{4z}\vert
00\frac{1}{4}
\right\} \\
 e_1 &  e^{-\frac{1}{4} \mathrm{i} \pi  (3+t)} \\
 e_2 &  e^{-\frac{1}{4} \mathrm{i} \pi (-3+t)} \\
 e_3 &  e^{-\frac{1}{4} \mathrm{i} \pi  (1+t)} \\
 e_4 &  e^{-\frac{1}{4} \mathrm{i} \pi  (-1+t)} \\
\end{array}
$ &$g\ \ \left(\pi ,\pi t,\pi t\right)$ & $
\begin{array}{cc}
 \text{} & \left\{C_{4z}\vert
00\frac{1}{4}
\right\} \\
 g_1 &   e^{-\frac{1}{4} \mathrm{i} \pi  (3+t)} \\
 g_2 &   e^{-\frac{1}{4} \mathrm{i} \pi  (-3+t)} \\
 g_3 &   e^{-\frac{1}{4} \mathrm{i} \pi  (1+t)} \\
 g_4 &   e^{-\frac{1}{4} \mathrm{i} \pi  (-1+t)} \\
\end{array}
$ \\
\hline
$f\ \ \left(\pi ,0,\pi t\right)$ & $
\begin{array}{cc}
 \text{} & \left\{C_{2z}\vert
00\frac{1}{2}
\right\} \\
 f_1 & -\mathrm{i} e^{-\frac{1}{2} \mathrm{i} \pi  t} \\
 f_2 & \mathrm{i} e^{-\frac{1}{2} \mathrm{i} \pi  t} \\
\end{array}
$ \\
\hline
\hline
\end{tabular}
\end{table}
\\
Next, we calculate all the gapless and gapped topological invariants for all representative models for these phases.
\subsection{1D TSCs: $(1,0)\in \mZ_2\times\mZ_2$}
As shown in Fig.~\ref{fig:SM_AU}(b), the patchwork is equivalent to a wire construction, which is putting a 1D Kitaev chain on the rotation axis $x=y=1/2$.
In each unit cell, we place four sublattices at positions $({1/2, 1/2, 0})$,$ ({1/2, 1/2, 1/4})$,$ ({1/2, 1/2, 1/2})$, and $ ({1/2, 1/2, 3/4})$.
Then, the matrix form of screw symmetry $\mathcal{S}_4=\{C_{4z}|00\frac{1}{4}\}$ is
\begin{equation}
	U_\bk(\mathcal{S}_4)=\left(
\begin{array}{cccccccc}
 0 & 0 & 0 & 0 & 0 & 0 & e^{\frac{-  i \pi }{4}-i (k_y+k_z)} & 0 \\
 0 & 0 & 0 & 0 & 0 & 0 & 0 & e^{\frac{i \pi }{4}-i (k_y+k_z)} \\
 e^{\frac{- i \pi }{4}-i k_y} & 0 & 0 & 0 & 0 & 0 & 0 & 0 \\
 0 & e^{\frac{i \pi }{4}-i k_y} & 0 & 0 & 0 & 0 & 0 & 0 \\
 0 & 0 & e^{\frac{-  i \pi }{4}-i k_y} & 0 & 0 & 0 & 0 & 0 \\
 0 & 0 & 0 & e^{\frac{i \pi }{4}-i k_y} & 0 & 0 & 0 & 0 \\
 0 & 0 & 0 & 0 & e^{\frac{- i \pi }{4}-i k_y} & 0 & 0 & 0 \\
 0 & 0 & 0 & 0 & 0 & e^{\frac{i \pi }{4}-i k_y} & 0 & 0 \\
\end{array}
\right)
\end{equation}
The corresponding normal state tight-binding Hamiltonian is
\begin{equation}
	h_\bk=\left(
	\renewcommand\arraystretch{1.2}
\begin{array}{cccccccc}
 -\mu  & 0 & -\frac{t}{2} & 0 & 0 & 0 & -\frac{1}{2}t e^{-i k_z}  & 0 \\
 0 & -\mu  & 0 & -\frac{t}{2} & 0 & 0 & 0 & -\frac{1}{2} e^{-i k_z} t \\
 -\frac{t}{2} & 0 & -\mu  & 0 & -\frac{t}{2} & 0 & 0 & 0 \\
 0 & -\frac{t}{2} & 0 & -\mu  & 0 & -\frac{t}{2} & 0 & 0 \\
 0 & 0 & -\frac{t}{2} & 0 & -\mu  & 0 & -\frac{t}{2} & 0 \\
 0 & 0 & 0 & -\frac{t}{2} & 0 & -\mu  & 0 & -\frac{t}{2} \\
 -\frac{1}{2}t e^{i k_z}  & 0 & 0 & 0 & -\frac{t}{2} & 0 & -\mu  & 0 \\
 0 & -\frac{1}{2}t e^{i k_z}  & 0 & 0 & 0 & -\frac{t}{2} & 0 & -\mu  \\
\end{array}
\right),
\end{equation}
and the pairing term is
\begin{equation}
	\Delta_{\bk}=\Delta_{{\text{sc}}}\left(
		\renewcommand\arraystretch{1.2}
\begin{array}{cccccccc}
 0 & 0 & 0 & -\frac{i}{2} & 0 & 0 & 0 & \frac{1}{2} i e^{-i k_z} \\
 0 & 0 & -\frac{i}{2} & 0 & 0 & 0 & \frac{1}{2} i e^{-i k_z} & 0 \\
 0 & \frac{i}{2} & 0 & 0 & 0 & -\frac{i}{2} & 0 & 0 \\
 \frac{i}{2} & 0 & 0 & 0 & -\frac{i}{2} & 0 & 0 & 0 \\
 0 & 0 & 0 & \frac{i}{2} & 0 & 0 & 0 & -\frac{i}{2} \\
 0 & 0 & \frac{i}{2} & 0 & 0 & 0 & -\frac{i}{2} & 0 \\
 0 & -\frac{1}{2} i e^{i k_z} & 0 & 0 & 0 & \frac{i}{2} & 0 & 0 \\
 -\frac{1}{2} i e^{i k_z} & 0 & 0 & 0 & \frac{i}{2} & 0 & 0 & 0 \\
\end{array}
\right)
\end{equation}
\begin{figure}[b]
	\begin{center}
		\includegraphics[width=0.9\columnwidth]{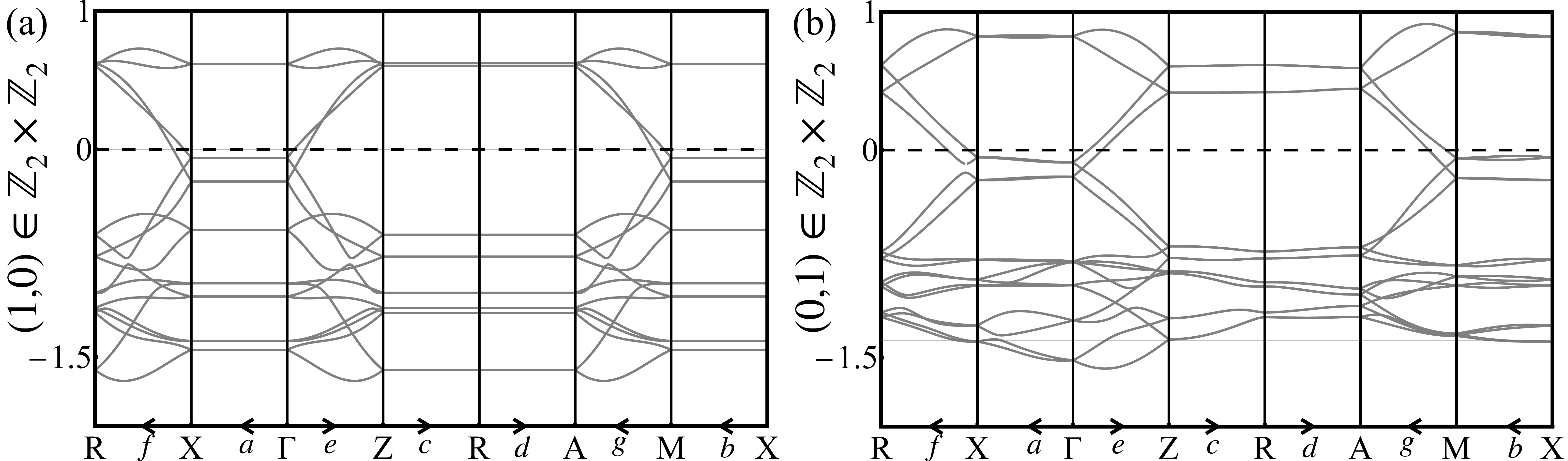}
		\caption{\label{fig:SM_1dTSC}
		 (a) The band structure of patchwork in Fig.~\ref{fig:SM_AU}(b) for the topological superconductor constructed using 1D TSCs. (b) The band structure of patchwork in Fig.~\ref{fig:SM_AU}(c) for the topological superconductor constructed using 1D TSCs.
		 The parameters are set to be $\{t,\mu\}=\{1,0.2\}$.
		}
	\end{center}
\end{figure}

After adding some trivial degrees of freedom and symmetry-allowed perturbations, the corresponding band structure is shown in Fig.~\ref{fig:SM_1dTSC}(a).
From the band structure, we can get that there are Fermi points on the line segments $f,\ e,\ g$:
\begin{itemize}
  \item On the line segment $f$, there are two distinct Fermi points with the same Fermi velocities $\sgn(v_f)=+1$. One Fermi point with irrep $f_1$ has the negative sign of the pairing $\sgn(\delta_\bk)=-1$, while another Fermi point with irrep $f_2$ has the positive sign of the pairing $\sgn(\delta_\bk)=+1$.
  \item On the line segment $e$, there are two distinct Fermi points with the same Fermi velocities $\sgn(v_f)=+1$. One Fermi point with irrep $e_1$ has the negative sign of the pairing $\sgn(\delta_\bk)=-1$, while another Fermi point with irrep $e_3$ has the positive sign of the pairing $\sgn(\delta_\bk)=+1$.
  \item On the line segment $g$, there are two distinct Fermi points with the same Fermi velocities $\sgn(v_f)=+1$. One Fermi point with irrep $g_4$ has the negative sign of the pairing $\sgn(\delta_\bk)=-1$, while another Fermi point with irrep $g_2$ has the positive sign of the pairing $\sgn(\delta_\bk)=+1$.
  \end{itemize}
Substituting into Eqs.~\eqref{eq:SMgapless}-\eqref{eq:SMZ8}, we obtain
\begin{align}
	\mathcal{W}^{\text{gapless}}_1=0,\quad\mathcal{W}^{\text{gapless}}_2=0,\quad\mathcal{X}_1=1,\quad\mathcal{X}_2=2,\quad\mathcal{X}_3=4.
\end{align}

\subsection{1D TSCs: $(0,1)\in \mZ_2\times\mZ_2$}
As shown in Fig.~\ref{fig:SM_AU}(c), the patchwork is equivalent to a wire construction, which is putting a 1D Kitaev chain on the rotation axis $x=y=0$.
In each unit cell, we place four sublattices at positions $({0,0, 0})$,$ ({0, 0, 1/4})$,$ ({0, 0, 3/4})$, and $ ({0, 0, 1/2})$.
Then, the matrix form of screw symmetry $\mathcal{S}_4=\{C_{4z}|00\frac{1}{4}\}$ is
\begin{equation}
	U_\bk(\mathcal{S}_4)=\left(
\begin{array}{cccccccc}
 0 & 0 & 0 & 0 & 0 & 0 &  e^{\frac{-  i \pi }{4}-i k_z} & 0 \\
 0 & 0 & 0 & 0 & 0 & 0 & 0 &   e^{\frac{  i \pi }{4}-i k_z} \\
e^{\frac{-  i \pi }{4} } & 0 & 0 & 0 & 0 & 0 & 0 & 0 \\
 0 & e^{\frac{  i \pi }{4} }& 0 & 0 & 0 & 0 & 0 & 0 \\
 0 & 0 & e^{\frac{-  i \pi }{4} } & 0 & 0 & 0 & 0 & 0 \\
 0 & 0 & 0 & e^{\frac{  i \pi }{4} } & 0 & 0 & 0 & 0 \\
 0 & 0 & 0 & 0 & e^{\frac{-  i \pi }{4} } & 0 & 0 & 0 \\
 0 & 0 & 0 & 0 & 0 & e^{\frac{  i \pi }{4} } & 0 & 0 \\
\end{array}
\right),
\end{equation}
The corresponding normal state tight-binding Hamiltonian is
\begin{equation}
	h_\bk=\left(
	\renewcommand\arraystretch{1.2}
\begin{array}{cccccccc}
 -\mu  & 0 & -\frac{t}{2} & 0 & 0 & 0 & -\frac{1}{2} e^{-i k_z} t & 0 \\
 0 & -\mu  & 0 & -\frac{t}{2} & 0 & 0 & 0 & -\frac{1}{2} e^{-i k_z} t \\
 -\frac{t}{2} & 0 & -\mu  & 0 & -\frac{t}{2} & 0 & 0 & 0 \\
 0 & -\frac{t}{2} & 0 & -\mu  & 0 & -\frac{t}{2} & 0 & 0 \\
 0 & 0 & -\frac{t}{2} & 0 & -\mu  & 0 & -\frac{t}{2} & 0 \\
 0 & 0 & 0 & -\frac{t}{2} & 0 & -\mu  & 0 & -\frac{t}{2} \\
 -\frac{1}{2} e^{i k_z} t & 0 & 0 & 0 & -\frac{t}{2} & 0 & -\mu  & 0 \\
 0 & -\frac{1}{2} e^{i k_z} t & 0 & 0 & 0 & -\frac{t}{2} & 0 & -\mu  \\
\end{array}
\right),
\end{equation}
and the pairing term is
\begin{equation}
	\Delta_{\bk}=\Delta_{{\text{sc}}}\left(
		\renewcommand\arraystretch{1.2}
\begin{array}{cccccccc}
 0 & 0 & 0 & -\frac{i}{2} & 0 & 0 & 0 & \frac{1}{2} i e^{-i k_z} \\
 0 & 0 & -\frac{i}{2} & 0 & 0 & 0 & \frac{1}{2} i e^{-i k_z} & 0 \\
 0 & \frac{i}{2} & 0 & 0 & 0 & -\frac{i}{2} & 0 & 0 \\
 \frac{i}{2} & 0 & 0 & 0 & -\frac{i}{2} & 0 & 0 & 0 \\
 0 & 0 & 0 & \frac{i}{2} & 0 & 0 & 0 & -\frac{i}{2} \\
 0 & 0 & \frac{i}{2} & 0 & 0 & 0 & -\frac{i}{2} & 0 \\
 0 & -\frac{1}{2} i e^{i k_z} & 0 & 0 & 0 & \frac{i}{2} & 0 & 0 \\
 -\frac{1}{2} i e^{i k_z} & 0 & 0 & 0 & \frac{i}{2} & 0 & 0 & 0 \\
\end{array}
\right)
\end{equation}

After adding some trivial degrees of freedom and symmetry-allowed perturbations, the corresponding band structure is shown in Fig.~\ref{fig:SM_1dTSC}(b).
From the band structure, we can get that there are Fermi points on the line segments $f,\ e,\ g$:
\begin{itemize}
  \item On the line segment $f$, there are two distinct Fermi points with the same Fermi velocities $\sgn(v_f)=+1$. One Fermi point with irrep $f_2$ has the negative sign of the pairing $\sgn(\delta_\bk)=-1$, while another Fermi point with irrep $f_1$ has the positive sign of the pairing $\sgn(\delta_\bk)=+1$.
  \item On the line segment $e$, there are two distinct Fermi points with the same Fermi velocities $\sgn(v_f)=+1$. One Fermi point with irrep $e_1$ has the negative sign of the pairing $\sgn(\delta_\bk)=-1$, while another Fermi point with irrep $e_3$ has the positive sign of the pairing $\sgn(\delta_\bk)=+1$.
  \item On the line segment $g$, there are two distinct Fermi points with the same Fermi velocities $\sgn(v_f)=+1$. One Fermi point with irrep $g_1$ has the negative sign of the pairing $\sgn(\delta_\bk)=-1$, and another Fermi point with irrep $g_3$ has the negative sign of the pairing $\sgn(\delta_\bk)=-1$.
  \end{itemize}
Substituting into Eqs.~\eqref{eq:SMgapless}-\eqref{eq:SMZ8}, we obtain
\begin{align}
	\mathcal{W}^{\text{gapless}}_1=0,\quad\mathcal{W}^{\text{gapless}}_2=0,\quad\mathcal{X}_1=1,\quad\mathcal{X}_2=0,\quad\mathcal{X}_3=4.
\end{align}

\subsection{2D TSCs: $(1,0)\in \mZ_2\times\mZ_2$}
As shown in Fig.~\ref{fig:SM_AU}(d), the patchwork is equivalent to the layer construction, which is putting 2D DIII class TSCs on the planes $x=1/2$ and $y=1/2$.
In each unit cell, we place four sublattices at positions $({1/2, 1/2, 0})$,$ ({1/2, 1/2, 1/2})$,$ ({1/2, 1/2, 1/4})$, and $ ({1/2, 1/2, 3/4})$.
Then, the matrix form of screw symmetry $\mathcal{S}_4=\{C_{4z}|00\frac{1}{4}\}$ is
\begin{equation}
	U_\bk(\mathcal{S}_4)=\left(
\begin{array}{cccccccc}
 0 & 0 & 0 & 0 & 0 & 0 &  e^{\frac{-i\pi}{4}-i (k_y+k_z)} & 0 \\
 0 & 0 & 0 & 0 & 0 & 0 & 0 &  e^{\frac{i\pi}{4}-i (k_y+k_z)} \\
 0 & 0 & 0 & 0 &  e^{ \frac{-i\pi}{4}-i k_y} & 0 & 0 & 0 \\
 0 & 0 & 0 & 0 & 0 &   e^{\frac{i\pi}{4}-i k_y} & 0 & 0 \\
 e^{\frac{-i\pi}{4}-i k_y} & 0 & 0 & 0 & 0 & 0 & 0 & 0 \\
 0 &   e^{\frac{i\pi}{4}-i k_y} & 0 & 0 & 0 & 0 & 0 & 0 \\
 0 & 0 &   e^{\frac{-i\pi}{4}-i k_y} & 0 & 0 & 0 & 0 & 0 \\
 0 & 0 & 0 &   e^{\frac{i\pi}{4}-i k_y} & 0 & 0 & 0 & 0 \\
\end{array}
\right)
\end{equation}
The corresponding normal state tight-binding Hamiltonian is
\begin{equation}
	h_\bk=\left(
	\renewcommand\arraystretch{1.2}
\begin{array}{cccc}
 -t \cos k_x-\mu +1 & 0 & -\frac{1}{2} \left(1+e^{-i k_z}\right) t & 0 \\
 0 & -t \cos k_x-\mu +1 & 0 & -\frac{1}{2} \left(1+e^{-i k_z}\right) t \\
 -\frac{1}{2} \left(1+e^{i k_z}\right) t & 0 & -t \cos k_x-\mu +1 & 0 \\
 0 & -\frac{1}{2} \left(1+e^{i k_z}\right) t & 0 & -t \cos k_x-\mu +1 \\
\end{array}
\right),
\end{equation}
and the pairing term is
\begin{equation}
	\Delta_{\bk}=\Delta_{{\text{sc}}}\left(
		\renewcommand\arraystretch{1.2}
\begin{array}{cccc}
 \sin k_x & 0 & 0 & \frac{1}{2} (\sin k_z+i (\cos k_z-1)) \\
 0 & -\sin k_x & \frac{1}{2} (\sin k_z+i (\cos k_z-1)) & 0 \\
 0 & -\frac{1}{2} i \left(-1+e^{i k_z}\right) & \sin k_x & 0 \\
 -\frac{1}{2} i \left(-1+e^{i k_z}\right) & 0 & 0 & -\sin k_x \\
\end{array}
\right)
\end{equation}

\begin{figure}[t]
	\begin{center}
		\includegraphics[width=0.9\columnwidth]{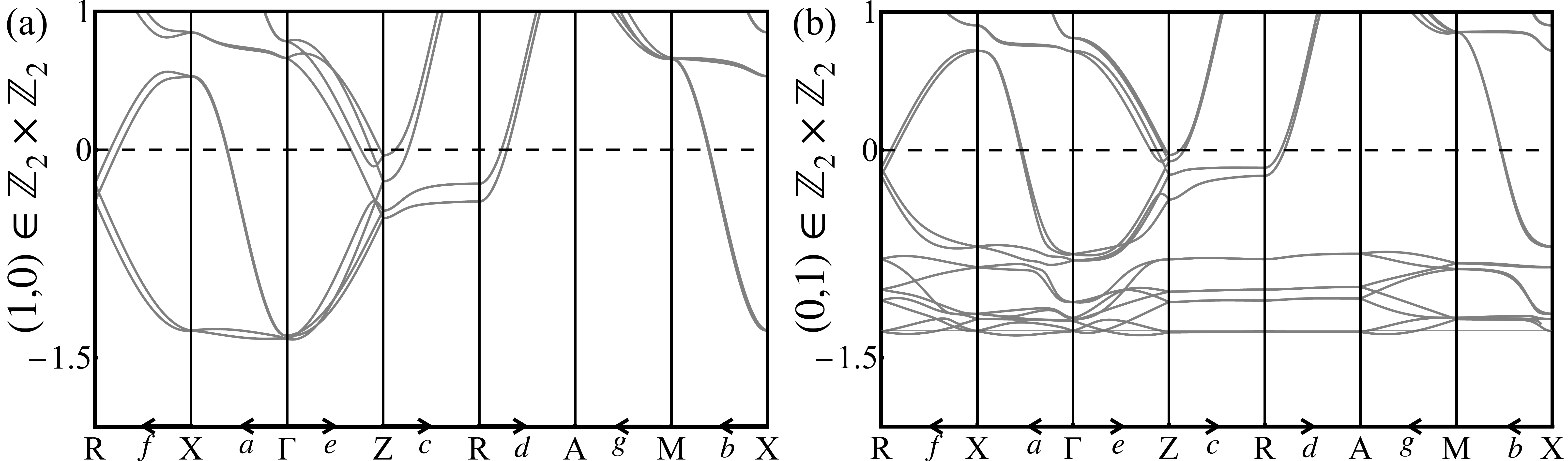}
		\caption{\label{fig:SM_2dTSC}
		 (a) The band structure of patchwork in Fig.~\ref{fig:SM_AU}(d) for the topological superconductor constructed using 2D TSCs. (b) The band structure of patchwork in Fig.~\ref{fig:SM_AU}(e) for the topological superconductor constructed using 2D TSCs.
	The parameters are set to be $\{t ,\mu\}=\{1 ,0.2\}$.
		}
	\end{center}
\end{figure}

After adding some trivial degrees of freedom and symmetry-allowed perturbations, the corresponding band structure is shown in Fig.~\ref{fig:SM_2dTSC}(a).
From the band structure, we can get that there are Fermi points on the line segments $f,\ a,\ e,\ c,\ d,\ b$:
\begin{itemize}
  \item On the line segment $f$, there are two distinct Fermi points with the same Fermi velocities $\sgn(v_f)=-1$. One Fermi point with irrep $f_1$ has the positive sign of the pairing $\sgn(\delta_\bk)=+1$, while another Fermi point with irrep $f_2$ has the negative sign of the pairing $\sgn(\delta_\bk)=-1$.
  \item On the line segment $a$, there are two distinct Fermi points with the same Fermi velocities $\sgn(v_f)=+1$. One Fermi point has the positive sign of the pairing $\sgn(\delta_\bk)=+1$, while another Fermi point has the negative sign of the pairing $\sgn(\delta_\bk)=-1$.
  \item On the line segment $e$, there are four distinct Fermi points with the same Fermi velocities $\sgn(v_f)=-1$. One Fermi point with irrep $e_1$ has the positive sign of the pairing $\sgn(\delta_\bk)=+1$, one Fermi point with irrep $e_2$ has the positive sign of the pairing $\sgn(\delta_\bk)=-1$, one Fermi point with irrep $e_3$ has the positive sign of the pairing $\sgn(\delta_\bk)=-1$, and one Fermi point with irrep $e_4$ has the positive sign of the pairing $\sgn(\delta_\bk)=+1$.
  \item On the line segment $c$, there are two distinct Fermi points with the same Fermi velocities $\sgn(v_f)=+1$. One Fermi point has the positive sign of the pairing $\sgn(\delta_\bk)=+1$, while another Fermi point has the negative sign of the pairing $\sgn(\delta_\bk)=-1$.
  \item On the line segment $d$, there are two distinct Fermi points with the same Fermi velocities $\sgn(v_f)=+1$. One Fermi point has the positive sign of the pairing $\sgn(\delta_\bk)=+1$, while another Fermi point has the negative sign of the pairing $\sgn(\delta_\bk)=-1$.
  \item On the line segment $b$, there are two distinct Fermi points with the same Fermi velocities $\sgn(v_f)=+1$. One Fermi point has the positive sign of the pairing $\sgn(\delta_\bk)=+1$, while another Fermi point has the negative sign of the pairing $\sgn(\delta_\bk)=-1$.
  \end{itemize}
Substituting into Eqs.~\eqref{eq:SMgapless}-\eqref{eq:SMZ8}, we obtain
\begin{align}
	\mathcal{W}^{\text{gapless}}_1=0,\quad\mathcal{W}^{\text{gapless}}_2=0,\quad\mathcal{X}_1=0,\quad\mathcal{X}_2=3,\quad\mathcal{X}_3=4.
\end{align}

\subsection{2D TSCs: $(0,1)\in \mZ_2\times\mZ_2$}
As shown in Fig.~\ref{fig:SM_AU}(f), the patchwork is equivalent to the layer construction, which is putting 2D DIII class TSCs on the planes $x=0$ and $y=0$.
In each unit cell, we place four sublattices at positions $({0,0, 0})$,$ ({0, 0, 1/2})$,$ ({0, 0, 1/4})$, and $ ({0, 0, 3/4})$.
Then, the matrix form of screw symmetry $\mathcal{S}_4=\{C_{4z}|00\frac{1}{4}\}$ is
\begin{equation}
	U_\bk(\mathcal{S}_4)=\left(
\begin{array}{cccccccc}
 0 & 0 & 0 & 0 & 0 & 0 &  e^{\frac{-i\pi}{4}-i k_z} & 0 \\
 0 & 0 & 0 & 0 & 0 & 0 & 0 &  e^{\frac{i\pi}{4}-i k_z} \\
 0 & 0 & 0 & 0 & e^{\frac{-i\pi}{4}} & 0 & 0 & 0 \\
 0 & 0 & 0 & 0 & 0 & e^{\frac{i\pi}{4}} & 0 & 0 \\
e^{\frac{-i\pi}{4}} & 0 & 0 & 0 & 0 & 0 & 0 & 0 \\
 0 & e^{\frac{i\pi}{4}} & 0 & 0 & 0 & 0 & 0 & 0 \\
 0 & 0 & e^{\frac{-i\pi}{4}} & 0 & 0 & 0 & 0 & 0 \\
 0 & 0 & 0 & e^{\frac{i\pi}{4}} & 0 & 0 & 0 & 0 \\
\end{array}
\right)
\end{equation}
The corresponding normal state tight-binding Hamiltonian is

\begin{equation}
	h_\bk=\left(
	\renewcommand\arraystretch{1.2}
\begin{array}{cccccccc}
 -t \cos k_x  & 0 &  h_z t & 0 & 0 & 0 & 0 & 0 \\
 0 & -t \cos k_x  & 0 &  h_z t & 0 & 0 & 0 & 0 \\
  h_z^* t & 0 & -t \cos k_x  & 0 & 0 & 0 & 0 & 0 \\
 0 &  h_z^* t & 0 & -t \cos k_x  & 0 & 0 & 0 & 0 \\
 0 & 0 & 0 & 0 & -t \cos k_y  & 0 &  h_z t & 0 \\
 0 & 0 & 0 & 0 & 0 & -t \cos k_y  & 0 &  h_z t \\
 0 & 0 & 0 & 0 &  h_z^* t & 0 & -t \cos k_y  & 0 \\
 0 & 0 & 0 & 0 & 0 &  h_z^* t & 0 & -t \cos k_y  \\
\end{array}
\right)-(\mu-1)\mathds{1},
\end{equation}
where $h_z=-\frac{1}{2} \left(1+e^{i k_z}\right)$.
The pairing term is
\begin{equation}
	\Delta_{\bk}=\Delta_{{\text{sc}}}\begin{small}\left(
		\renewcommand\arraystretch{1.2}
\begin{array}{cccccccc}
 \sin k_x & 0 & 0 & \frac{i}{2}   (e^{-i k_z}-1) & 0 & 0 & 0 & 0 \\
 0 & -\sin k_x & \frac{i}{2}   (e^{-i k_z}-1) & 0 & 0 & 0 & 0 & 0 \\
 0 & -\frac{i}{2}  \left(e^{i k_z}-1\right) & \sin k_x & 0 & 0 & 0 & 0 & 0 \\
 -\frac{i}{2}  \left(e^{i k_z}-1\right) & 0 & 0 & -\sin k_x & 0 & 0 & 0 & 0 \\
 0 & 0 & 0 & 0 & -i \sin k_y & 0 & 0 & \frac{i}{2}    (e^{-i k_z}-1) \\
 0 & 0 & 0 & 0 & 0 & -i \sin k_y & \frac{i}{2}   (e^{-i k_z}-1) & 0 \\
 0 & 0 & 0 & 0 & 0 & -\frac{i}{2}  \left(e^{i k_z}-1\right) & -i \sin k_y & 0 \\
 0 & 0 & 0 & 0 & -\frac{i}{2}  \left(e^{i k_z}-1\right) & 0 & 0 & -i \sin k_y \\
\end{array}
\right)\end{small}
\end{equation}

After adding some trivial degrees of freedom and symmetry-allowed perturbations, the corresponding band structure is shown in Fig.~\ref{fig:SM_2dTSC}(a).
From the band structure, we can get that there are Fermi points on the line segments $f,\ a,\ e,\ c,\ d,\ b$:
\begin{itemize}
  \item On the line segment $f$, there are two distinct Fermi points with the same Fermi velocities $\sgn(v_f)=-1$. One Fermi point with irrep $f_1$ has the negative sign of the pairing $\sgn(\delta_\bk)=-1$, while another Fermi point with irrep $f_2$ has the positive sign of the pairing $\sgn(\delta_\bk)=+1$.
  \item On the line segment $a$, there are two distinct Fermi points with the same Fermi velocities $\sgn(v_f)=+1$. One Fermi point has the positive sign of the pairing $\sgn(\delta_\bk)=+1$, while another Fermi point has the negative sign of the pairing $\sgn(\delta_\bk)=-1$.
  \item On the line segment $e$, there are four distinct Fermi points with the same Fermi velocities $\sgn(v_f)=-1$. One Fermi point with irrep $e_1$ has the positive sign of the pairing $\sgn(\delta_\bk)=+1$, one Fermi point with irrep $e_2$ has the positive sign of the pairing $\sgn(\delta_\bk)=-1$, one Fermi point with irrep $e_3$ has the positive sign of the pairing $\sgn(\delta_\bk)=-1$, and one Fermi point with irrep $e_4$ has the positive sign of the pairing $\sgn(\delta_\bk)=+1$.
  \item On the line segment $c$, there are two distinct Fermi points with the same Fermi velocities $\sgn(v_f)=+1$. One Fermi point has the positive sign of the pairing $\sgn(\delta_\bk)=+1$, while another Fermi point has the negative sign of the pairing $\sgn(\delta_\bk)=-1$.
  \item On the line segment $d$, there are two distinct Fermi points with the same Fermi velocities $\sgn(v_f)=+1$. One Fermi point has the positive sign of the pairing $\sgn(\delta_\bk)=+1$, while another Fermi point has the negative sign of the pairing $\sgn(\delta_\bk)=-1$.
  \item On the line segment $b$, there are two distinct Fermi points with the same Fermi velocities $\sgn(v_f)=+1$. One Fermi point has the positive sign of the pairing $\sgn(\delta_\bk)=+1$, while another Fermi point has the negative sign of the pairing $\sgn(\delta_\bk)=-1$.
  \end{itemize}
Substituting into Eqs.~\eqref{eq:SMgapless}-\eqref{eq:SMZ8}, we obtain
\begin{align}
	\mathcal{W}^{\text{gapless}}_1=0,\quad\mathcal{W}^{\text{gapless}}_2=0,\quad\mathcal{X}_1=0,\quad\mathcal{X}_2=1,\quad\mathcal{X}_3=0.
\end{align}

\subsection{3D TSC: $1\in \mZ$}
As shown in main text, in each unit cell, we place four sublattices at positions $({0,0, 0})$,$ ({0, 0, 1/2})$,$ ({0, 0, 1/4})$, and $ ({0, 0, 3/4})$.
Then, the matrix form of screw symmetry $\mathcal{S}_4=\{C_{4z}|00\frac{1}{4}\}$ is
\begin{equation}
	U_\bk(\mathcal{S}_4)=\left(
\begin{array}{cccccccc}
 0 & 0 & 0 & 0 &   e^{\frac{-i\pi}{4}-i k_z} & 0 & 0 & 0 \\
 0 & 0 & 0 & 0 & 0 &   e^{\frac{i\pi}{4}-i k_z} & 0 & 0 \\
 e^{\frac{-i\pi}{4}} & 0 & 0 & 0 & 0 & 0 & 0 & 0 \\
 0 & e^{\frac{i\pi}{4}} & 0 & 0 & 0 & 0 & 0 & 0 \\
 0 & 0 & 0 & 0 & 0 & 0 &  e^{\frac{-i\pi}{4}} & 0 \\
 0 & 0 & 0 & 0 & 0 & 0 & 0 & e^{\frac{i\pi}{4}} \\
 0 & 0 &  e^{\frac{-i\pi}{4}} & 0 & 0 & 0 & 0 & 0 \\
 0 & 0 & 0 & e^{\frac{i\pi}{4}} & 0 & 0 & 0 & 0 \\
\end{array}
\right),
\end{equation}
The corresponding normal state tight-binding Hamiltonian in momentum space is

\begin{equation}
	h_\bk=\left(
	\renewcommand\arraystretch{1.2}
		\begin{array}{cccccccc}
h_x+h_y  & 0 & -\frac{t}{2} & 0 & -\frac{1}{2} e^{-i k_z} t & 0 & 0 & 0 \\
 0 &h_x+h_y  & 0 & -\frac{t}{2} & 0 & -\frac{1}{2} e^{-i k_z} t & 0 & 0 \\
 -\frac{t}{2} & 0 &h_x+h_y  & 0 & 0 & 0 & -\frac{t}{2} & 0 \\
 0 & -\frac{t}{2} & 0 &h_x+h_y  & 0 & 0 & 0 & -\frac{t}{2} \\
 -\frac{1}{2} e^{i k_z} t & 0 & 0 & 0 &h_x+h_y  & 0 & -\frac{t}{2} & 0 \\
 0 & -\frac{1}{2} e^{i k_z} t & 0 & 0 & 0 &h_x+h_y  & 0 & -\frac{t}{2} \\
 0 & 0 & -\frac{t}{2} & 0 & -\frac{t}{2} & 0 &h_x+h_y  & 0 \\
 0 & 0 & 0 & -\frac{t}{2} & 0 & -\frac{t}{2} & 0 &h_x+h_y  \\
\end{array}
\right)-\mu\mathds{1},
\end{equation}
where $h_x=-t\cos k_x$ and  $h_y=-t\cos k_y$.
The pairing term is
\begin{equation}
	\Delta_{\bk}=\Delta_{{\text{sc}}}\begin{small}\left(
		\renewcommand\arraystretch{1.2}
\begin{array}{cccccccc}
 -\Delta^*_{xy} & 0 & 0 & -\frac{i}{2} & 0 & \frac{1}{2} e^{-ik_z} & 0 & 0 \\
 0 & \Delta_{xy} & -\frac{i}{2} & 0 & \frac{1}{2} e^{-ik_z} & 0 & 0 & 0 \\
 0 & \frac{i}{2} & -\Delta^*_{xy} & 0 & 0 & 0 & 0 & -\frac{i}{2} \\
 \frac{i}{2} & 0 & 0 & \Delta_{xy} & 0 & 0 & -\frac{i}{2} & 0 \\
 0 & -\frac{1}{2} i e^{i k_z} & 0 & 0 & -\Delta^*_{xy} & 0 & 0 & \frac{i}{2} \\
 -\frac{1}{2} i e^{i k_z} & 0 & 0 & 0 & 0 & \Delta_{xy} & \frac{i}{2} & 0 \\
 0 & 0 & 0 & \frac{i}{2} & 0 & -\frac{i}{2} & -\Delta^*_{xy} & 0 \\
 0 & 0 & \frac{i}{2} & 0 & -\frac{i}{2} & 0 & 0 & \Delta_{xy} \\
\end{array}
\right)\end{small}
\end{equation}
where $\Delta_{xy}=\sin k_x+i\sin k_y$.
\begin{figure}[t]
	\begin{center}
		\includegraphics[width=0.5\columnwidth]{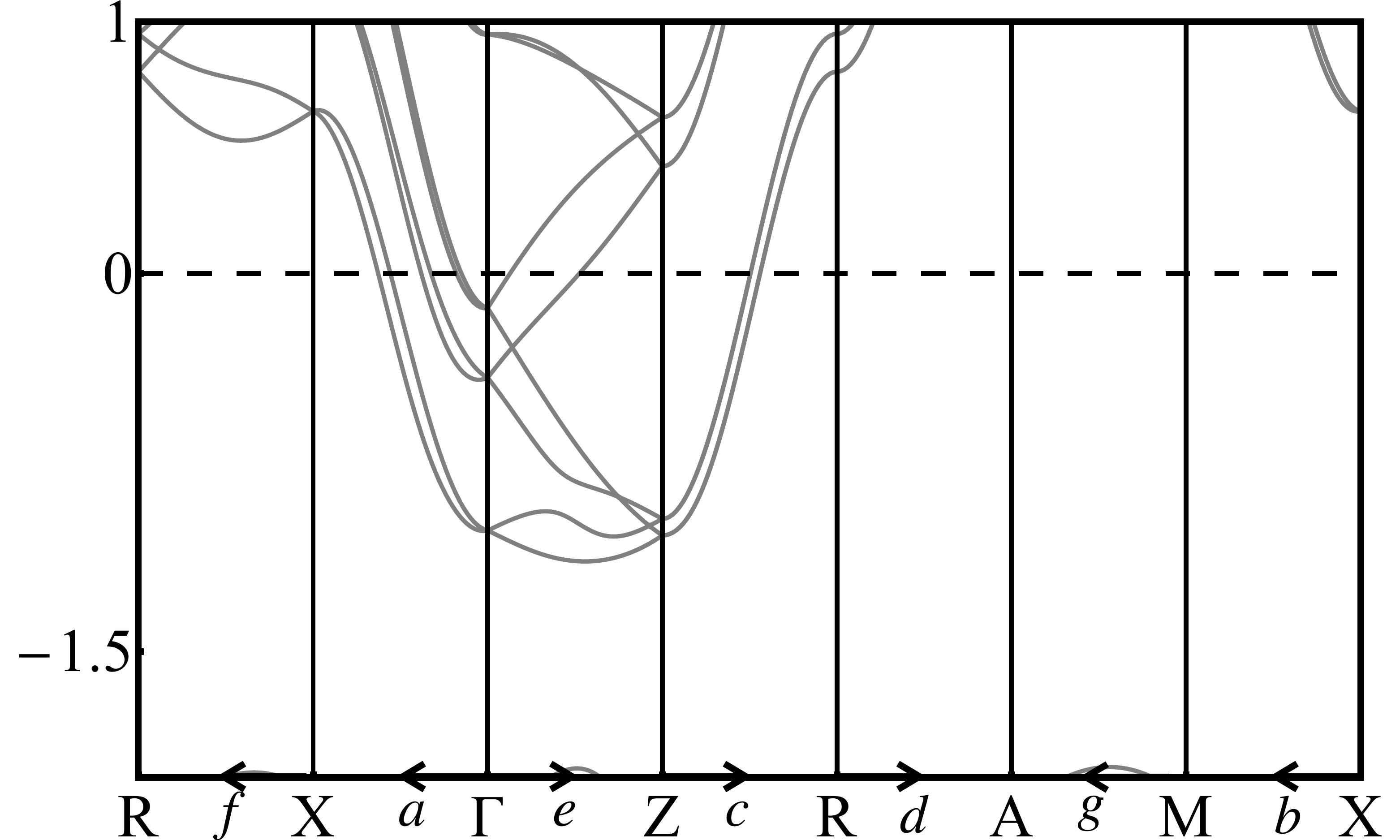}
		\caption{\label{fig:SM_3dTSC}
		  The band structure of the topological superconductor with nonzero three dimensional winding number.
	The parameters are set to be $\{t,\mu\}=\{1,1.8\}$.
		}
	\end{center}
\end{figure}
After adding some trivial degrees of freedom and symmetry-allowed perturbations, the corresponding band structure is shown in Fig.~\ref{fig:SM_3dTSC}.
From the band structure, we can get that there are Fermi points on the line segments $ a,\ e,\ c$:
\begin{itemize}
  \item On the line segment $a$, there are six distinct Fermi points with the same Fermi velocities $\sgn(v_f)=+1$. Three Fermi points have the positive sign of the pairing $\sgn(\delta_\bk)=+1$, while another three Fermi points have the negative sign of the pairing $\sgn(\delta_\bk)=-1$.
  \item On the line segment $e$, there are two distinct Fermi points with the same Fermi velocities $\sgn(v_f)=+1$. One Fermi point with irrep $e_1$ has the negative sign of the pairing $\sgn(\delta_\bk)=11$, while one Fermi point with irrep $e_3$ has the positive sign of the pairing $\sgn(\delta_\bk)=-1$.
  \item On the line segment $c$, there are two distinct Fermi points with the same Fermi velocities $\sgn(v_f)=+1$. One Fermi point has the positive sign of the pairing $\sgn(\delta_\bk)=+1$, while another Fermi point has the negative sign of the pairing $\sgn(\delta_\bk)=-1$.
  \end{itemize}
Substituting into Eqs.~\eqref{eq:SMgapless}-\eqref{eq:SMZ8}, we obtain
\begin{align}
\mathcal{W}^{\text{gapless}}_1=0,\quad\mathcal{W}^{\text{gapless}}_2=0,\quad\mathcal{X}_1=1,\quad\mathcal{X}_2=1,\quad\mathcal{X}_3=3.
\end{align}

\section{Gapless and gapped topological invariants for all representative models in SG $Pn_1$}

\subsection{Gapless and gapped topological invariants for all representative models in SG $P2_1$}
In this section, we provide all the gapless and gapped topological invariants for all representative models of the real-space construction in SG $P2_1$.
In addition to the strong TSC, we have $(\mZ_2)^3$ phases constructed from one-dimensional TSCs and $(\mZ_2)^3$ phases constructed from two-dimensional TSCs.

The patchwork for those phases constructed from one-dimensional TSCs are equivalent to the wire constructions, which are
\begin{itemize}
	\item placing the 1D DIII class TSCs along lines $x=z=0$, $y=z=0$, $y-0.5=z=0$, $x=z-0.5=0$.
	\item placing the 1D DIII class TSC along line $x=z=0$.
	\item placing the 1D DIII class TSC along line $x=z-0.5=0$.
\end{itemize}
The patchwork for those phases constructed from two-dimensional TSCs are equivalent to the layer constructions, which are
\begin{itemize}
	\item placing the 2D DIII class TSC on the plane $x=0$.
	\item placing the 2D DIII class TSC on the plane $ z=0$.
	\item placing the 2D DIII class TSC on the plane $ z-0.5=0$.
\end{itemize}

As shown in the Section.~\textcolor{blue}{S4} of the Supplemental Materials II, there are three $\mZ$-valued gapless invariants $\mathcal{W}_1^{\text{gapless}},\mathcal{W}_2^{\text{gapless}},\mathcal{W}_3^{\text{gapless}}$, one $\mZ_2$-valued gapped invariant $\mathcal{X}_1$ and three $\mZ_4$-valued gapped invariants $\mathcal{X}_2,\mathcal{X}_3,\mathcal{X}_4$:
The quantitative mappings between all representative models of real-space construction and the topological invariants $(\mathcal{W}_1^{\text{gapless}},\mathcal{W}_2^{\text{gapless}},\mathcal{W}_3^{\text{gapless}}, \mathcal{X}_1,\mathcal{X}_2,\mathcal{X}_3,\mathcal{X}_4)$ computed by our Fermi surface formula are shown in the Table.~\ref{tab:SMp21}.

We find that $w_{\text{3D}} \equiv {\cal X}_2+{\cal X}_3+{\cal X}_4 \bmod 2$ holds. Here, we derive this $\mZ_2$-valued indicator.
Any gapped phase can be expressed by a direct some of the above six patchworks and a strong TSC, which is represented by
\begin{align}
	&(\text{1D TSCs  (a)})^{\oplus N_{\text{1D}, a}} \oplus (\text{1D TSCs  (b)})^{\oplus N_{\text{1D}, b}} \oplus (\text{1D TSCs  (c)})^{\oplus N_{\text{1D}, c}} \nonumber \\
	&\oplus (\text{2D TSCs  (a)})^{\oplus N_{\text{2D}, a}} \oplus (\text{2D TSCs  (b)})^{\oplus N_{\text{2D}, b}} \oplus (\text{1D TSCs  (c)})^{\oplus N_{\text{2D}, c}} \nonumber \\
	&\oplus (\text{3D TSC})^{\oplus N_{\text{3D}}}.
\end{align}
From Table~\ref{tab:SMp21}, we find a set of topological invariants is given by
\begin{align}
	&\mathcal{X}_1 = N_{\text{1D}, b} +  N_{\text{1D}, c} + N_{\text{2D}, a} + N_{\text{2D}, b} + N_{\text{2D}, c}  + N_{\text{3D}} \mod 2; \\
	 &\mathcal{X}_2 = 2(N_{\text{1D}, a} + N_{\text{1D}, b} +  N_{\text{1D}, c}) + 2N_{\text{2D}, a} + 3N_{\text{2D}, b} + 3N_{\text{2D}, c}  + 2N_{\text{3D}} \mod 4;\\
	 &\mathcal{X}_3 = 2(N_{\text{1D}, a} +  N_{\text{1D}, c}) + 3(N_{\text{2D}, a} + N_{\text{2D}, b} + N_{\text{2D}, c})  + 3N_{\text{3D}} \mod 4;\\
	 &\mathcal{X}_4 = 2N_{\text{1D}, b} + 3N_{\text{2D}, a} + 2N_{\text{2D}, b} + 2N_{\text{3D}} \mod 4;\\
	 &w_{\text{3D}} = N_{\text{3D}}.
\end{align}
We compute 
\begin{align}
    \mathcal{X}_2 + \mathcal{X}_3 + \mathcal{X}_4 &= 2N_{\text{2D}, c}  + 3N_{\text{3D}} \mod 4 \nonumber \\
    &\equiv N_{\text{3D}} \mod 2. 
\end{align}
Therefore, we conclude $w_{\text{3D}} \equiv {\cal X}_2+{\cal X}_3+{\cal X}_4 \bmod 2$.

\begin{table}[H]
    \caption{\label{tab:SMp21}
    The quantitative mappings between all representative models and the topological invariants in SG $P2_1$.
    }
    \centering
    \renewcommand\arraystretch{1.3}
    \begin{tabular}{c|c|c|c}
    \hline\hline
	Classification in Ref.~\cite{Ono-Shiozaki-Watanabe2022} & $w_{\text{3D}}$ & $(\mathcal{W}^{\text{gapless}}_1,\mathcal{W}^{\text{gapless}}_2,\mathcal{W}^{\text{gapless}}_3)$& $(\mathcal{X}_1,\mathcal{X}_2,\mathcal{X}_3,\mathcal{X}_4)$\\ \hline
	1D TSCs (a): $(1, 0, 0) \in \mZ_2 \times \mZ_2\times \mZ_2$ & $0$ & $(0,0,0)$ &$( 0,2,2,0)$\\ \hline
	1D TSCs (b): $(0, 1, 0) \in \mZ_2 \times \mZ_2\times \mZ_2$ & $0$ & $(0,0,0)$ &$( 1,2,0,2)$\\ \hline
	1D TSCs (c): $(0, 0, 1) \in \mZ_2 \times \mZ_2\times \mZ_2$ & $0$ & $(0,0,0)$ &$( 1,2,2,0)$\\ \hline
	2D TSCs (a): $(1, 0, 0) \in \mZ_2 \times \mZ_2\times \mZ_2$ & $0$ & $(0,0,0)$ &$( 1,2,3,3)$\\ \hline
	2D TSCs (b): $(0, 1, 0) \in \mZ_2 \times \mZ_2\times \mZ_2$ & $0$ & $(0,0,0)$ &$( 1,3,3,2)$\\ \hline
	2D TSCs (c): $(0, 0, 1) \in \mZ_2 \times \mZ_2\times \mZ_2$ & $0$ & $(0,0,0)$ &$( 1,3,3,0)$\\ \hline
	Strong TSC: $1 \in \mZ$ & $1$ & $(0,0,0)$ &$(1,2,3,2)$\\ \hline\hline
	\end{tabular}
    \end{table}

\subsection{Gapless and gapped topological invariants for all representative models in SG $P3_1$}
In this section, we provide all the gapless and gapped topological invariants for all representative models of the real-space construction in SG $P3_1$.
In addition to the strong TSC, we have $\mZ_2$ phase constructed from one-dimensional TSCs and $\mZ_2$ phase constructed from two-dimensional TSCs.
The patchwork for this $\mZ_2$ phase constructed from one-dimensional TSCs is equivalent to the wire constructions, which is placing the 1D DIII class TSC along line $x=y=0$.
The patchwork for this $\mZ_2$ phase constructed from two-dimensional TSCs is equivalent to the layer construction, which is placing the 2D DIII class TSC on the planes $z=0$, $z=1/3$, and $z=2/3$.

As shown in the Section.~\textcolor{blue}{S144} of the Supplemental Materials II, there are one $\mZ$-valued gapless invariant $\mathcal{W}_1^{\text{gapless}}$, two $\mZ_2$-valued gapped invariants $\mathcal{X}_1$, $\mathcal{X}_2$ and one $\mZ_6$-valued gapped invariant $\mathcal{X}_3$:
The quantitative mappings between all representative models of real-space construction and the topological invariants $(\mathcal{W}_1^{\text{gapless}}, \mathcal{X}_1,\mathcal{X}_2,\mathcal{X}_3)$ computed by our Fermi surface formula are shown in the Table.~\ref{tab:SMp31}.

In the same way to the case of $P2_1$, we show that there exists a $\mZ_6$-valued indicator of $w_{\text{3D}}$.
Since 1D, 2D, and strong 3D TSCs listed in Table~\ref{tab:SMp31} span all possible gapped phases, any gapped state can be expressed as a direct sum of them as $({\rm 1D\ TSC})^{\oplus N_{\rm 1D}} \oplus ({\rm 2D\ TSC})^{\oplus N_{\rm 2D}} \oplus ({\rm 3D\ TSC})^{\oplus N_{\rm 3D}}$, whose invariants are 
\begin{align}
    {\cal X}_1 \equiv N_{\rm 2D} \bmod 2, \quad 
    {\cal X}_2 \equiv N_{\rm 1D} + N_{\rm 2D} \bmod 2, \quad 
    {\cal X}_3 \equiv 3N_{\rm 1D} + N_{\rm 3D} \bmod 6, \quad 
    w_{\text{3D}} = N_{\rm 3D}. 
\end{align}
We find that $w_{\text{3D}} \equiv 3{\cal X}_1 + 3{\cal X}_2 + {\cal X}_3 \bmod 6$ holds.


\begin{table}[H]
    \caption{\label{tab:SMp31}
    The quantitative mappings between all representative models and the topological invariants in SG $P3_1$.
    }
    \centering
    \renewcommand\arraystretch{1.3}
    \begin{tabular}{c|c|c|c}
    \hline\hline
	Classification in Ref.~\cite{Ono-Shiozaki-Watanabe2022} & $w_{\text{3D}}$ & $\mathcal{W}^{\text{gapless}}_1$& $(\mathcal{X}_1,\mathcal{X}_2,\mathcal{X}_3)$\\ \hline
	1D TSCs: $1 \in \mZ_2$ & $0$ & $0$ &$( 0,1,3)$\\ \hline
	2D TSCs: $1 \in \mZ_2$ & $0$ & $0$ &$( 1,1,0)$\\ \hline
	Strong TSC: $1 \in \mZ$ & $1$ & $0$ &$(0,0,1)$\\ \hline\hline
	\end{tabular}
    \end{table}

\subsection{Gapless and gapped topological invariants for all representative models in SG $P6_1$}
In this section, we provide all the gapless and gapped topological invariants for all representative models of the real-space construction in SG $P6_1$.
In addition to the strong TSC, we have $\mZ_2$ phase constructed from one-dimensional TSCs and $\mZ_2$ phase constructed from two-dimensional TSCs.
The patchwork for this $\mZ_2$ phase constructed from one-dimensional TSCs is equivalent to the wire constructions, which is placing the 1D DIII class TSC along line $x=y=0$.
The patchwork for this $\mZ_2$ phase constructed from two-dimensional TSCs is equivalent to the layer construction, which is placing the 2D DIII class TSC on the planes $z=0$, $z=1/6$, $z=1/3$, $z=1/2$, $z=2/3$ and $z=5/6$.

As shown in the Section.~\textcolor{blue}{S169} of the Supplemental Materials II, there are two $\mZ$-valued gapless invariants $\mathcal{W}_1^{\text{gapless}},\mathcal{W}_2^{\text{gapless}}$, one $\mZ_2$-valued gapped invariant $\mathcal{X}_1$ and one $\mZ_{12}$-valued gapped invariant $\mathcal{X}_2$:
The quantitative mappings between all representative models of real-space construction and the topological invariants $(\mathcal{W}_1^{\text{gapless}},\mathcal{W}_2^{\text{gapless}}, \mathcal{X}_1,\mathcal{X}_2)$ computed by our Fermi surface formula are shown in the Table.~\ref{tab:SMp61}.
Similar to the case of $P4_1$, $\mathcal{X}_2 \mod 6$ is a $\mZ_6$-valued indicator of $w_{\text{3D}}$. 
From Table~\ref{tab:SMp61}, we find $ w_{\text{3D}} \equiv -{\cal X}_2 \bmod 6$.

\begin{table}[H]
    \caption{\label{tab:SMp61}
    The quantitative mappings between all representative models and the topological invariants in SG $P6_1$.
    }
    \centering
    \renewcommand\arraystretch{1.3}
    \begin{tabular}{c|c|c|c}
    \hline\hline
	Classification in Ref.~\cite{Ono-Shiozaki-Watanabe2022} & $w_{\text{3D}}$ & $(\mathcal{W}_1^{\text{gapless}},\mathcal{W}_2^{\text{gapless}})$& $(\mathcal{X}_1,\mathcal{X}_2)$\\ \hline
	1D TSCs: $1 \in \mZ_2$& $0$ & $(0,0)$  &$(1,6)$\\ \hline
	2D TSCs: $1 \in \mZ_2$& $0$ & $(0,0)$  &$( 0,6)$\\ \hline
	Strong TSC: $1 \in \mZ$ & $1$& $(0,0)$  &$(0,5)$\\ \hline\hline
	\end{tabular}
    \end{table}
